%% file: burst-box-sigproc.tex
\documentclass[journal]{IEEEtran}

\usepackage{cite}
\usepackage{amsmath}
\usepackage{array}
\usepackage[caption=false,font=footnotesize]{subfig}
\usepackage{url}

\ifCLASSINFOpdf
\usepackage[pdftex]{graphicx}
\else
\usepackage[dvips]{graphicx}
\fi

\newif\ifshowfixmes
\showfixmesfalse

\newif\ifshowoutline
\showoutlinefalse

\input{burst-box-sigproc-macros-fix}
\input{burst-box-sigproc-macros-cite}

\begin{document}
%
%
%
\title{A Scalable Real-Time Architecture for Neural Oscillation Detection
and Phase-Specific Stimulation}
%
%
%
\author{%
Christopher~Thomas,~\IEEEmembership{Member,~IEEE,}
and~Thilo~Womelsdorf%
\thanks{C. Thomas and T. Womelsdorf are with the Department of Psychology,
Vanderbilt University, Nashville, TN.}%
\thanks{Manuscript received \textbf{(date)}; revised \textbf{(date)}.}%
\thanks{This work has been submitted to the IEEE for possible publication.
Copyright may be transferred without notice, after which this version
may no longer be accessible.}
}
\fixmenote{Manuscript ``received'' and ``revised'' dates.}
\fixmenote{Remove IEEE arxiv boilerplate when submitting. May have to remove
the preprint from arxiv too, and/or add new boilerplate, depending.}
%
%
%
%
\markboth{IEEE Transactions on Signal Processing,~\textbf{volume/number/date}}%
{\textbf{Author and title here.}}
\fixmenote{Volume/number/date info, author and title info. Do I add this or
does IEEE?}
%
%
\maketitle
%
%
%
\begin{abstract}
\input{burst-box-sigproc-abstract}
\end{abstract}
%
%
%
%
%
%
%
%
%
\ifCLASSOPTIONpeerreview
%
\begin{center} \bfseries EDICS Category: BIO-BCI \end{center}
\fi
%
\IEEEpeerreviewmaketitle
%
%
%
\input{burst-box-sigproc-intro}
\input{burst-box-sigproc-background}
\input{burst-box-sigproc-design}
\input{burst-box-sigproc-validation}
\input{burst-box-sigproc-conc}
%
%
%
\bibliographystyle{IEEEtran}
\bibliography{burst-box-sigproc-refs}
%
%
%
%
\begin{IEEEbiography}
[{\includegraphics[width=1in,height=1.25in,clip,keepaspectratio]
{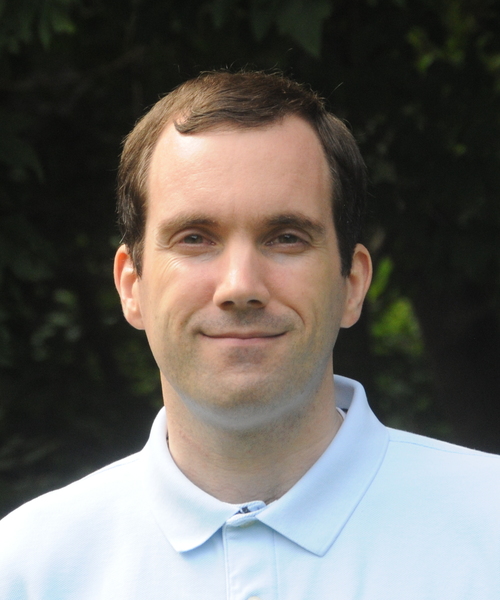}}]
{Christopher Thomas}
Christopher Thomas is a research scientist with the Attention Circuits
Control Laboratory in the Department of Psychology at Vanderbilt University,
where he specializes in signal processing and embedded systems. He received
his PhD from York University in Toronto, Canada for his work on image sensors.
\end{IEEEbiography}
\begin{IEEEbiography}
[{\includegraphics[width=1in,height=1.25in,clip,keepaspectratio]
{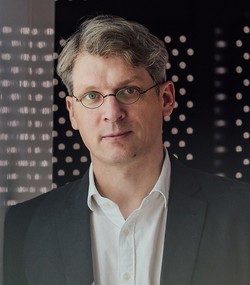}}]
{Thilo Womelsdorf}
Thilo Womelsdorf is Associate Professor in the Departments of Psychology
and Biomedical Engineering at Vanderbilt University, where he leads the
Attention Circuits Control Laboratory.
His research investigates how neural circuits learn and control attentional
allocation in non-human primates and humans.
Before arriving at Vanderbilt he led a systems neuroscience lab in Toronto
(York University), receiving in 2017 the prestigious E.W.R. Steacie Memorial
Fellowship for his work bridging the cell- and network- levels of
understanding how brain activity dynamics relate to behavior.
\end{IEEEbiography}
%
%
%
\ifshowfixmes
\onecolumn
\listfixcites
\listfixmes
\twocolumn
\clearpage
\fi
%
%
%
\end{document}

%% file: burst-box-sigproc-macros-fix.tex
\ifshowfixmes

\newcounter{fixmecounter}

\newwrite\fixmewritefile
\immediate\openout\fixmewritefile=texput.fix
\newread\fixmereadfile

\newcommand{\fixme}[2]{%
\addtocounter{fixmecounter}{1}%
\textbf{FIXME \thefixmecounter: #1}%
\immediate\write\fixmewritefile{\thefixmecounter}%
\immediate\write\fixmewritefile{#1}%
\immediate\write\fixmewritefile{#2}%
}

\newcommand{\fixmenote}[1]{%
\addtocounter{fixmecounter}{1}%
\immediate\write\fixmewritefile{\thefixmecounter}%
\immediate\write\fixmewritefile{}%
\immediate\write\fixmewritefile{#1}%
}

\newcommand{\listfixmes}{%
\immediate\write\fixmewritefile{-1}%
\immediate\closeout\fixmewritefile%
\clearpage%
\section{FIXME Notes}
~\\ 
\immediate\openin\fixmereadfile=texput.fix%
\loop%
\read\fixmereadfile to \fixtempindex%
\typeout{FIXME \fixtempindex}
\setcounter{fixmecounter}{\fixtempindex}%
\ifnum\value{fixmecounter}<0%
\else%
\read\fixmereadfile to \fixtempmessage%
\read\fixmereadfile to \fixtempnotes%
\typeout{Message: \fixtempmessage}
\typeout{Notes: \fixtempnotes}
\begin{tabular}{p{0.5in}|p{3in}|p{3in}}
\fixtempindex & \fixtempmessage & \fixtempnotes \\
\end{tabular}
\repeat%
\immediate\closein\fixmereadfile%
}

\else

\newcommand{\fixme}[2]{}
\newcommand{\fixmenote}[1]{}
\newcommand{\listfixmes}{}

\fi

%% file: burst-box-sigproc-macros-cite.tex
\ifshowfixmes

\newcounter{fixcitecounter}

\newwrite\fixcitewritefile
\immediate\openout\fixcitewritefile=texput.cit
\newread\fixcitereadfile

\newcommand{\fixcite}[1]{%
\addtocounter{fixcitecounter}{1}%
\textsuperscript{\bfseries CITE HERE \thefixcitecounter}%
\immediate\write\fixcitewritefile{\thefixcitecounter}%
\immediate\write\fixcitewritefile{#1}%
}

\newcommand{\listfixcites}{%
\immediate\write\fixcitewritefile{-1}%
\immediate\closeout\fixcitewritefile%
\clearpage%
\section{Citation Notes}
~\\ 
\immediate\openin\fixcitereadfile=texput.cit%
\loop%
\read\fixcitereadfile to \fixtempindex%
\setcounter{fixcitecounter}{\fixtempindex}%
\ifnum\value{fixcitecounter}<0%
\else%
\read\fixcitereadfile to \fixtempnotes%
\begin{tabular}{p{0.5in}|p{6in}}
\fixtempindex & \fixtempnotes \\
\end{tabular}
\repeat%
\immediate\closein\fixcitereadfile%
}

\else

\newcommand{\fixcite}[1]{}
\newcommand{\listfixcites}{}

\fi

%% file: burst-box-sigproc-abstract.tex
%
Oscillations in the local field potential (LFP) of the brain are key
signatures of neural information processing. Perturbing these oscillations
at specific phases in order to alter neural information processing is an
area of active research. Existing systems for phase-specific brain stimulation
typically either do not offer real-time timing guarantees (desktop computer
based systems) or require extensive programming of vendor-specific
equipment. This work presents a real-time detection system architecture that
is platform-agnostic and that scales to thousands of recording channels,
validated using a proof-of-concept microcontroller-based implementation.
%

%% file: burst-box-sigproc-intro.tex
%
\section{Introduction}
\label{sect-intro}
Recording of electrical signals from neurons in human and animal brains is
a well-established field\cite{buszaki:scaling}.
Processing these signals reveals two related components: ``spikes'',
representing the firing of individual neurons near the pickup electrodes,
and the ``local field potential'' (LFP), representing the aggregate activity
of the larger population of neurons surrounding the electrode
site\cite{buszaki:largescale}.
Both of these signal components carry information: spikes via firing rate
and timing\cite{womelsdorf:gamma2012}\cite{kayser:spikephase}
\cite{voloh:spikephase}\cite{turesson:spikephase},
and the LFP via the presence or absence of transient oscillations
representing coherent activity of a large group of
neurons\cite{lundqvist:betagamma}\cite{vanede:oscillations}
\cite{sherman:beta}\cite{voloh:thetagamma}\cite{feingold:beta}.
The relative timing of spikes with respect to LFP oscillation phase has also
been shown to encode
information\cite{huxter:spikephase}\cite{siegel:phasememory}
\cite{turesson:spikephase}\cite{vinck:gammabook}\cite{vinck:gamma2010}.

Artificial stimulation of human and animal brains (via electrical, optical,
or other means) is also a field of active
study\cite{polania:stim}\cite{lebedev:interfaces}
\cite{qiao:neuromodulation}\cite{grosenick:optoloop}.
It has recently been shown that if LFP oscillations are present near a
stimulation site, the timing of stimulation with respect to the LFP phase is
important\cite{zanos:curve-fit-stim}\cite{siegle:optoloop}
\cite{cagnan:stimphase}\cite{weerasinghe:stimphase}.
In order to study this, it is necessary to perform ``on-line'' detection of
transient LFP oscillations and to extract phase in real-time.

Existing experiments studying phase-specific stimulation can be divided into
those that use a desktop computer to perform their signal
processing\cite{zanos:curve-fit-stim}\cite{blackwood:phasemodel}
\cite{chen:phasemodel}\cite{siegle:optoloop}
and those which perform some or all of their signal processing on dedicated
hardware\cite{karvat:loop}\cite{talakoub:loopripple}.
Both types of system have signal processing latency that must be compensated
for (typically 20-100~ms)\cite{karvat:loop}\cite{siegle:optoloop} but
desktop computer based systems usually have substantial random variation
(jitter) in processing and communications
latency (typically 5-10~ms)\cite{siegle:optoloop},
which is avoided in systems that keep the stimulation trigger processing
entirely in dedicated hardware.

Low-latency signal processing systems running on dedicated hardware may be
implemented in software running on dedicated digital signal processing (DSP)
platforms\cite{karvat:loop}\cite{tdt:rz2guide}
or implemented using a field-programmable gate array (FPGA) tightly coupled
to the recording system\cite{talakoub:loopripple}\cite{neuralynx:hppguide}.
Signal processing on dedicated hardware is widely used for processing of
neural signals but is typically implemented ad-hoc.

The goal of this work is to present and validate an open architecture for
``on-line'' LFP oscillation detection and for phase-aligned stimulation that
is suitable for instantiation on conventional FPGA-based electrophysiology
equipment and that is scalable to thousands of recording channels.
The purpose of this architecture is to make it easier and faster to implement
experiment-specific closed-loop stimulation systems (using FPGA-based
equipment or using embedded software), as most of the implementation and
debugging will already have been done.

%
%

%% file: burst-box-sigproc-background.tex
%
\section{Background}
\label{sect-background}

\subsection{Electrophysiology Measurements}
\label{sect-background-ephys}

\begin{figure}[!t]
\centering
\includegraphics[width=0.95\columnwidth]{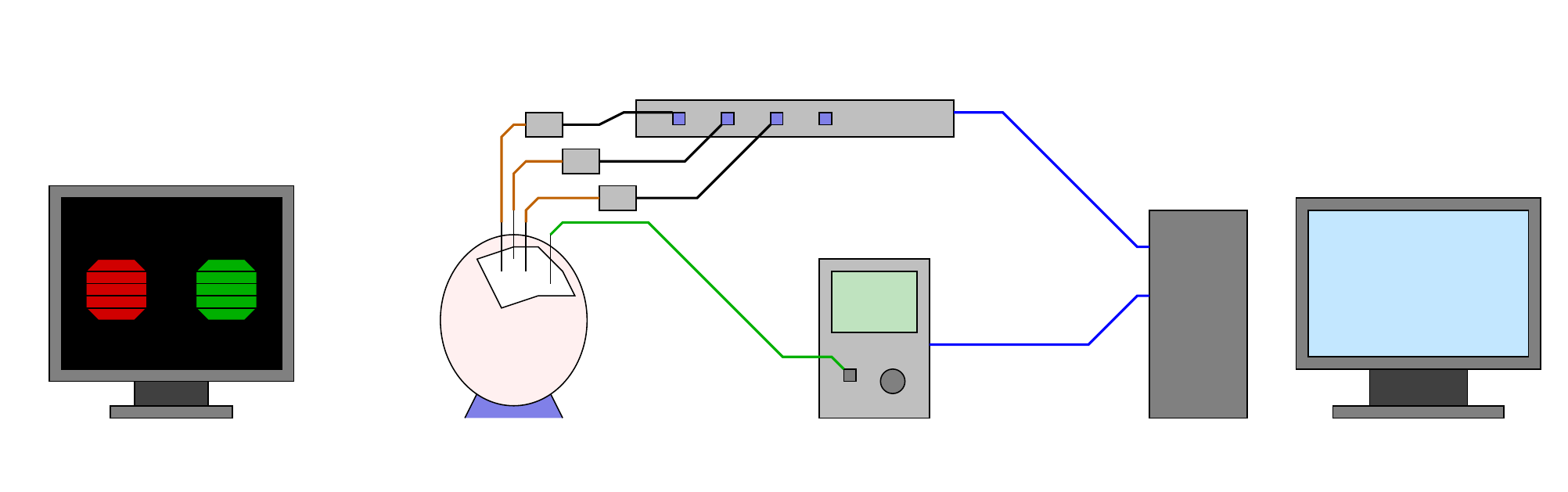}
\caption{Typical electrophysiology recording and stimulation setup using
wire probes or linear silicon probes. A recorder captures electrical signals
from probes, which are processed by a control computer, which commands a
stimulation controller.}
\label{fig-background-setup}
\end{figure}

A diagram of a typical electrophysiology recording and stimulation setup is
shown in Figure \ref{fig-background-setup}. One or more probes, typically
containing multiple electrical contacts per probe, are inserted into the
brain. A ``headstage'' and a recording controller amplify and digitize the
analog signals and forward them to a host computer. Electrical stimulation
is performed using either a dedicated controller and probes or auxiliary
functions of the controller, headstage, and probes used for recording.
Recording and stimulation are typically performed while the subject performs
some consistently-structured activity.

\begin{figure}[!t]
\centering
\includegraphics[width=0.95\columnwidth]{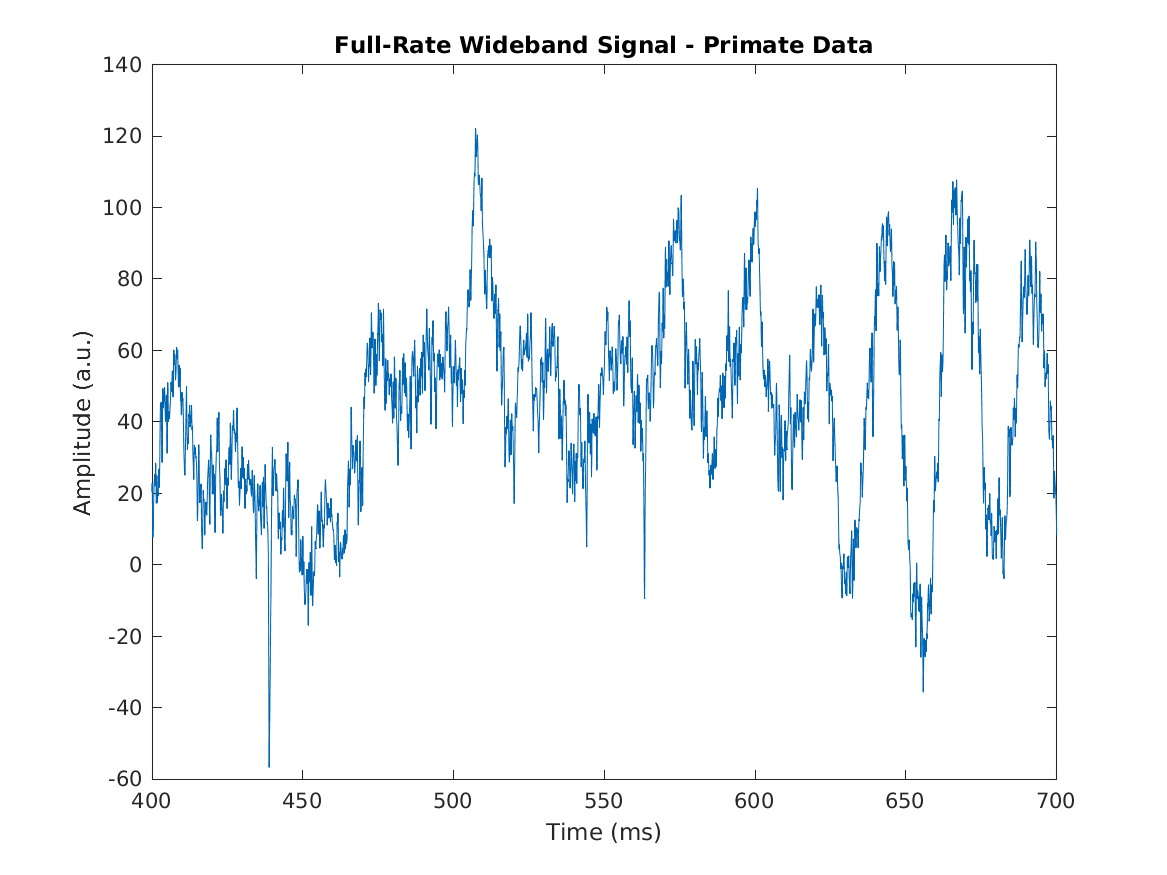}
\caption{Typical wide-band-signal recorded from a primate brain using
tungsten wire probes\cite{oemisch:predictionerrors}.
Noteworthy features in this signal are
sub-millisecond spikes and 20-25~Hz oscillations.}
\label{fig-background-signal}
\end{figure}

A typical single-channel recorded waveform is shown in Figure
\ref{fig-background-signal}\cite{oemisch:predictionerrors}.
Noteworthy features are spikes (sub-millisecond
duration)\cite{bean:actionpotential}\cite{boroujeni:spikeartifacts},
local field potential oscillations (typically 4-50~Hz and lasting for a
small number of cycles\cite{buzsaki:originlfp}),
and background noise (typically $\frac{1}{f^2}$ power-law noise at LFP
frequencies\cite{miller:powerlaw}\cite{milstein:powerlaw}).
Spiking and LFP oscillation patterns vary widely depending on the region
of the brain being
measured\cite{steriade:oscillations}\cite{womelsdorf:motifs},
and LFP oscillation duration (absolute and number of cycles) also depends
strongly on the oscillation frequency\cite{buzsaki:rhythmsbook}.
At high frequencies (50-200~Hz), oscillations occur with durations of many
cycles that are are modulated by co-occurring low-frequency
oscillations\cite{voloh:thetagamma}.

\begin{figure}[!t]
\centering
\includegraphics[width=0.95\columnwidth]{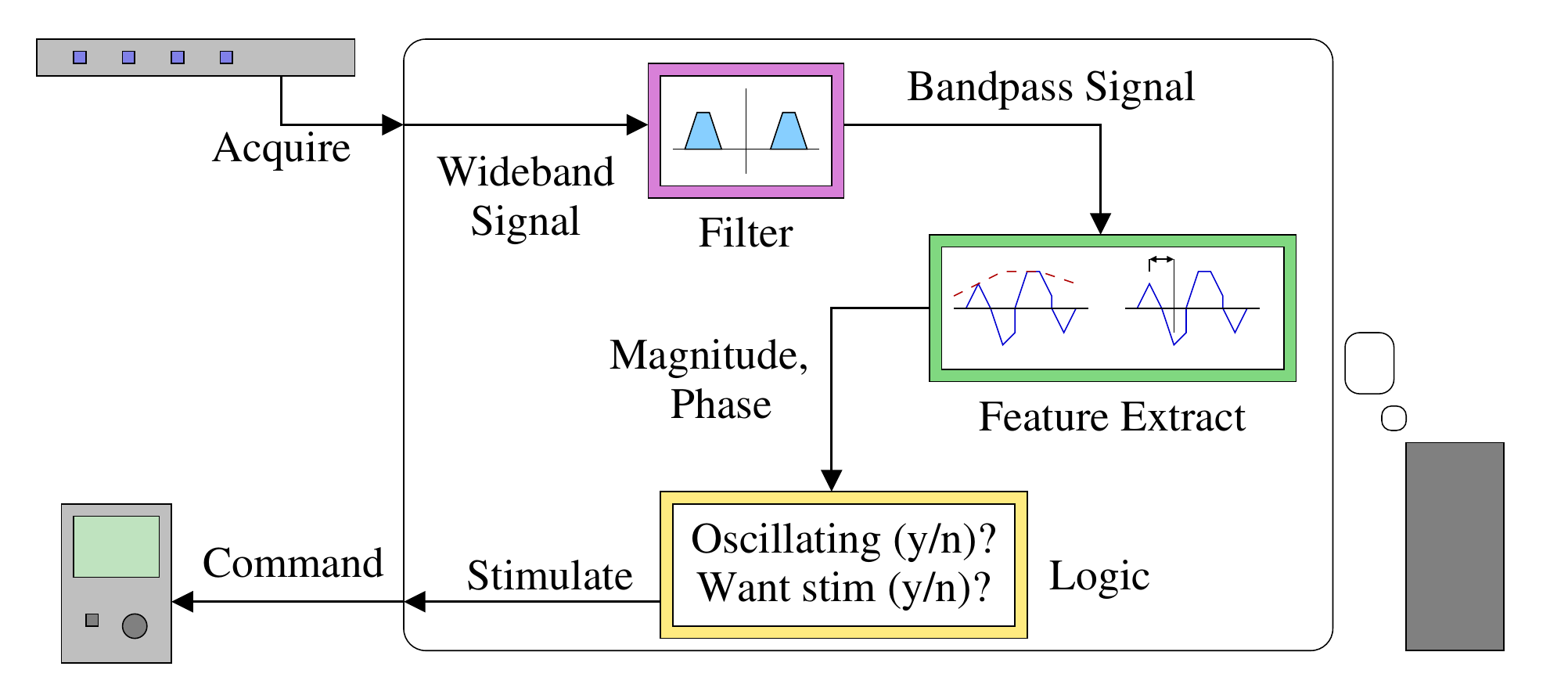}
\caption{Data processing flow within a typical electrophysiology recording
and stimulation setup using a desktop computer for phase-aligned
stimulation.
Acquired signals are band-pass filtered, and magnitude and phase are
estimated either by feature extraction, template matching, or by using a
windowed Hilbert transform. If a request for stimulation is queued, and
an oscillation is occurring, stimulation is performed at the desired phase.}
\label{fig-background-pcloop}
\end{figure}

A typical closed-loop phase-aligned stimulation setup based on a desktop
computer is shown in Figure \ref{fig-background-pcloop}. Signals are acquired
using the recording controller and processed using ``on-line'' algorithms that
are intended to function in real-time. When a transient oscillation occurs and
stimulation is commanded during the experiment, the desktop computer waits
until the appropriate oscillation phase before commanding the stimulation
controller to activate.

Typical ``on-line'' oscillation detection and characterization algorithms
are variants of a widely-used ``offline'' algorithm. In this particular
offline algorithm, the LFP frequency band of interest is isolated and the
analytic signal is computed, with the imaginary component provided by the
Hilbert transform of the band-pass-filtered signal. The analytic signal
encodes the magnitude and phase of the original narrow-band
signal\cite{ktonas:hilbert-etc}\cite{chen:phasemodel}.
Oscillation events are identified by looking for magnitude excursions,
with $2\sigma$ or $3\sigma$ from baseline magnitude being
typical\cite{zanos:curve-fit-stim}\cite{lundqvist:betagamma}
\cite{karvat:loop}\cite{voloh:spikephase}.
Oscillation phase at any given instant is taken to be the analytic signal
phase at the time of interest. For ease of reference, this will be referred
to as the ``offline Hilbert algorithm''.

\fixme{Diagram illustrating the ``Offline Hilbert'' algorithm.}
{Figure needed.}

For ``on-line'' implementation, band-pass filtering is typically performed
using a finite impulse response filter (FIR)\cite{karvat:loop}. Magnitude
and phase may be extracted using template-fitting\cite{zanos:curve-fit-stim}
using interpolation between peaks, troughs, and zero-crossings in the
narrow-band signal\cite{ktonas:hilbert-etc}\cite{voytek:peaktrough}.
Oscillation period may be estimated by using template fitting, by using the
locations of peaks, troughs, and zero-crossings, or by using a filter bank
with densely-spaced center frequencies and looking for the filter with the
strongest response\cite{karvat:loop}.

``Offline'' algorithms for oscillation detection and parameterization are
more varied\cite{principe:decomposing}, as they do not need to meet response
time constraints and they can consider both the
past and future signal around a point of interest. Typical approaches that
do not use the Hilbert transform involve decomposing the signal using
either a fixed dictionary such as Gabor
wavelets\cite{canolty:gabor}
or an optimized dictionary via sparse coding
approaches\cite{brockmeier:dictionary}\cite{hitziger:dictionary}.

The desired goal for real-time closed-loop experiments is to detect local
field potential oscillations while they are still happening (within 1-2
oscillation periods), and to accurately determine the oscillation phase so
that phase-specific stimulation may be performed. The accuracy needed can be
inferred from the number of phase bins used for spike-phase coding analyses;
4--10 phase bins are typical, with diminishing returns past
6 bins\cite{voloh:spikephase} \cite{womelsdorf:gamma2012}. This indicates
that the full-width half-maximum of the phase error distribution should
be 60~degrees or less.

\subsection{Signal Processing Hardware}
\label{sect-background-hardware}

Hardware-based signal processing of electrophysiology signals typically
involves electrophysiology controllers that expose digital signal processors
(DSPs) or field-programmable gate arrays (FPGAs) to the user. These are
programmable (DSP) or configurable (FPGA) hardware devices capable of
running specialized computing operations much faster than general-purpose
microprocessors. User-supplied code is written to these DSP or FPGA
components, which then becomes a part of the signal processing pipeline
within these devices.

\fixme{Hardware signal processing figure.}{Figure needed.}

The performance metric that determines filtering and signal processing
capability is the number of multiply-accumulate operations (MACs) that a
given platform can perform per second. For a given number of channels, this
determines the number of multiply-accumulate operations per channel per
second, and for a given sampling rate, this determines the number of
multiply-accumulate operations that may be performed per sample. The number
of MAC operations available per sample is a design constraint for the
implementation of signal processing pipelines.

A typical electrophysiology controller that exposes DSP
features to the user is the Tucker-Davis RZ2 BioAmp
Processor\cite{tdt:rz2guide} (based on the SHARC series of DSP processors).
For DSP-based systems, the number of
MACs is usually equivalent to the number of floating point operations per
second (FLOPS). The SHARC DSP processors used by Tucker-Davis can perform
2.4~GFLOPS per core (at 400~MHz), for an aggregate maximum processing power
of about 77~GMAC/sec (8 quad-core boards). These processors are typically
connected to digitizing pre-amplifiers supplying up to
256 channels\cite{tdt:pz2guide}, resulting in a budget of
300~MMAC/sec$\cdot$channel.

Typical electrophysiology controllers that expose FPGA features to the user
are the Open Ephys acquisition
board\cite{siegle:openephys}\cite{openephys:acqboard}
and the related Intan RHD recording controller\cite{intan:rhdfpga}
(both based on the Xilinx Spartan 6 LX45 FPGA), and the NeuraLynx Hardware
Processing Platform expansion board\cite{neuralynx:hppguide}
(based on the Xilinx Zynq 7045 SoC which integrates a Kintex 7 FPGA).
For Xilinx-family FPGA-based systems, the number of MACs per second is
determined by the number of ``digital signal processing slices'' and the
rate at which these slices maybe clocked.
The XC6LX45 chip used in the Open Ephys and Intan controllers
provides 5.8~GMAC/sec (58 units clocked at 100~MHz); some of this capacity
is used for the controller's built-in filtering operations. These controllers
can acquire data from up to 1024 recording channels, resulting in a budget
of 5.8~MMAC/sec$\cdot$channel, minus overhead for built-in filtering.
The XC7Z045 chip used
in the NeuraLynx processing board provides 900~GMAC/sec (900 units clocked
at 1~GHz), all of which is available for signal processing. The controller in
which this board is installed can acquire data from up to 512 recording
channels, resulting in a budget of 1.76~GMAC/sec$\cdot$channel.

While these examples are not exhaustive, it is reasonable to
assume a processing budget of at least 3~MMAC/sec per channel, with up to
2~GMAC/sec per channel available in systems with more hardware resources
available. LFP signal processing is typically performed at
1~ksps, with signals acquired at 25~ksps--40~ksps\cite{karvat:loop}.
%
%

%% file: burst-box-sigproc-design.tex
%
\section{Implementation}
\label{sect-design}
\subsection{Architecture}
\label{sect-design-arch}
\begin{figure}[!t]
\centering
\includegraphics[width=0.95\columnwidth]{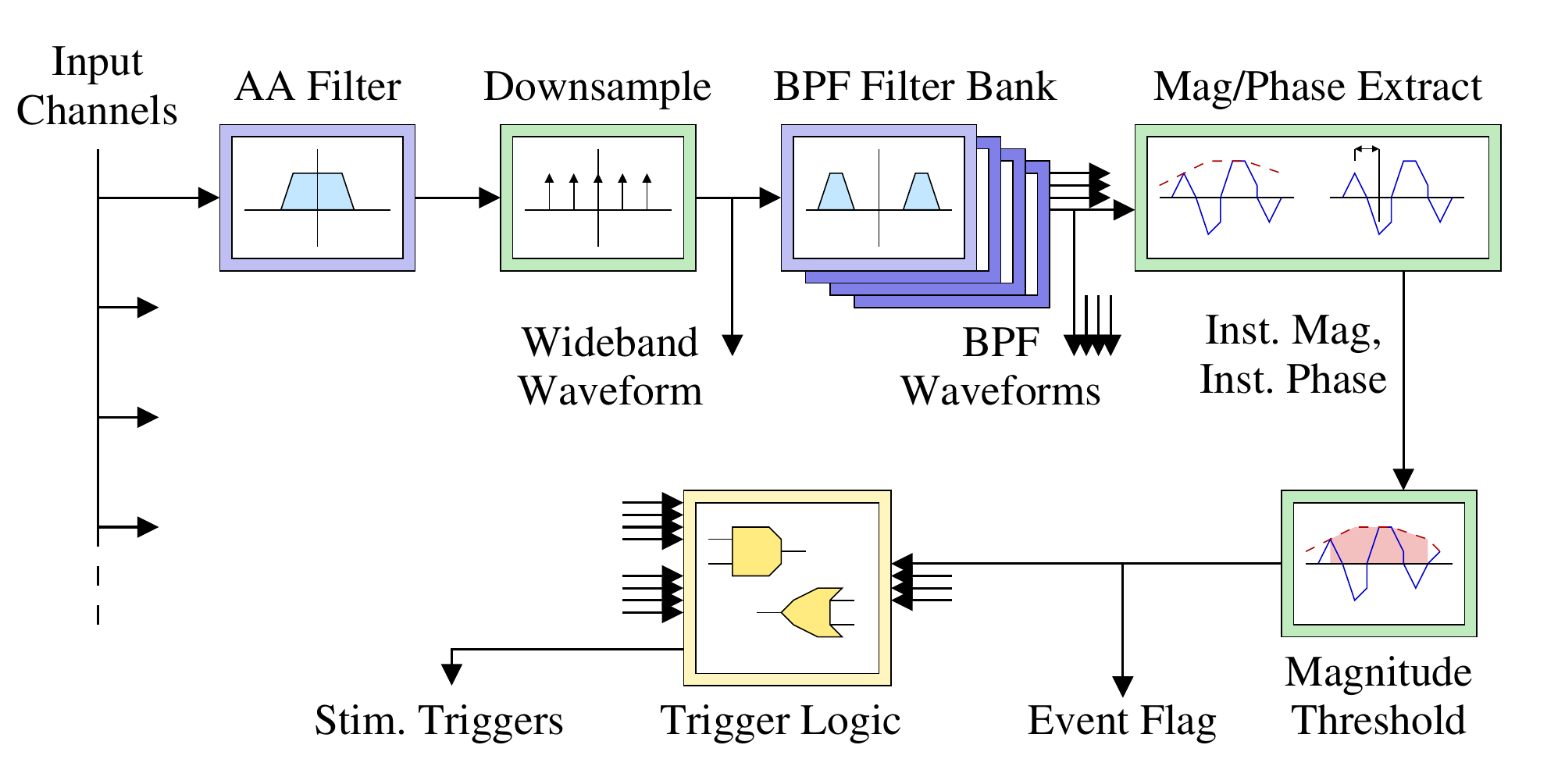}
\caption{Top-level oscillation detector architecture. Input signals are
downsampled and band-pass filtered; instantaneous magnitude, frequency, and
phase are then estimated. Oscillations are detected by magnitude thresholding,
and stimulation triggers are generated based on oscillation detector outputs.}
\label{fig-design-arch}
\end{figure}

A block diagram of the oscillation detection architecture is shown in
Figure \ref{fig-design-arch}. Full-rate data is passed through an
anti-aliasing filter and downsampled. Downsampled data is passed through a
filter bank that performs band-pass filtering, and an approximation of its
instantaneous magnitude and phase is extracted. Oscillation detection in
each band is performed by magnitude thresholding, event detection logic
building event triggers.

Several features of this architecture require detailed discussion: Filter
implementation and the associated corrections for delays introduced by the
filters; estimation of instantaneous magnitude, frequency, and phase; and
event detection logic.

Anti-aliasing filtering and band-pass filtering are implemented using either
finite impulse response filters (FIR filters) or infinite impulse response
filters (IIR filters). Both of these introduce delay (a fixed delay for FIR
filters, and a frequency-dependent delay for IIR filters). Time shifts are
added to the phase estimate to correct for these delays. For finite impulse
response filters, a fixed time shift is used; for infinite impulse response
filters, the time shift is read from a lookup table indexed by the estimated
instantaneous frequency. Errors in the frequency estimate result in errors
in the phase shift estimate when using lookup table time shifts.

Estimates of instantaneous magnitude, frequency, and phase are obtained
using time-domain methods. The baseline implementation used for this
architecture uses peaks and troughs to estimate magnitude and zero-crossings
to estimate phase\cite{ktonas:hilbert-etc}. Potential extensions include
using flank midpoints to estimate phase\cite{voytek:peaktrough} and using
complex demodulation to estimate magnitude and phase\cite{ktonas:hilbert-etc},
with components provided by in-phase and quadrature FIR filters in the
band-pass filter bank.

Transient oscillation detection is performed using magnitude thresholding.
Two thresholds are used: a higher ``turn-on'' threshold and a lower
``turn-off' threshold, providing hysteresis. Threshold comparator inputs
may optionally be required to remain high or low for a certain time interval
before changing state, to suppress ``glitching'' in comparator output. The
intention is to provide a number of tuning methods sufficient to suppress
spurious detections and ``drop-outs'' in noisy input, despite the character
of this input varying widely between use-cases.

Stimulation trigger generation logic also varies with use-case. The baseline
implementation used for this architecture links stimulation triggers to the
outputs of individual event detectors. A planned extension is contingent
triggering based on the output of multiple detectors, to cover use-cases
where triggering is to be performed during co-occurring
oscillations\cite{voloh:thetagamma}. Triggering is constrained to a
user-specified time window and maximum number of trigger assertions, to
ensure that spurious event detections do not result in unsafe stimulation.

The individual signal processing blocks in this architecture were implemented
as modules, with the intention being that application-specific signal
processing architectures would be built by assembling modules with a minimum
of new code needed. While the desired output from a system implementation
is typically the stimulation trigger flags, the outputs of all signal
processing modules are potentially accessible for debugging or diagnostic
purposes.

Each of the signal processing modules was implemented in C++ and in Matlab,
with FPGA-based implementations in development. The intention is to allow
rapid prototyping via Matlab, embedded software and workstation-based
implementations via C++, and full-scale hardware implementations via
hardware description languages, with confidence that all three types of
implementation would produce comparable output if given the same input.
As FPGA implementation is the end-goal, the C++ modules are written to
operate in a pipelined manner on a sample-by-sample basis to facilitate
translation to hardware and cycle-by-cycle comparison with hardware.
\fixme{Offering pipelined Matlab too would make this useful for
Matlab/Simulink implementations.}{Want pipelined Matlab implementation.}

Three closed-loop systems were assembled as reference implementations: One
off-line Matlab-based implementation\fixme{Name of Matlab implementation}
{Name of Matlab implementation.}, one off-line C++-based running
on a desktop workstation (the ``Burst Station'' implementation), and one
on-line embedded C++ implementation running on a proof-of-concept
microcontroller-based hardware (the ``Burst Box'' implementation).
The Matlab implementation was used to verify that the architecture
is conceptually sound; it was not otherwise resource-constrained (no memory
or processing time limits, double-precision floating-point arithmetic).
The workstation-based ``Burst Station'' C++ implementation was used to
verify that the architecture's integer-arithmetic implementation produced
output acceptably close to that of the Matlab implementation. While the
workstation-based implementation was not explicitly memory-constrained,
care was taken to keep internal structure sizes small enough to be
instantiated on FPGAs.
\fixme{Mention C++ for an Open Ephys plugin if we implement one by
press time.}{Open Ephys plugin implementation.}
The embedded ``Burst Box'' prototype was used to verify that the
architecture was capable of performing closed-loop stimulation in
real-time with limited memory and a limited amount of processing power
available.

Module library code and the closed-loop system reference implementations
were made freely available under an open-source
license\cite{acclab:neuroloop}.

\subsection{Workstation Implementation}
\label{sect-design-bstat}

The workstation-based C++ implementation (the ``Burst Station'') was run as
an ``off-line'' system: input signals were loaded from disk, rather than
captured in real-time. The ``Burst Station'' was implemented as a test-bed
for detection architectures and as a prototyping tool for building embedded
implementations. The output of the ``Burst Station'' for a given input
signal should be identical to the output of an ``on-line'' embedded system
running with the same configuration.

\begin{figure}[!t]
\centering
\includegraphics[width=0.95\columnwidth]{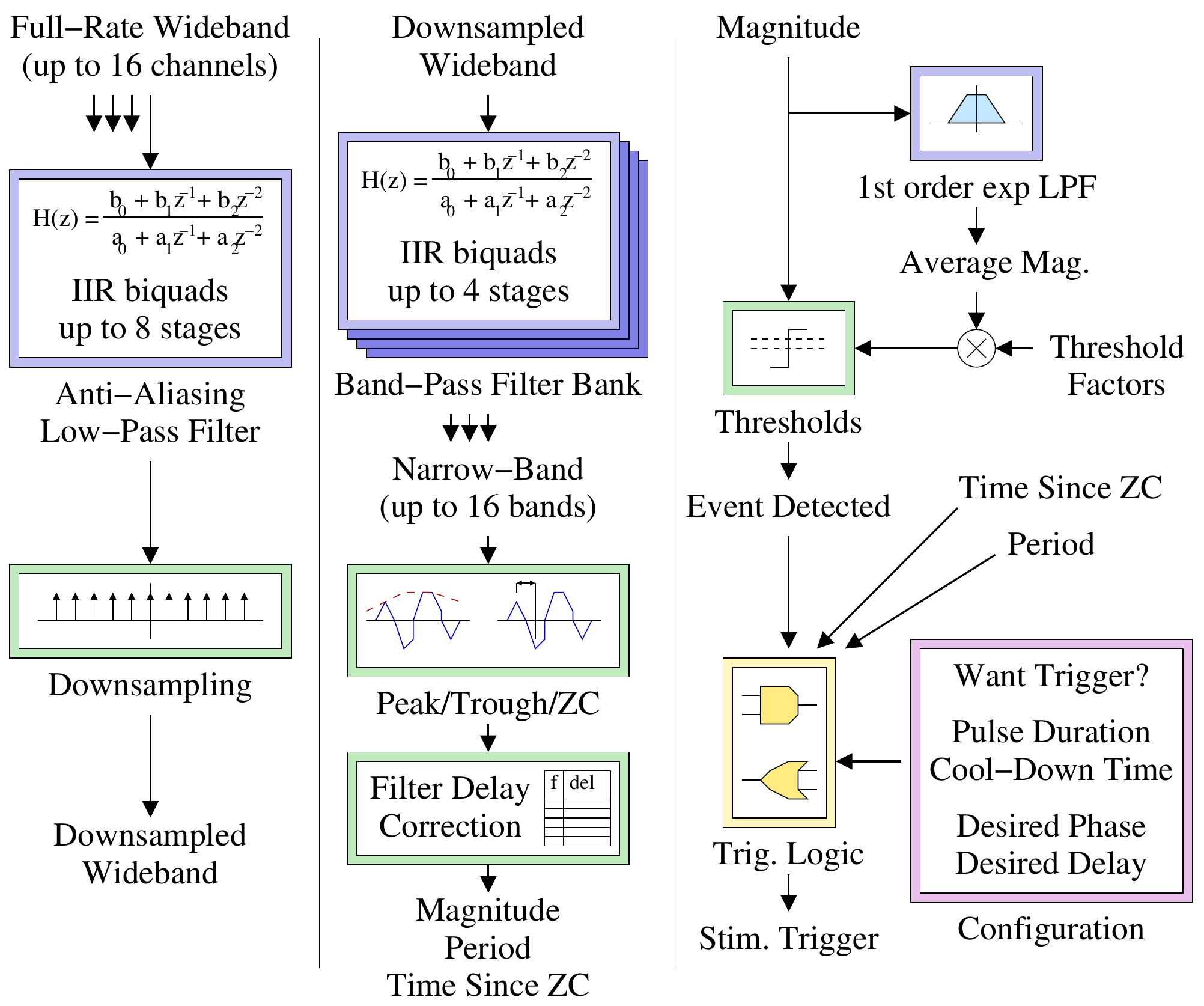}
\fixme{Placeholder figure. Need to add FIR and dense winner-take-all.}
{Burst station signal processing figure.}
\caption{Workstation-based implementation of the oscillation detector
architecture. Input is passed through an anti-aliasing filter, downsampled,
and then passed through widely-spaced or densely-spaced band-pass filters.
For densely-spaced filters, ``winner-take-all'' logic identifies the
filter with the strongest response. Running estimates of magnitude, phase,
and period are made using peak-and-trough estimators. Two-level magnitude
thresholding is used to detect transient oscillation events. If an event is
detected, phase- or delay-aligned stimulation trigger pulses may be
generated.}
\label{fig-design-bstat-sigproc}
\end{figure}

A block diagram of the ``Burst Station'' signal processing architecture is
shown in Figure \ref{fig-design-bstat-sigproc}. Input is passed through an
anti-aliasing filter (FIR or IIR) and then downsampled. The anti-aliasing
filter has a corner frequency higher than the highest LFP band edge of
interest and at least 5 times lower than the Nyquist frequency, to ensure
adequate suppression of aliased components. The downsampled signal is then
passed through a bank of band-pass filters. The pass bands of these filters
may be widely spaced and non-overlapping or densely-spaced and
highly-overlapping. In the non-overlapping case, any given transient
oscillation is expected to show up in a single filter's output. In the
highly-overlapping case, any given transient oscillation is expected to
show up in multiple filters' outputs, and a winner-take-all decision is
performed to identify the filter with the strongest response. This filter's
center frequency is taken to be the nominal frequency of the event.

A running estimate of signal magnitude, phase, and period is produced for
each band-pass filter's output by peak, trough, and zero-crossing
estimators. Oscillation
detection is performed using a two-threshold scheme (activating when
magnitude reaches the higher threshold and deactivating when magnitude
falls below the lower threshold). Activation after exceeding the rising
threshold is delayed by a fixed number of samples (typically half a period
at the mid-band frequency), to allow the estimate of the signal's period to
stabilize before the oscillation detection signal is asserted. The running
estimate of phase has a correction applied to compensate for the delay of
the filter. For FIR filters, this delay is fixed at design-time. For
IIR filters, this delay depends on frequency, and is fetched from a lookup
table indexed by the detected period. The detected period used for this
indexing may be either the running estimate or (for highly-overlapping
filters) the center frequency of the band identified by the winner-take-all
decision.

When an oscillation is detected, stimulation trigger logic compares the
running estimate of the signal phase, delay since rising zero-crossing, or
delay since falling zero-crossing against a user-specified target value.
When the running estimate crosses the target value, a trigger pulse is
generated. To ensure safety, a ``quota'' system is implemented, capping the
number of trigger pulses delivered to some user-specified maximum.
An activity time-out is also specified (typically several seconds); after
this time-out has expired, oscillations no longer generate stimulation
trigger pulses.

Calculations within the ``Burst Station'' were performed using 32-bit
integer arithmetic with a signal range of 14 bits (to ensure sufficient
head-room during multiply-accumulate operations). The C++ library's
IIR filters were implemented as cascaded biquad filters with Direct Form I
implementation, given by Equation \ref{eq-design-biquad}.
The $a_0$ biquad denominator coefficient was required to
be a power of two, so that the $\frac{1}{a_0}$ operation could be performed
as a bit-shift.

\begin{multline}
\label{eq-design-biquad}
y[n] = \left ( \frac{1}{a_0} \right )
( b_0 x[n] + b_1 x[n-1] + b_2 x[n-2] \\
- a_1 y[n-1] - a_2 y[n-2] )
\end{multline}

\fixme{Burst station filter configurations go here!}
{Burst station configurations.}

\subsection{Embedded Microcontroller Implementation}
\label{sect-design-bbox}

The embedded microcontroller-based implementation (the ``Burst Box'')
was run as an ``on-line'' system: input signals were
captured in real-time from analog input connectors, and stimulation trigger
pulses were emitted as TTL signals. Physical connections were via BNC
connectors, for compatibility with electrophysiology equipment. The
``Burst Box'' was implemented as a proof-of-concept prototype that can be
used in as part of a functioning electrophysiology experiment.

\begin{figure}[!t]
\centering
\includegraphics[width=0.95\columnwidth]{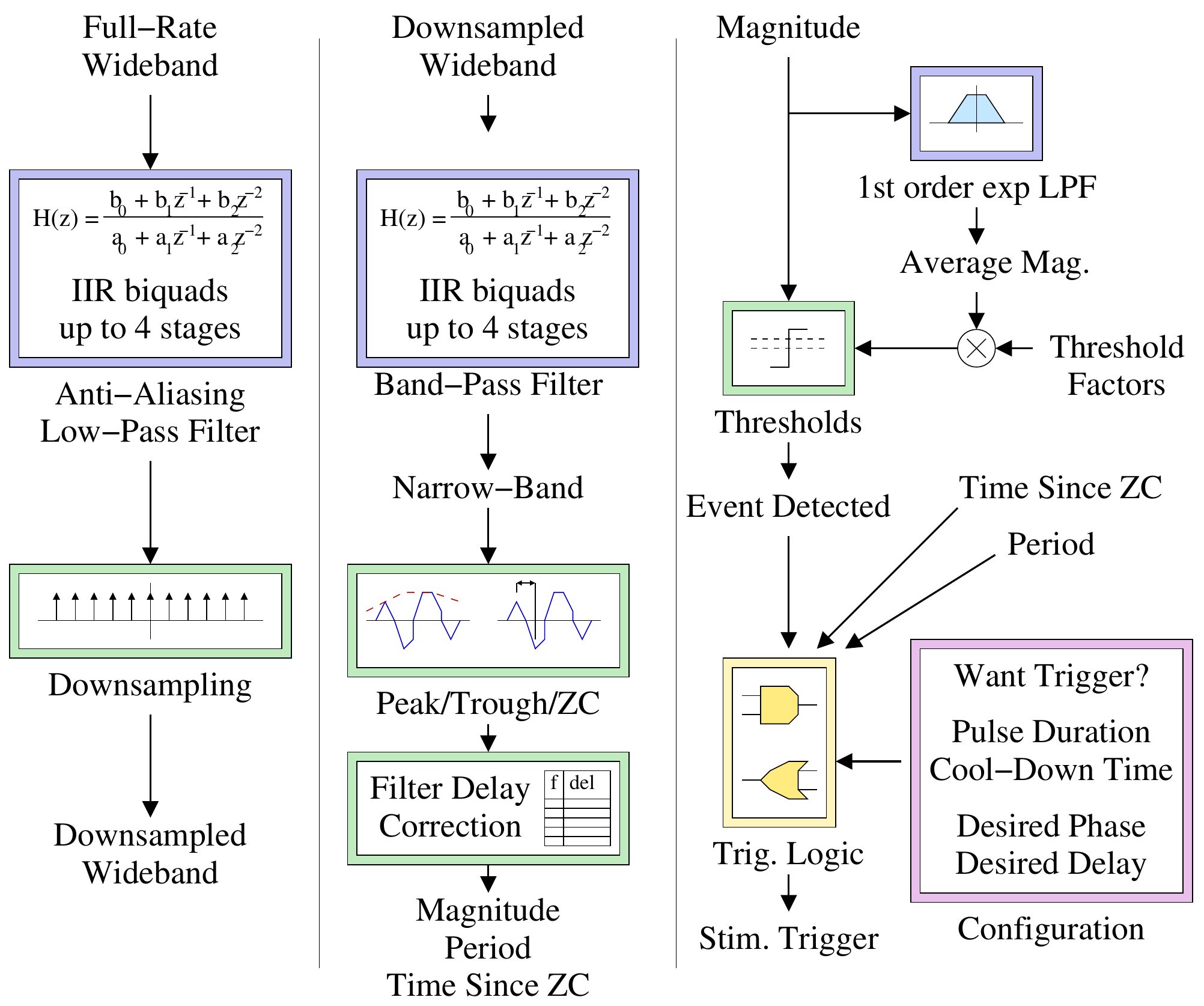}
\caption{Microcontroller-based implementation of the oscillation detector
architecture. Input is passed through an anti-aliasing filter, downsampled,
and then passed through a band-pass filter. Running estimates of magnitude,
phase, and period are made using a peak-and-trough estimator. Two-level
magnitude thresholding is used to detect transient oscillation events.
If an event is detected, phase- or delay-aligned stimulation trigger
pulses may be generated.}
\label{fig-design-bbox-sigproc}
\end{figure}

A block diagram of the ``Burst Box'' signal processing architecture is shown
in Figure \ref{fig-design-bbox-sigproc}. This is a subset of the ``Burst
Station'' workstation-based implementation's architecture described in
Section \ref{sect-design-bstat}.
Processing is restricted to one channel and
one frequency band. Input is passed through a hardware anti-aliasing filter
and a software anti-aliasing filter, downsampled, and then passed through a
band-pass filter. Anti-aliasing and band-pass filters are implemented as
infinite impulse response filters (IIR), to minimize processing load.

As with the ``Burst Station'', a running estimate of the magnitude, phase,
and period of the band-pass-filtered signal is produced using a peak, trough,
and zero-crossing estimator. Oscillation detection is performed using a
two-threshold scheme (activating when magnitude reaches the higher threshold
and deactivating when magnitude falls below the lower threshold). Activation
after exceeding the rising threshold is delayed by a fixed number of samples
(typically half a period at the mid-band frequency), to allow the estimate
of the signal's period to stabilize before the oscillation detection signal
is asserted. The running estimate of phase has a correction applied to
compensate for the delay of the IIR filter. As this delay depends on
frequency, the correction is fetched from a lookup table indexed by the
estimated period.

As with the ``Burst Station'', when an oscillation is detected, stimulation
trigger logic compares the running estimate of the signal phase, delay since
rising zero-crossing, or delay since falling zero-crossing against a
user-specified target value. When the running estimate crosses the target
value, a trigger pulse is generated. To ensure safety, the number of trigger
pulses that may be generated and the time window within which they may
be generated are both limited; oscillations that are detected after this
pulse quota has been exceeded or after the time-out window has expired do
not generate stimulation trigger pulses.

\begin{figure}[!t]
\centering
\includegraphics[width=0.95\columnwidth]{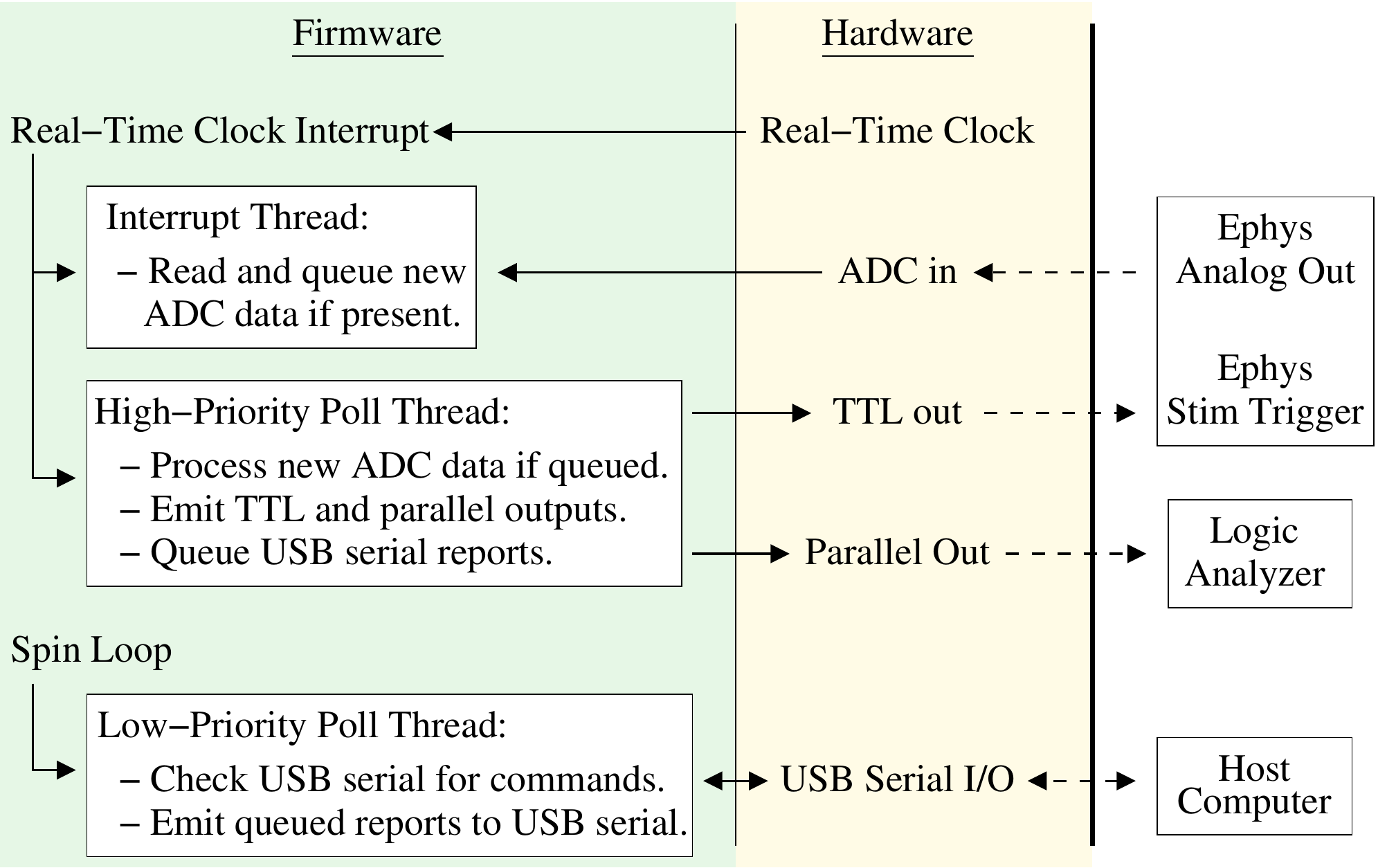}
\caption{Embedded oscillation detector firmware architecture. Analog to
digital conversion is handled once per real-time-clock tick. Signal
processing is handled in a high-priority polling loop; this reads new
ADC data if present, generates TTL output and diagnostic parallel output,
and queues USB reports. A low-priority polling thread reads USB commands
from the host computer and sends reports to the host computer.}
\label{fig-design-bbox-firmware}
\end{figure}

A block diagram of the firmware for the ``Burst Box'' is shown in Figure
\ref{fig-design-bbox-firmware}. Three concurrent execution threads are
running: an interrupt service thread, which handles events that must occur
with every real-time clock tick and complete within that timeslice; a
high-priority polling thread, which is woken up by the real-time clock
(preempting low-priority polling) but which may take multiple timeslices
to complete; and a low-priority polling thread, which handles operations
which do not require timing guarantees.

\begin{figure}[!t]
\centering
\includegraphics[width=0.95\columnwidth]{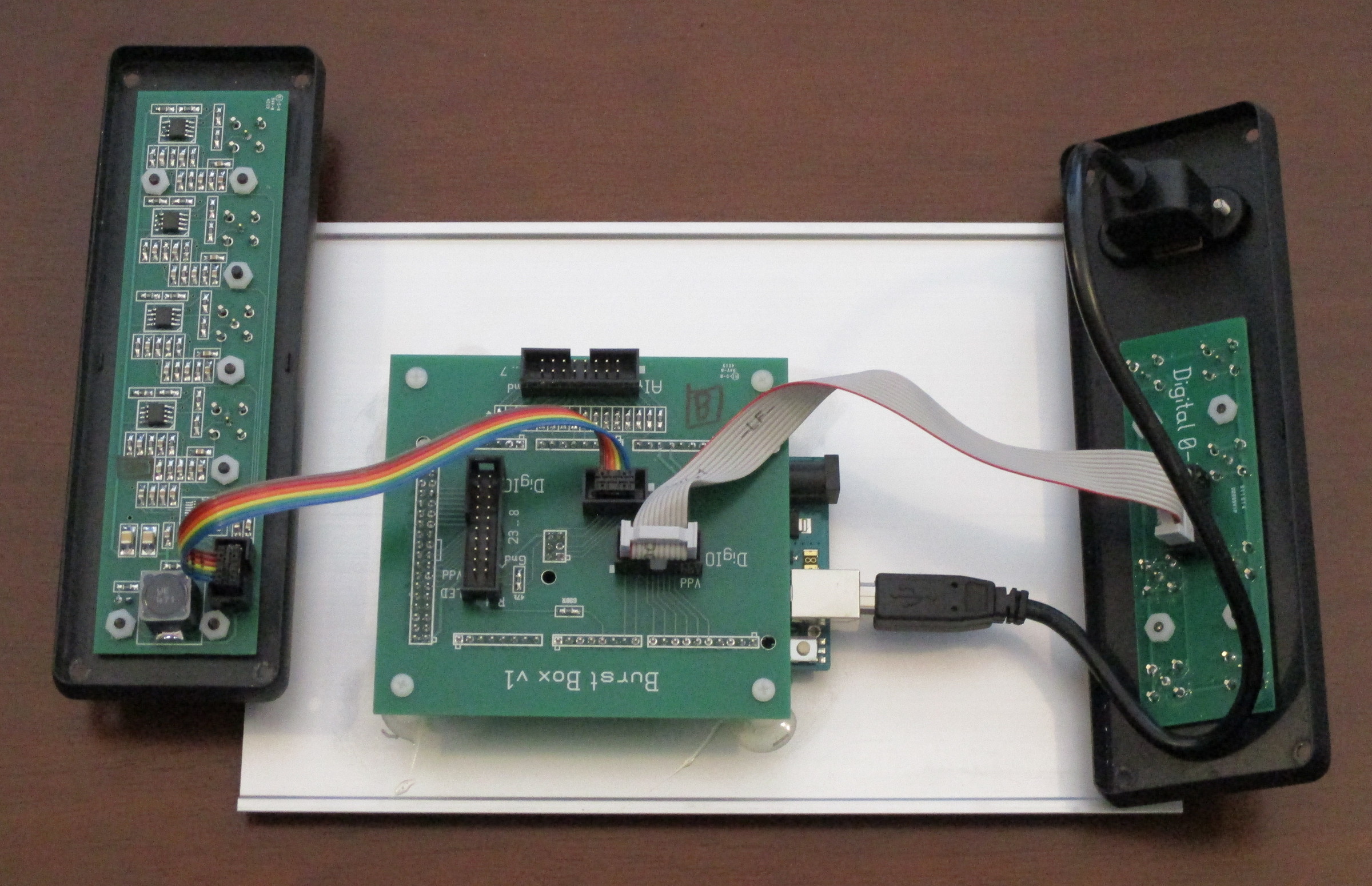}
\caption{Physical implementation of the microcontroller-based oscillation
detector (``Burst Box''). Analog filters and the analog-to-digital converter
are on the left; TTL outputs are on the right. The middle board is a
break-out board, with the microcontroller board underneath it.}
\label{fig-design-bbox-photo}
\end{figure}

The physical implementation of the ``Burst Box'' proof-of-concept prototype
is shown in Figure \ref{fig-design-bbox-photo}.
The microcontroller used is an Atmel ATmega2560 (8-bit, running at 16~MHz,
with 8~kiB of SRAM); DSP performance was benchmarked at approximately
100~kMAC/sec with 32-bit operands. ``Full-rate'' signals are sampled at
2.5~ksps using an external analog to digital converter and a hardware
anti-aliasing filter, and then downsampled to 500~sps internally (the
``DSP rate'') after the application of a digital anti-aliasing filter.
At 200~MAC/sample, this represents a worst-case lower bound to the processing
budget available in real implementations.
\fixme{Benchmark the Burst Box again.}{Burst Box benchmarking.}

\begin{figure}[!t]
\centering
\includegraphics[width=0.95\columnwidth]{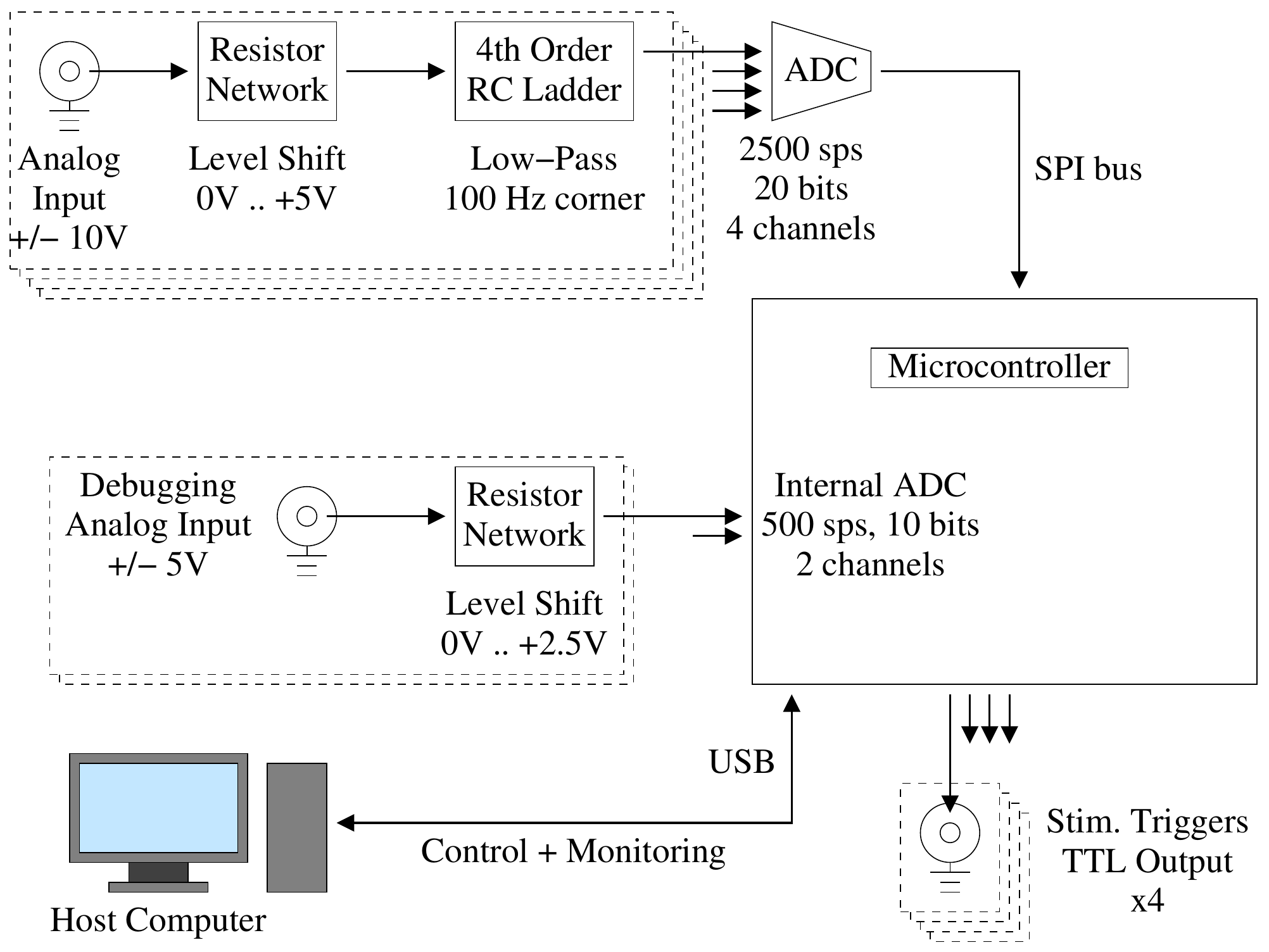}
\caption{Block diagram of the ``Burst Box'' hardware implementation.
Input channels are level-shifted and filtered using a passive analog network
before being digitized by the external analog to digital converter. Analog
input may optionally be fed to the microcontroller's internal analog to
digital converter without filtering.}
\label{fig-design-bbox-hardware}
\end{figure}

A diagram of the hardware implementation of the ``Burst Box'' is shown in
Figure \ref{fig-design-bbox-hardware}. There is hardware support for up
to 4 input channels and 4 TTL outputs. The hardware anti-aliasing filter
in this prototype was implemented as an RC ladder filter for simplicity and
to avoid any possibility of resonance from inductive components, with the
tradeoff of having poor roll-off compared to a Butterworth implementation.
For debugging purposes, the system can be configured to bypass the external
analog-to-digital converter and filters and use the microcontroller's
internal analog-to-digital converter at 500~sps without anti-aliasing.

%
%

%% file: burst-box-sigproc-validation.tex
%
\section{Validation}
\label{sect-validation}
%
\input{burst-box-sigproc-validation-methods}
\input{burst-box-sigproc-validation-results}
%
%

%% file: burst-box-sigproc-validation-methods.tex
%
\subsection{Datasets}
\label{sect-validation-data}

Two datasets were used for testing and validation of oscillation detector
implementations. The first (the ``synthetic'' dataset) consisted of 5 minutes
of $\frac{1}{f^2}$ noise (``red noise'') with tones overlaid. Tones had weak
frequency chirping and amplitude ramping (less than 5\% and 10\%
respectively), with cosine roll-off (Tukey window roll-off), and durations
of 3--5 periods between midpoints of the roll-off flanks. Tones had a
signal-to-noise ratio of 20~dB with respect to in-band noise; frequency
bands used for noise calculations are shown in
Table \ref{tab-validation-noise-bands}. The ``red noise'' spectrum spanned
from 2--200~Hz, with power concentrated at lower frequencies, so per-band
adjustment of tone amplitude was necessary in order to have consistent
signal-to-noise ratios.

\begin{table}[!t]
\caption{Synthetic Dataset Noise Bands}
\label{tab-validation-noise-bands}
\centering
\begin{tabular}{|c|cccccc|}\hline
\textbf{Freq (Hz)} &
4--7 & 7--12 & 12--21 & 21--36 & 36--63 & 63--108 \\
\textbf{Band} &
theta & alpha & beta & gamma & gamma & gamma \\
\hline
\end{tabular}
\end{table}

The second dataset (``biological'' dataset) consisted of a concatenated
selection of recordings from a primate
dataset\cite{boroujeni:interneuron}.
The raw dataset consisted of ``epochs'' that were typically less than 10
seconds long, taken during individual task trials within one extended
recording session. Signals from individual epochs were trimmed to time
periods within the task that showed consistent activity with few electrical
artifacts. Signals were evaluated on an epoch-by-epoch basis to reject
records that contained artifacts within the trimming interval (typically
large step transients caused by physical contact with equipment or 60~Hz
tones coupled from nearby equipment). Remaining ``clean'' epochs were
normalized to have consistent average power and were concatenated with an
overlap of 0.5~s with linear interpolation between signals within the
overlap interval. The intention was to produce an artifact-free signal of
several minutes' duration with biologically valid noise and oscillation
features.

\subsection{Test Procedure}
\label{sect-validation-setup}

Validation tests were intended to measure several things: the transfer
function of the band-pass filters, the accuracy of magnitude and phase
estimation with respect to the analytic signal's magnitude and phase,
and the timing accuracy of stimulation pulses with respect to the desired
stimulation times (specific phases or specific delays after a rising or
falling zero crossing). The goal is to demonstrate a closed-loop system
with stimulation phase accuracy of $\pm30^\circ$ or better (jitter
full-width half-maximum of $60^\circ$ or less).

Testing of the ``off-line'' Matlab-based and workstation-based oscillation
detectors was straightforward; both provide as output time series waveforms
for all internal signals in their processing pipelines, with a common time
reference between all signals. The challenge was to extract comparable
information from the ``on-line''  embedded microcontroller-based
implementation during real-time tests.

\begin{figure}[!t]
\centering
\includegraphics[width=0.95\columnwidth]{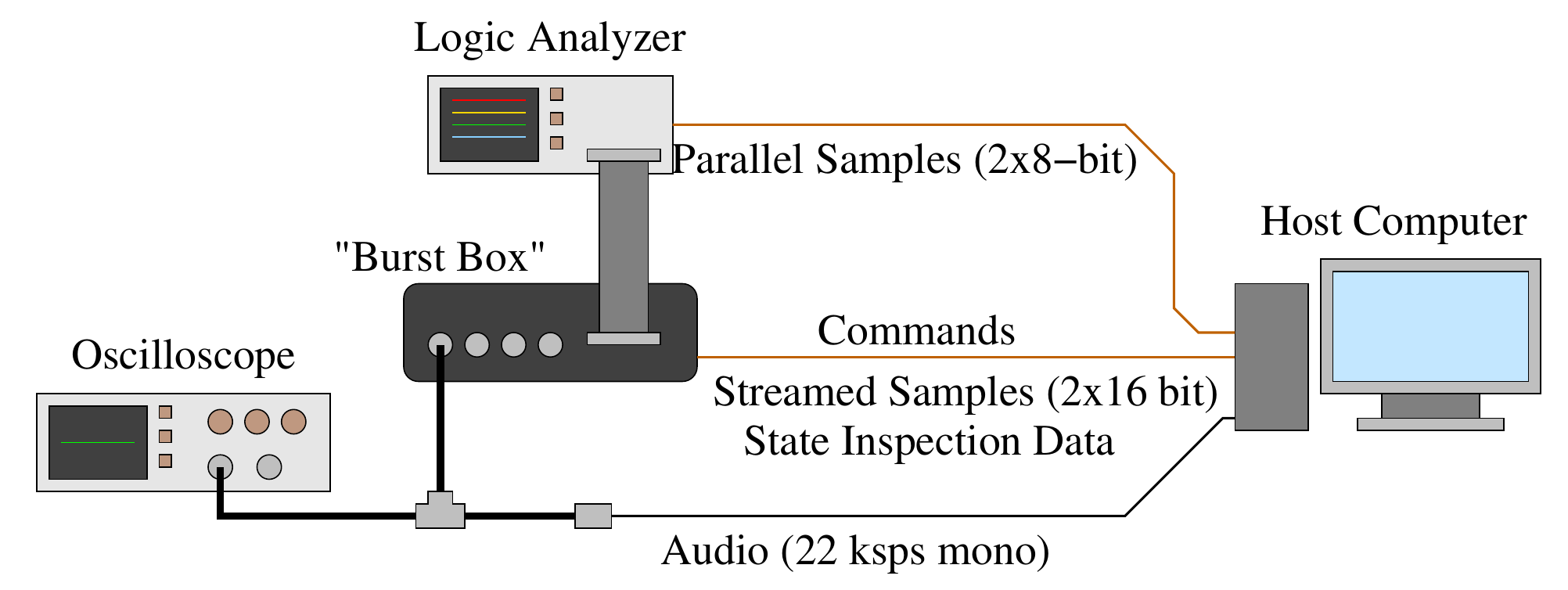}
\caption{Physical setup for validation tests.}
\label{fig-validation-setup-hardware}
\end{figure}

The physical setup for real-time testing is shown in Figure
\ref{fig-validation-setup-hardware}. Signal waveforms were converted to
sound files and played back to the ``Burst Box'' prototype via computer
audio output. Volume settings for playback were adjusted until the output
amplitude was approximately 3~V peak-to-peak, as measured using an
oscilloscope. The ``Burst Box'' is capable of providing monitoring streams
of two signals (typically the band-pass filtered waveform and one other
signal derived from it). Tests with a given input waveform were run
repeatedly, capturing different output waveform pairs, and these output
waveform pares were time-aligned using the band-pass filtered waveform
as a reference (which should remain consistent between successive trials).

Signals streamed from the ``Burst Box'' could be read via two methods:
parallel output via a logic header (8 bits per sample, precise timing and
no dropped samples), and diagnostic output via the USB serial command
interface (16 bits per sample, some dropped samples). Both capture methods
were used. Unless otherwise indicated, the logic header output was used to
generate plots.

Functionality exists for inspecting and modifying the internal state of the
``Burst Box'' using the serial command interface for single-stepped testing.
While this would provide all of the desired signals with high fidelity, it
was not practical to use for full-duration test signals, due to being far
slower than real-time testing.
%
%

%% file: burst-box-sigproc-validation-results.tex
%
%
\subsection{Filter Validation}
\label{sect-validation-filt}

The purpose of filter validation is to confirm that the integer math C++
implementations of the oscillation detector's filters match the behavior
of the Matlab implementation of the same filters. This tested by plotting
the inferred filter transfer functions measured during functionality tests
against the ideal transfer functions.

Filter gain, phase shift, phase delay, and group delay were characterized
by taking the Fourier transform of the time-aligned input and output
waveforms for each filter under test. Dividing spectrum elements gives the
frequency-domain transfer function directly, per
Equation \ref{eq-validation-filt-xfer}. This is smoothed, to reduce artifacts
due to noise, and the phase is unwrapped.
The phase delay and group delay are then computed per Equations
\ref{eq-validation-filt-pdelay} and \ref{eq-validation-filt-gdelay},
respectively. The derivative of $\phi(\omega)$ is approximated by taking the
first difference and performing additional smoothing.

\begin{equation}
\label{eq-validation-filt-xfer}
\begin{split}
H(\omega) &= \frac{\mathcal{F}\{y(t)\}}{\mathcal{F}\{x(t)\}} \\
G(\omega) &= ||H(\omega)|| \\
\phi(\omega) &= \mathrm{arg}(H(\omega))
\end{split}
\end{equation}

\begin{equation}
\tau_\phi(\omega) = - \frac{\phi(\omega)}{\omega}
\label{eq-validation-filt-pdelay}
\end{equation}

\begin{equation}
\tau_g(\omega) = - \frac{d\phi(\omega)}{d\omega}
\label{eq-validation-filt-gdelay}
\end{equation}

\begin{figure}[!t]
\centering
\includegraphics[width=0.95\columnwidth]
{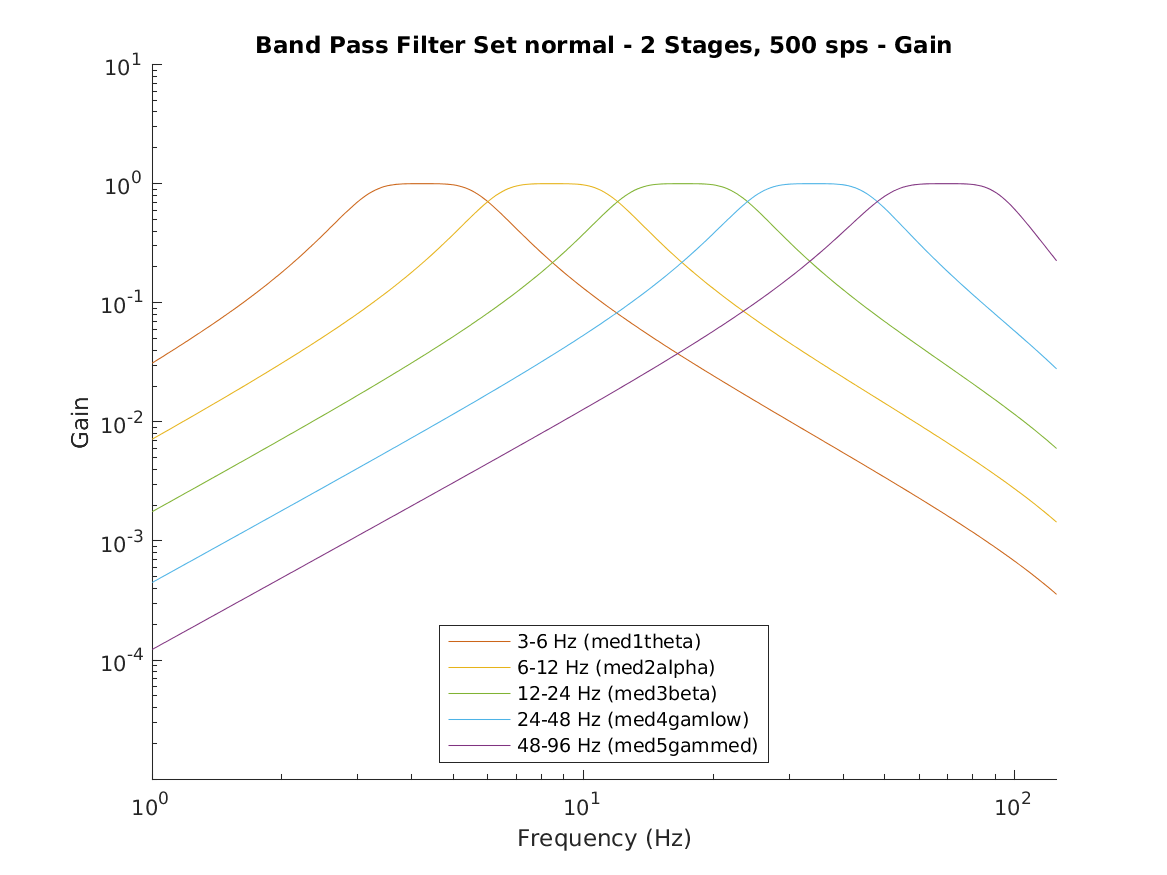}
\caption{Designed transfer functions for infinite impulse response band-pass
filters. These were Butterworth filters implemented as two-stage digital
biquad filters with 2nd-order roll-off.}
\label{fig-validation-filt-iirconfigs}
\end{figure}

The band-pass IIR filter configurations used by the reference implementations
are shown in Figure \ref{fig-validation-filt-iirconfigs}. These were
Butterworth filters implemented as two-stage digital biquad filters (with
second-order roll-off). The anti-aliasing filter (not shown) was a low-pass
Butterworth filter implemented as a two-stage digital biquad filter with
fourth-order roll-off.

\fixme{FIR filter configuration and representative plots go here.}
{FIR plots}

\begin{figure}[!t]
\centering
\includegraphics[width=0.95\columnwidth]
{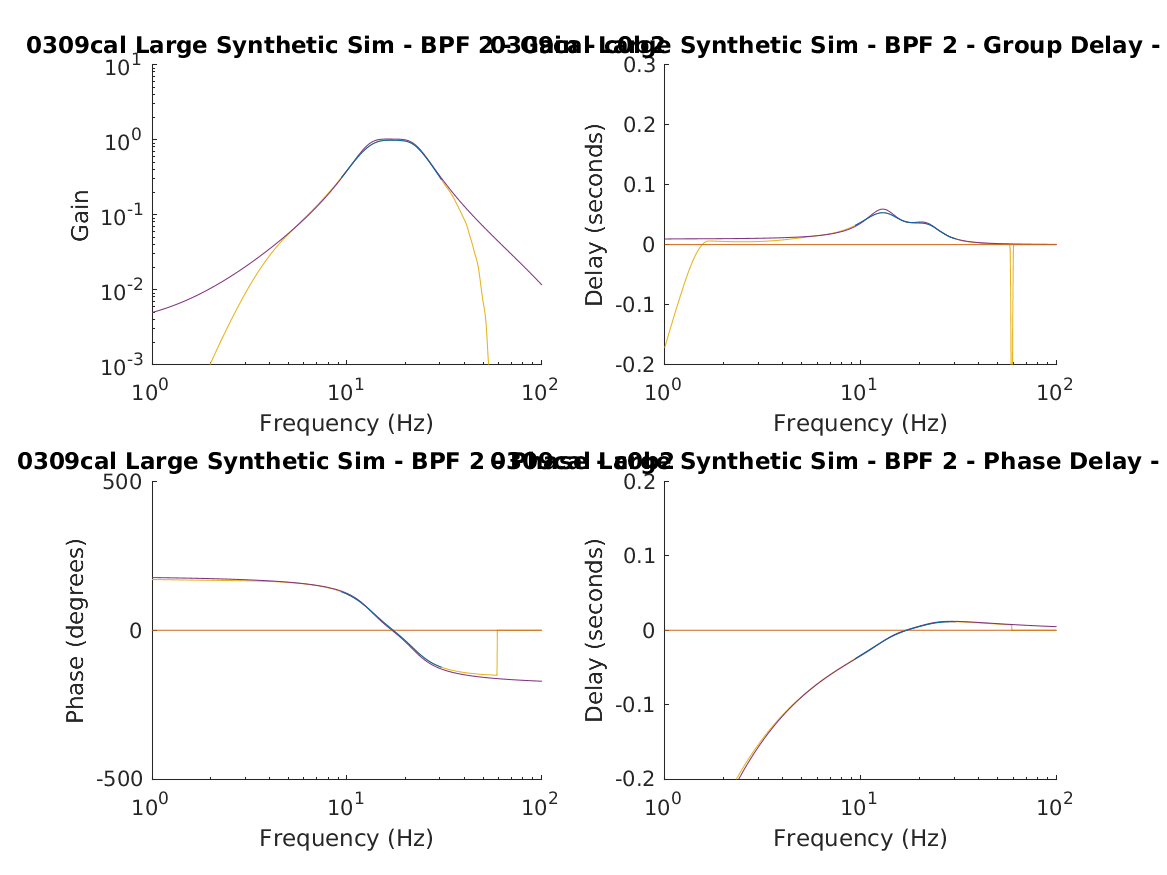}
\caption{Designed and measured transfer functions for the beta band infinite
impulse response filter. Upper left: gain response. Lower left: phase
response. Upper right: group delay. Lower right: phase delay.}
\label{fig-validation-filt-iirexample}
\end{figure}

A representative plot of the designed and measured transfer functions for
the ``beta band'' IIR filter is shown in Figure
\ref{fig-validation-filt-iirexample}, using the ``synthetic'' dataset as the
input signal. Within the regions of interest (blue in the plots), the
designed and measured transfer functions are virtually identical. The same
was observed in the measured transfer functions of IIR filters constructed
for other bands, and for the anti-aliasing IIR filter.
\fixme{Measured FIR plots go here.}{FIR plots.}
As a result, the filter implementation can be considered sound, and the
Matlab models of the filters may be used as proxies for the
real filter implementations without significant discrepancies expected.

All causal filters introduce delay into the filtered signal. For FIR filters,
this delay is constant, and for IIR filters, different frequency components
are delayed by different amounts. To allow later processing stages to
compensate for this, a calibration table of phase delay vs period was built.
This is discussed in Section \ref{sect-validation-cal}.

\subsection{Magnitude and Phase Estimation}
\label{sect-validation-feat}

Instantaneous magnitude and phase were estimated by looking for
zero-crossings in the band-pass-filtered waveform, inferring period and
phase from those zero crossings, and taking the maximum or minimum value of
the waveform between successive zero-crossings as the magnitude of the
signal.
\fixme{I-Q detection is an option too.}{Feature extraction.}
Estimation accuracy was characterized by comparing the oscillation
detector's estimates of instantaneous magnitude, phase, and frequency
(derived from period) to the instantaneous magnitude, phase, and (smoothed)
frequency computed from the band-pass-filtered signal by using Hilbert
transform to derive the imaginary component of the analytic signal.

\begin{figure}[!t]
\centering
\includegraphics[width=0.95\columnwidth]
{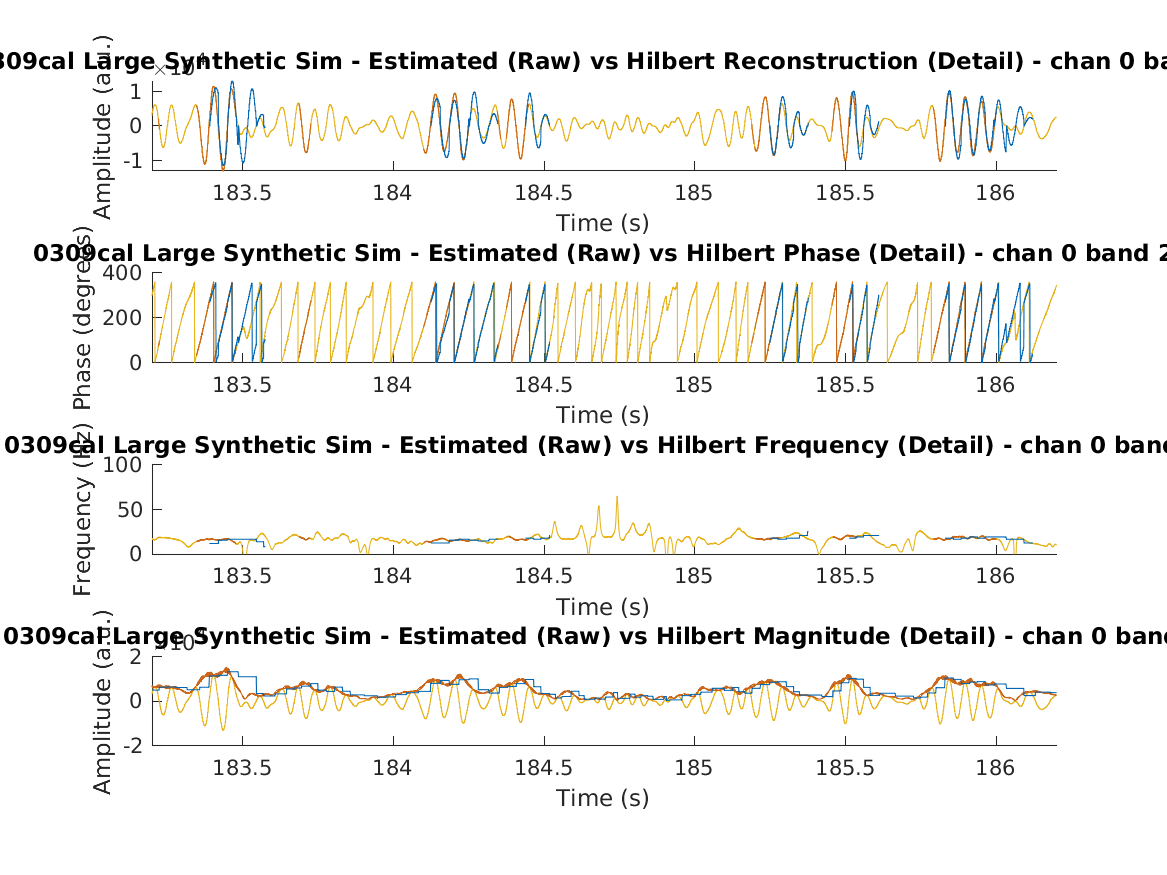}
\caption{Estimated vs true analytic magnitude, phase, frequency, and
waveform (real component). Estimated was performed using peak, trough, and
zero-crossing analysis of the band-pass filter output.
The estimated signal is shown in blue, during times when the oscillation
detector indicated an oscillation was present. The analytic signal is shown
in yellow, with orange regions indicating times when the instantaneous
magnitude was at least twice the average magnitude. Instantaneous frequency
of the analytic signal was computed as the smoothed derivative of phase.
``Synthetic'' dataset, IIR filters, beta band.}
\label{fig-validation-feat-ptrecon}
\end{figure}

Figure \ref{fig-validation-feat-ptrecon} shows a representative
reconstruction of magnitude, phase, frequency, and waveform using the
peak-trough-ZC feature extractor (blue) and using the analytic signal
(orange) (beta band signal, IIR filters, ``synthetic'' dataset). Feature
extractor reconstruction was performed in regions where the oscillation
detector described in \ref{sect-design-arch} indicated oscillation, and
overlaid on top of the analytic signal plots. Regions where the analytic
signal's instantaneous magnitude is greater than twice the average magnitude
are indicated in the plot for comparison.

\begin{figure}[!t]
\centering
\begin{tabular}{cc}
\includegraphics[width=0.45\columnwidth]
{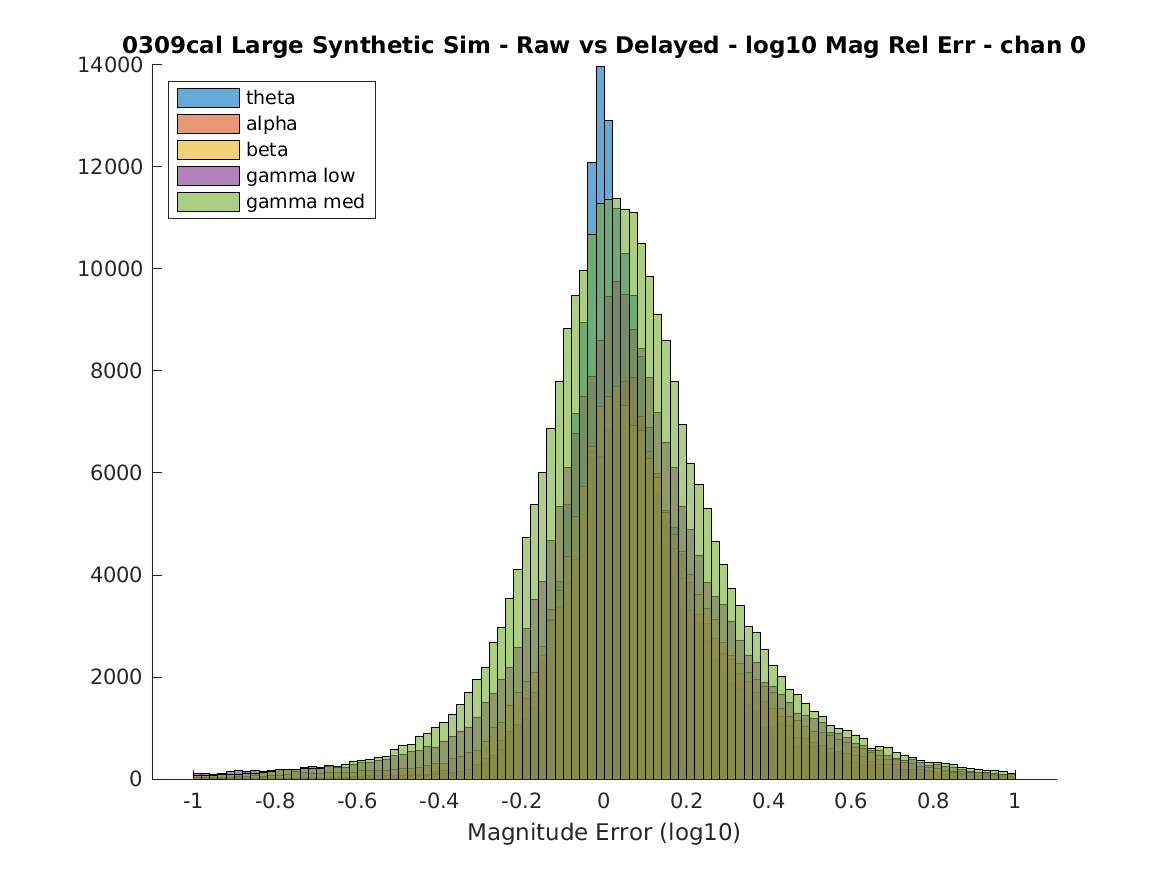}
&
\includegraphics[width=0.45\columnwidth]
{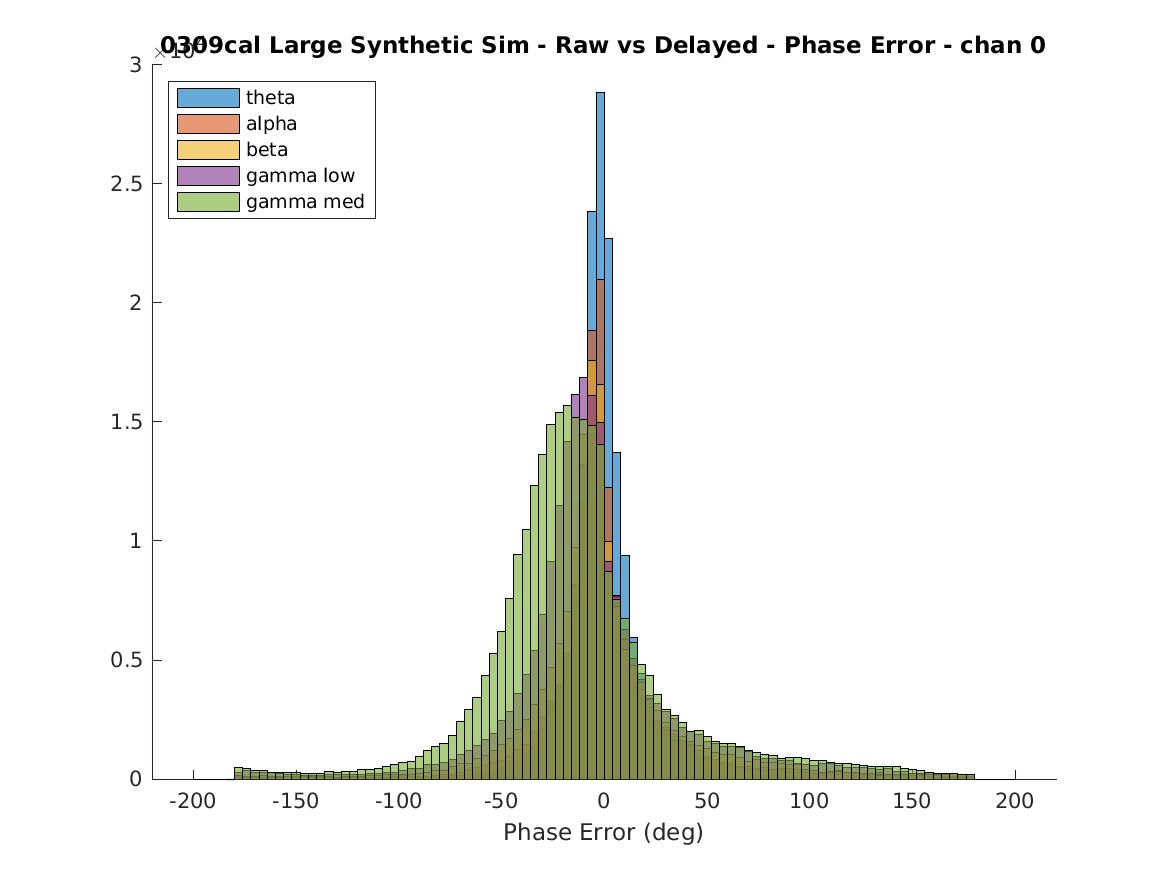}
\\
\includegraphics[width=0.45\columnwidth]
{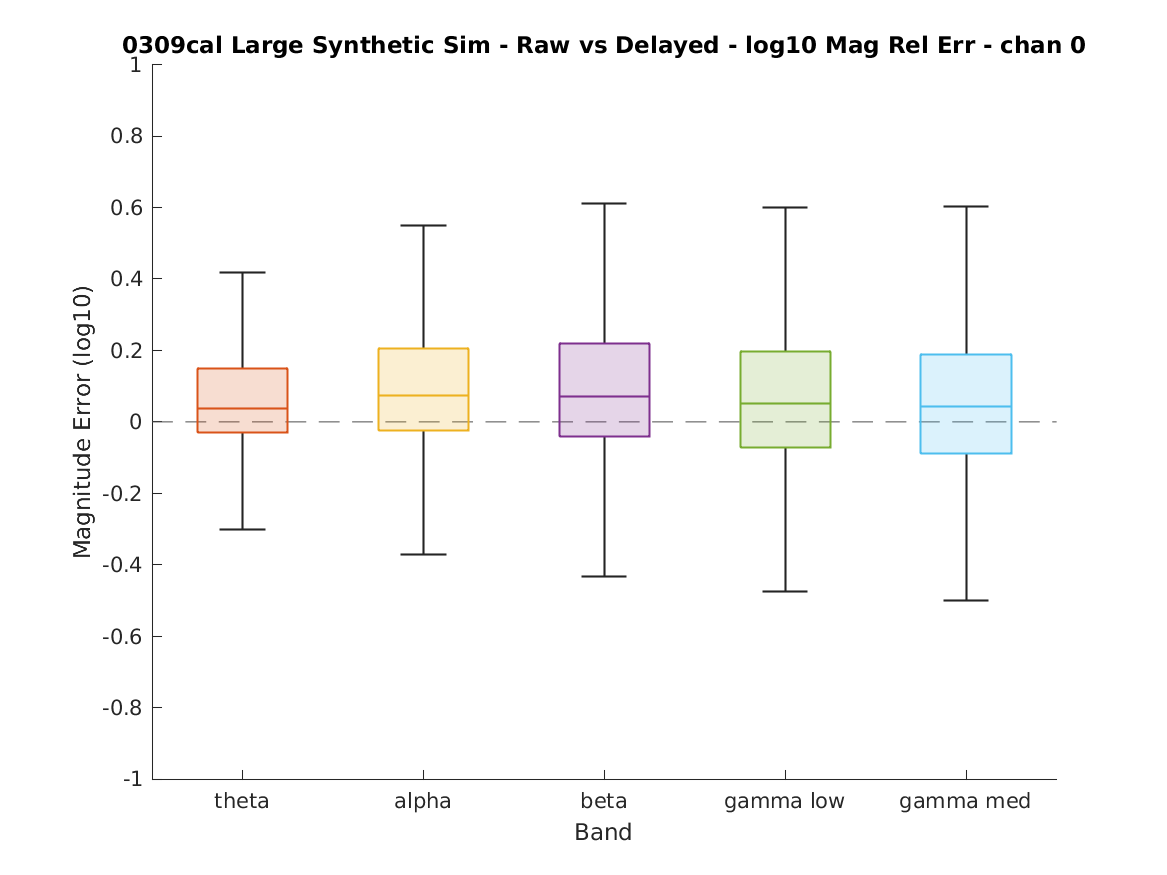}
&
\includegraphics[width=0.45\columnwidth]
{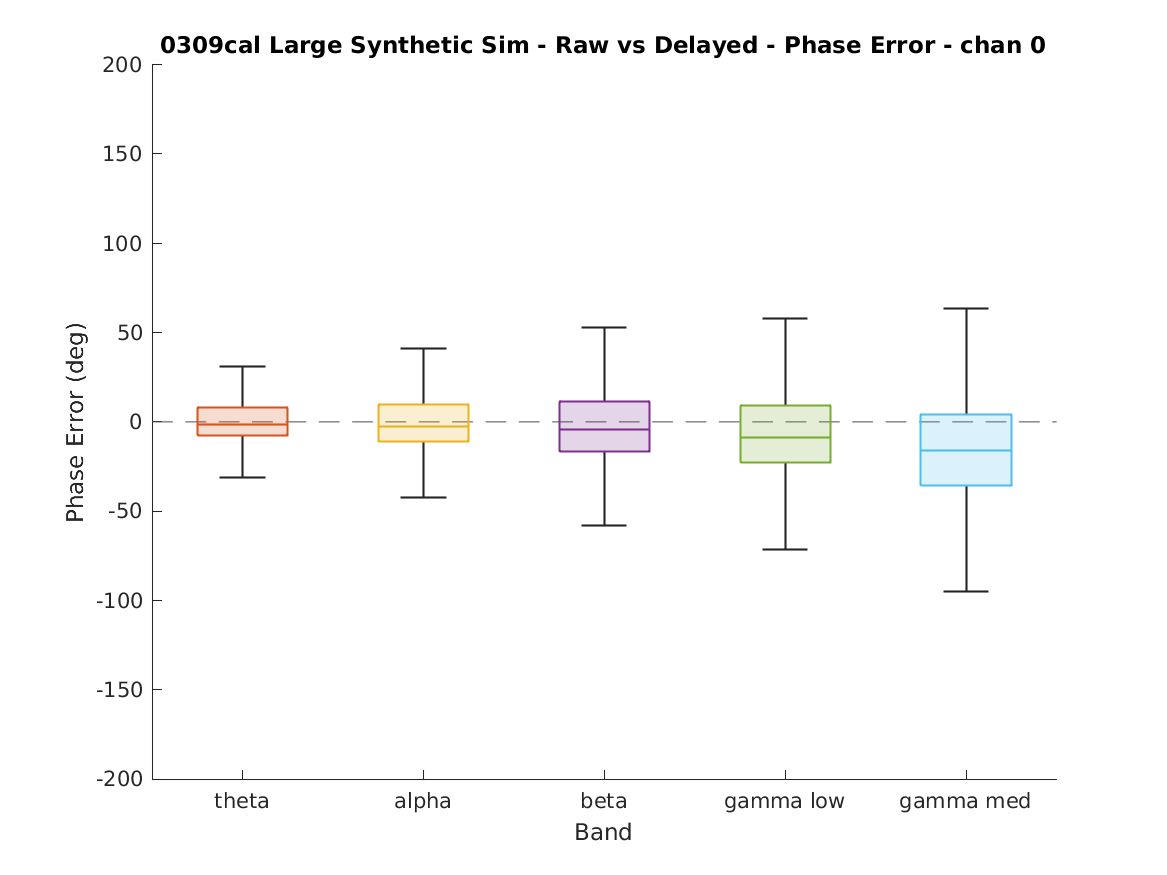}
\\
\includegraphics[width=0.45\columnwidth]
{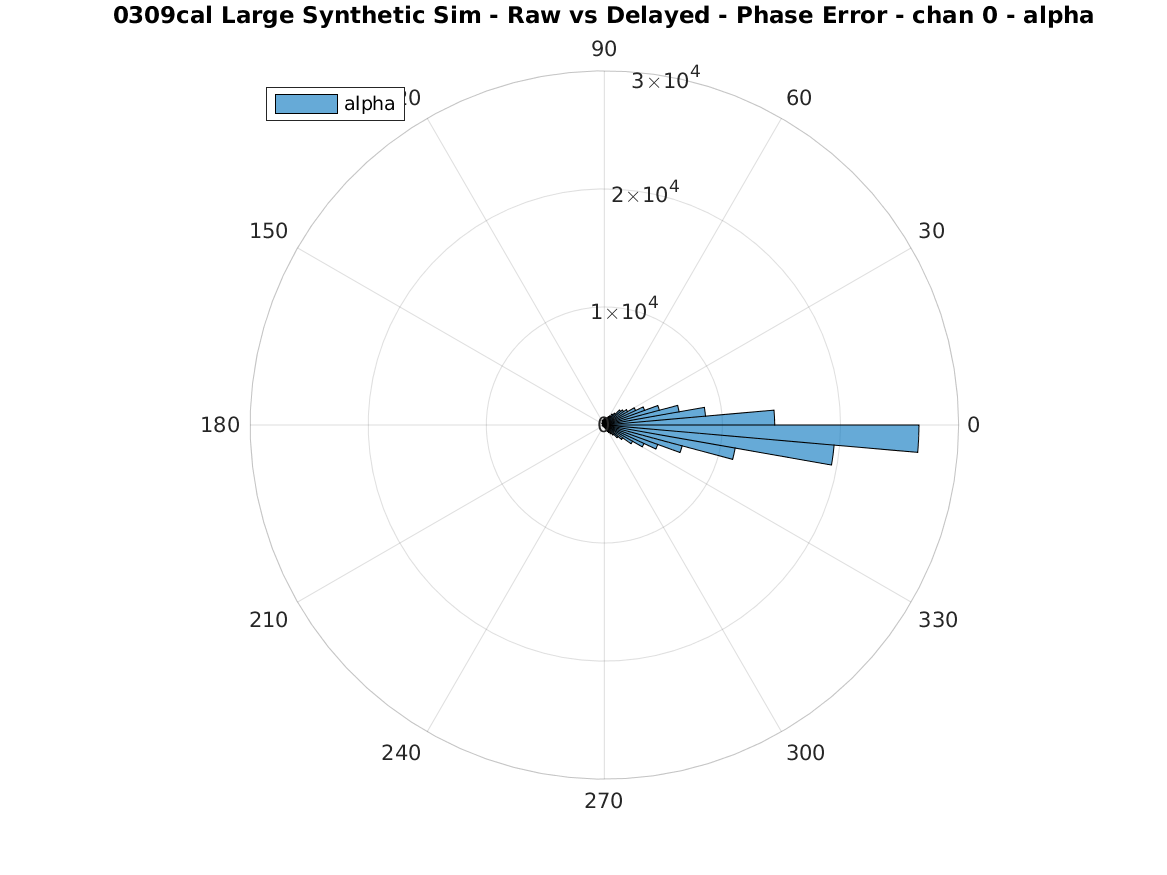}
&
\includegraphics[width=0.45\columnwidth]
{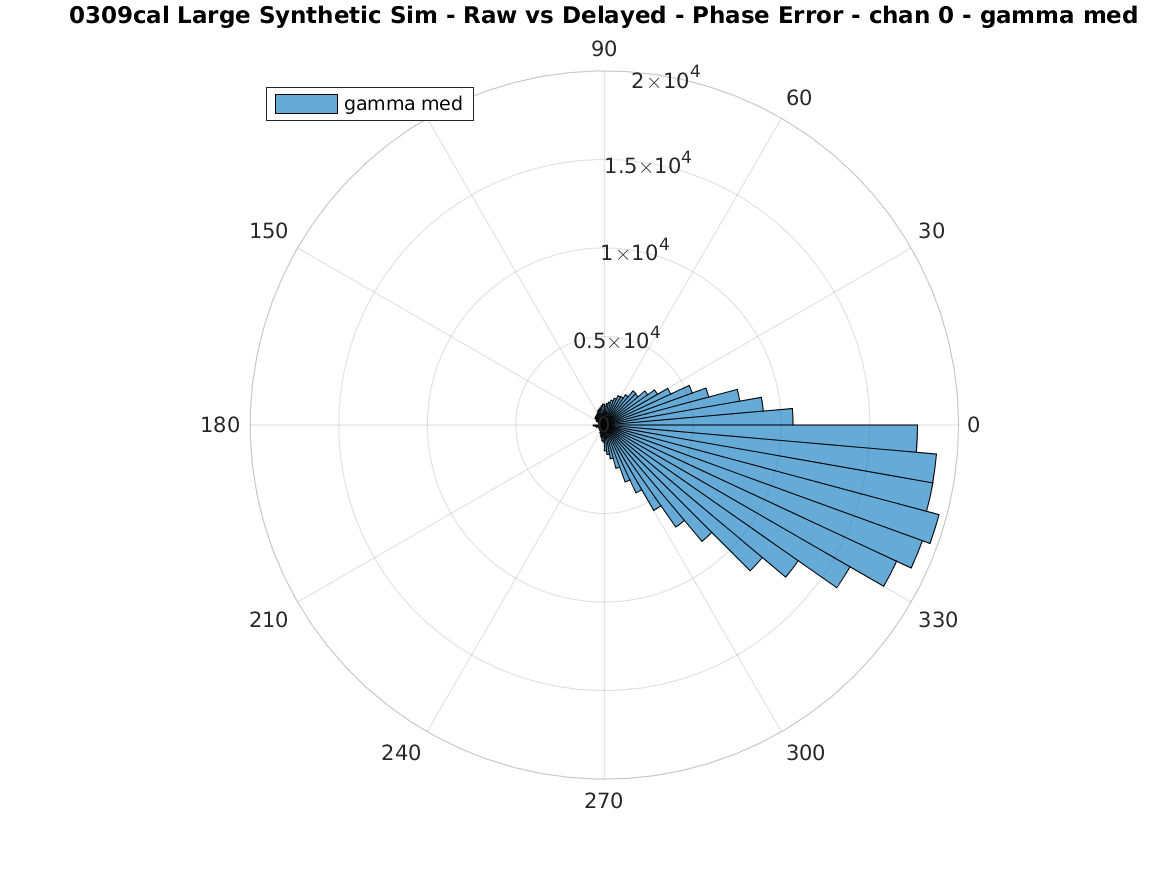}
\\
\end{tabular}
\caption{Normalized estimated magnitude error (left) and absolute
estimated phase error (right) with respect to analytic signal magnitude
and phase for the band-pass-filtered signal. The top row shows histograms
of sample-by-sample error during oscillations. The middle row shows
box plots of sample-by-sample error. The bottom row shows representative
rose plots of phase error for the alpha band (left) and middle gamma band
(right). IIR filters, ``synthetic'' dataset.}
\label{fig-validation-feat-magphase}
\end{figure}

Comparing estimated magnitude and phase to those of the analytic signal
computed from the band-pass-filtered signal shows whether the oscillation
detector's approximation of the instantaneous magnitude and phase are
accurate.
Figure \ref{fig-validation-feat-magphase} shows histograms and box plots of
magnitude error normalized to the analytic signal magnitude (relative error)
and of phase error with respect to the analytic signal phase.
Representative rose plots of phase estimation error are
shown for the alpha band (left) and medium gamma band (right).
This analysis was performed using the ``synthetic'' dataset and IIR filters.

\fixme{Need FIR plots.}{Peak and trough error plots (synthetic).}

Magnitude error distributions are broad in all cases. This is because the
envelopes of oscillations change on a timescale that is not substantially
longer than the analysis timescale (one half-period of the oscillation). As
the magnitude estimate is out of date by half a period, there may be a
considerable difference between the estimated and actual magnitudes. This
can be seen in the bottom strip in Figure \ref{fig-validation-feat-ptrecon};
the estimated envelope is time-shifted relative to the actual envelope.

Phase error with respect to the band-pass-filtered wave is tightly
clustered for theta, alpha, and beta bands ($\le 20^\circ$ FWHM,
$\le 5^\circ$ offset). This error range is primarily due to frequency shifts
during the oscillation. For higher-frequency bands (low and middle gamma),
additional noise
is present due to quantization of the detected half-period into an integer
number of samples and due to noise perturbing the detected locations of
zero-crossings (a single-sample shift at 100~Hz introduces a much larger
phase error than a single-sample shift at 10~Hz).

\begin{figure}[!t]
\centering
\begin{tabular}{cc}
\includegraphics[width=0.45\columnwidth]
{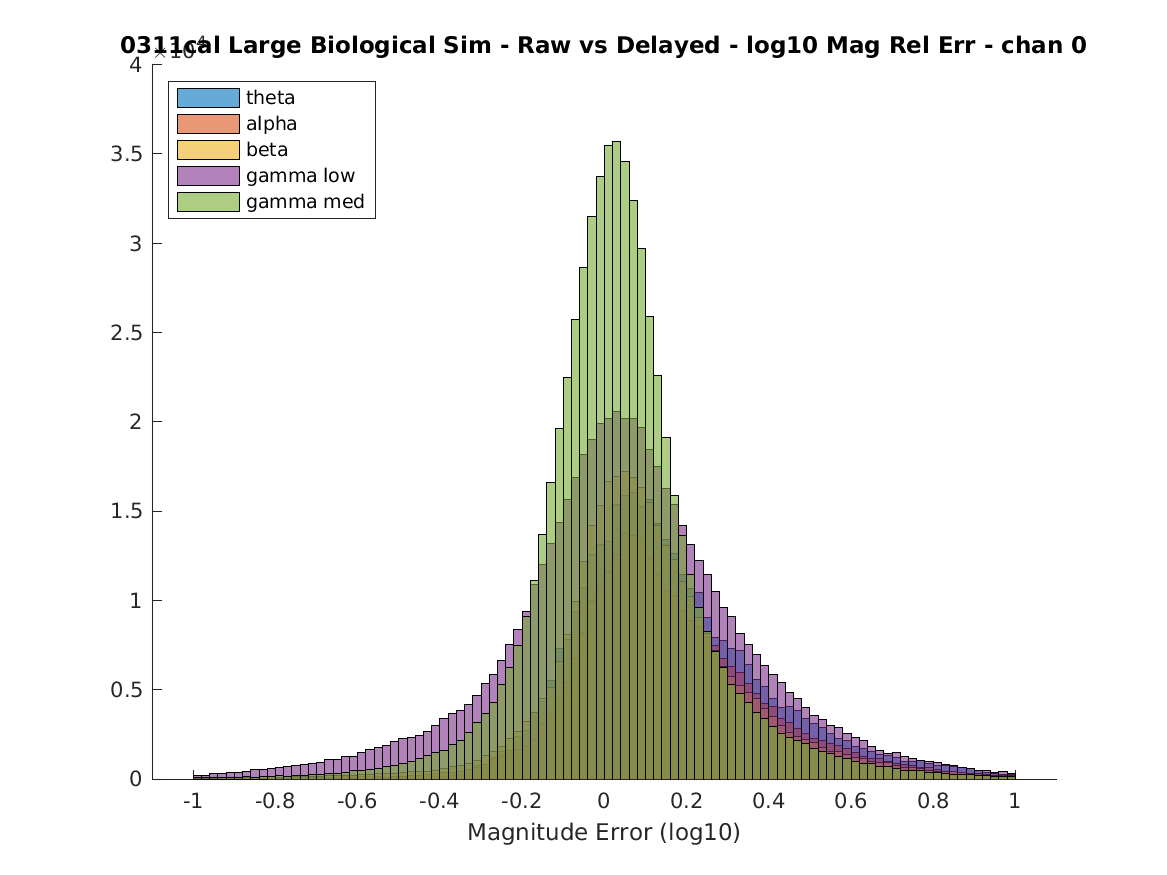}
&
\includegraphics[width=0.45\columnwidth]
{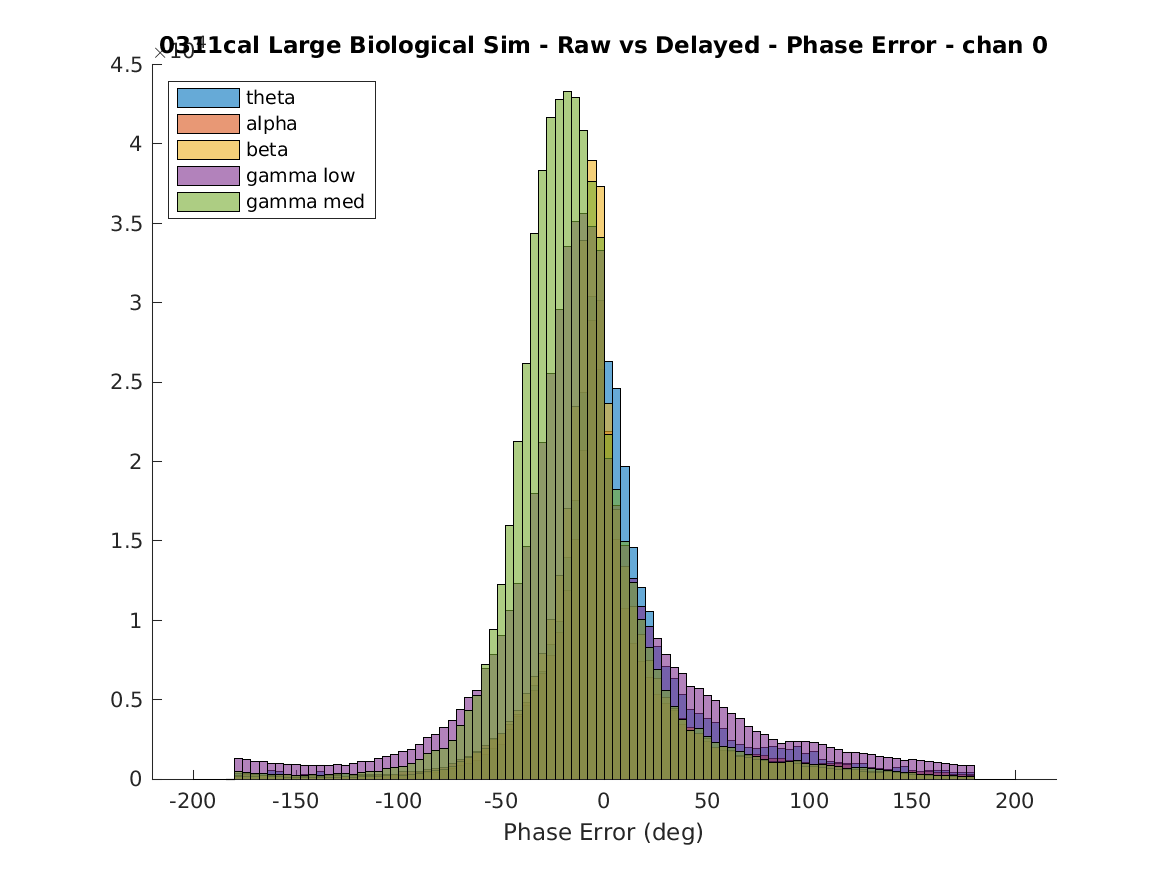}
\\
\includegraphics[width=0.45\columnwidth]
{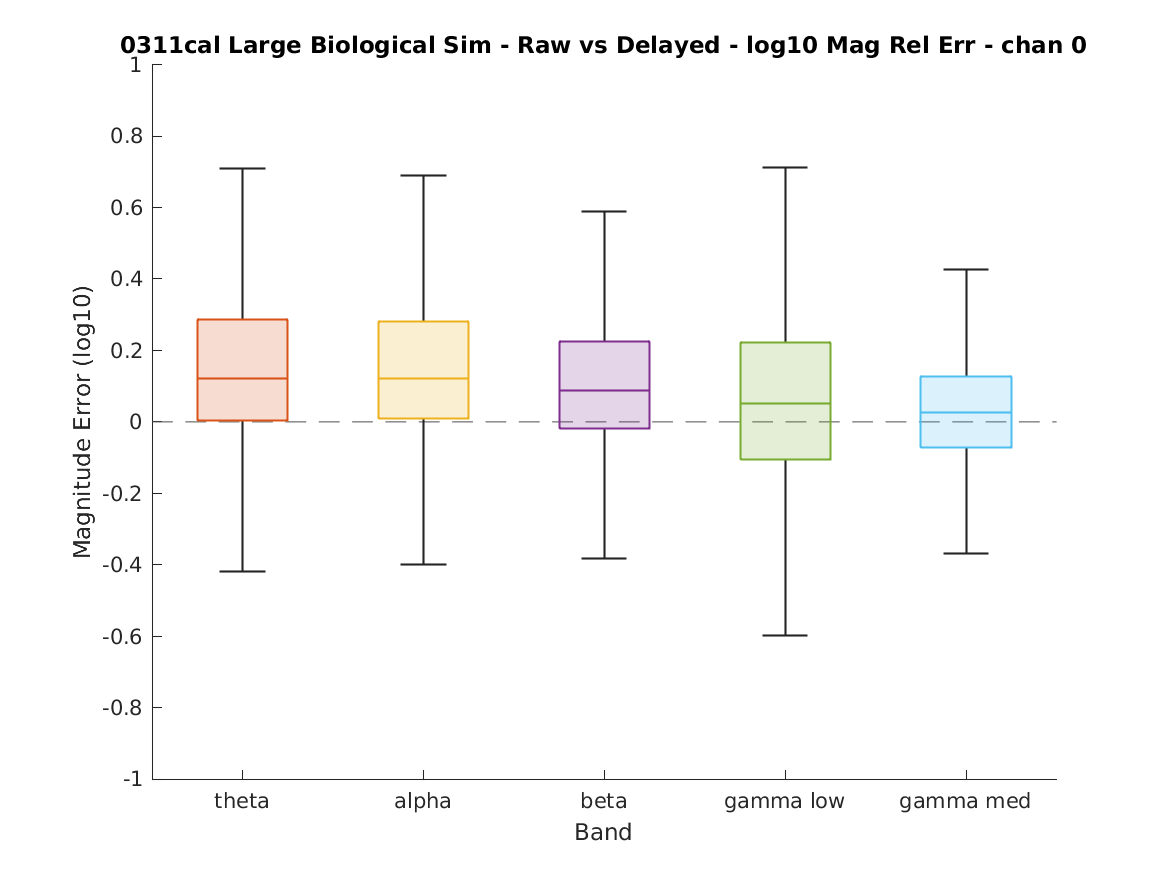}
&
\includegraphics[width=0.45\columnwidth]
{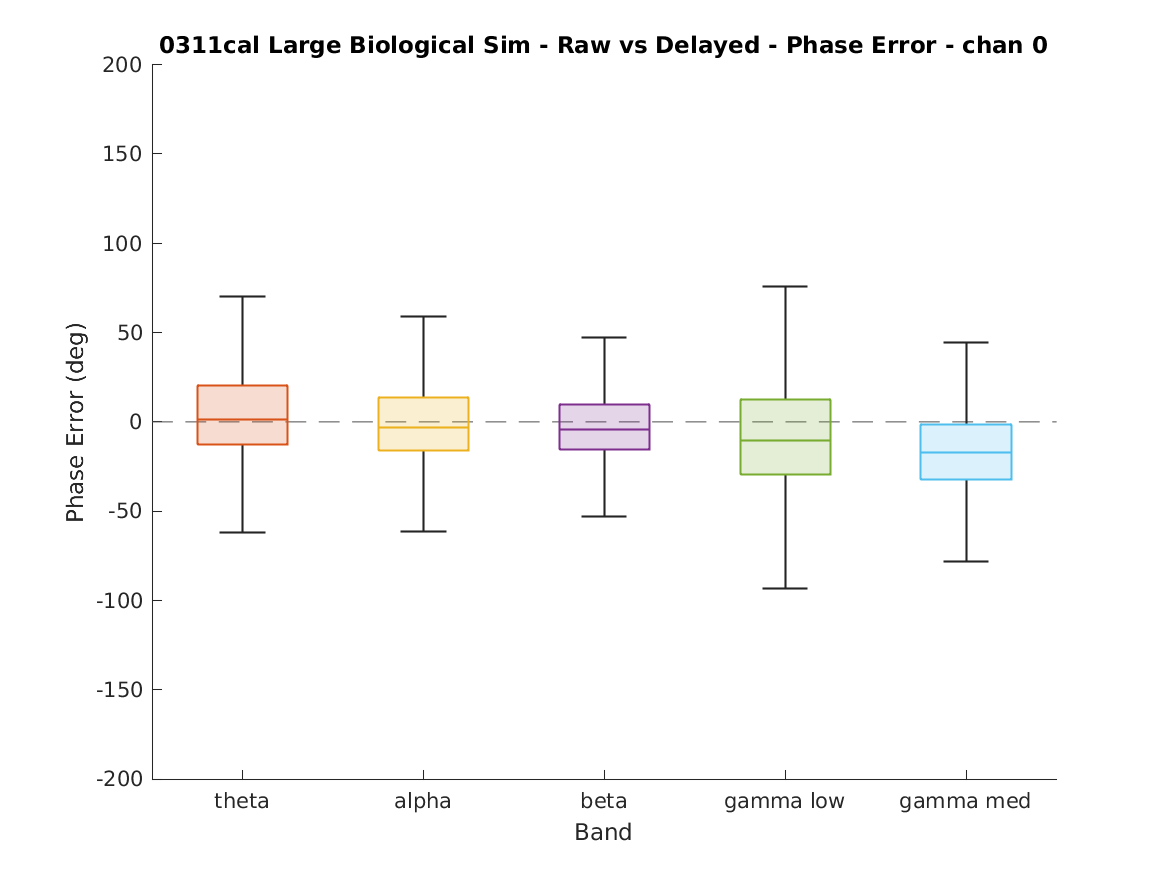}
\\
\includegraphics[width=0.45\columnwidth]
{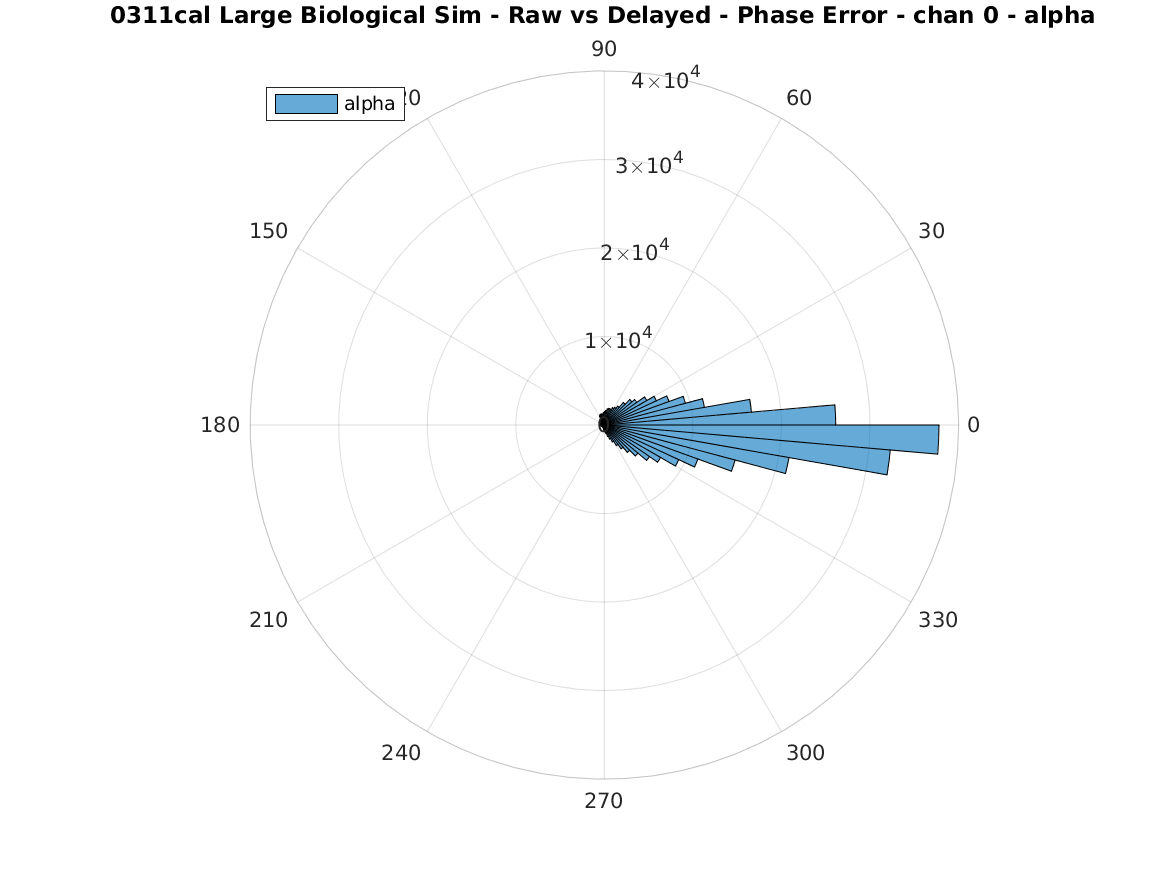}
&
\includegraphics[width=0.45\columnwidth]
{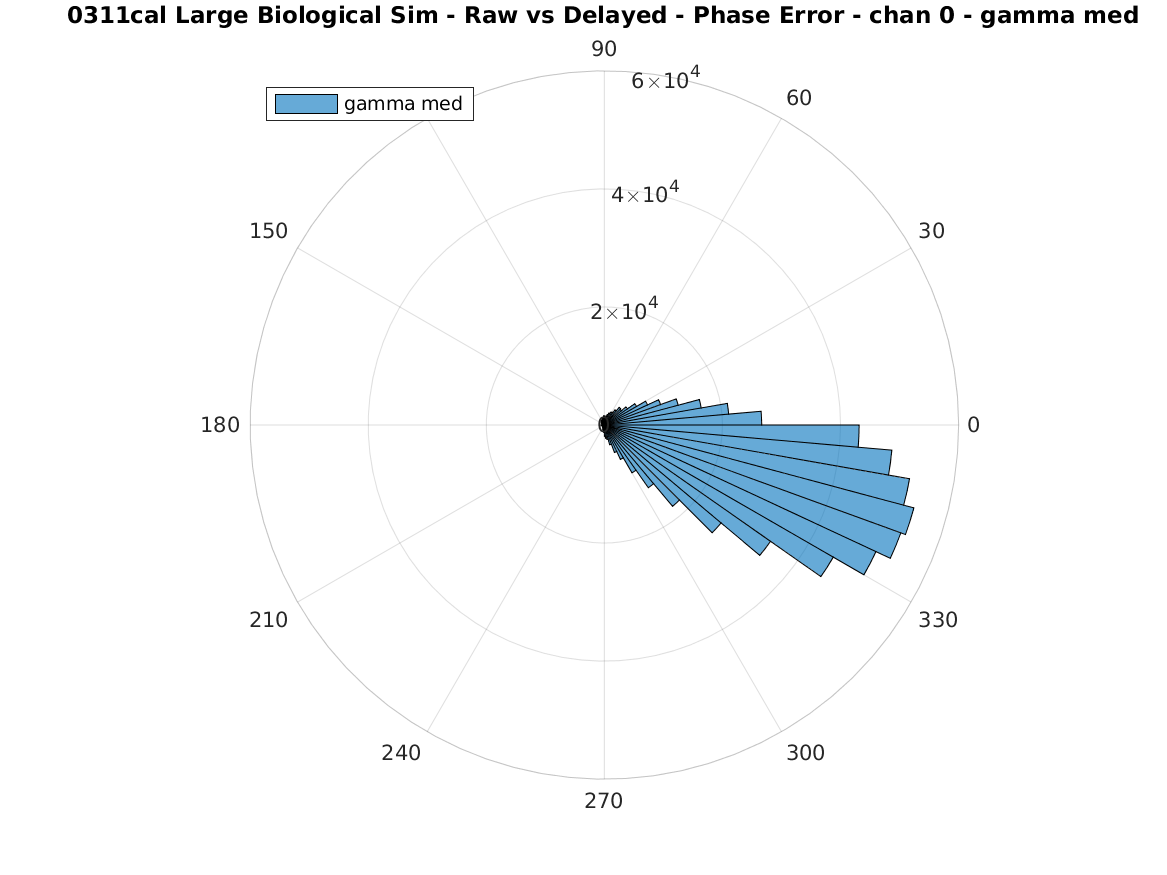}
\\
\end{tabular}
\caption{Normalized estimated magnitude error (left) and absolute
estimated phase error (right) with respect to analytic signal magnitude
and phase for the band-pass-filtered signal. The top row shows histograms
of sample-by-sample error during oscillations. The middle row shows
box plots of sample-by-sample error. The bottom row shows representative
rose plots of phase error for the alpha band (left) and middle gamma band
(right). IIR filters, ``biological'' dataset.}
\label{fig-validation-feat-monkey}
\end{figure}

Figure \ref{fig-validation-feat-monkey} shows histograms and box plots of
magnitude error normalized to the analytic signal magnitude
and of phase error with respect to the analytic signal phase
for the ``biological'' dataset, using the same IIR filters.
Representative rose plots of phase estimation error are again shown
for the alpha band (left) and medium gamma band (right).

\fixme{Need FIR plots.}{Peak and trough error plots (biological).}

Magnitude error distributions are again broad (typically $\pm 40\%$
variation). Phase error with respect to the band-pass-filtered wave is
less tightly clustered than with the ``synthetic'' dataset ($\le 30^\circ$
FWHM, $\le 5^\circ$ offset), but still well within the design requirements
from Section \ref{sect-background-ephys} ($\le 60^\circ$ FWHM). There is
again additional spread in the middle gamma band.

\subsection{Delay-Aligned and Phase-Aligned Triggering}
\label{sect-validation-trig}

Stimulation trigger alignment was characterized by specifying a desired
delay in milliseconds from the rising or falling zero-crossing, or a
desired phase angle, and measuring the distribution of delays and phase
angles at which stimulation trigger signals were actually generated.
This was evaluated with respect to two reference delays and phase angles:
The ``estimated'' set (the output of the peak-and-trough estimator
discussed in Section \ref{sect-validation-feat}), and the ``band-pass'' set
(the zero-crossings and instantaneous phase of the band-pass-filtered
signal). Testing against the ``estimated'' set characterizes errors
introduced by the trigger generation logic, while testing against the
``band-pass'' set characterizes the errors introduced by the combined
processing pipeline (in particular, phase perturbation caused by
mis-estimation of instantaneous frequency).

\begin{figure}[!t]
\centering
\begin{tabular}{cc}
\includegraphics[width=0.45\columnwidth]
{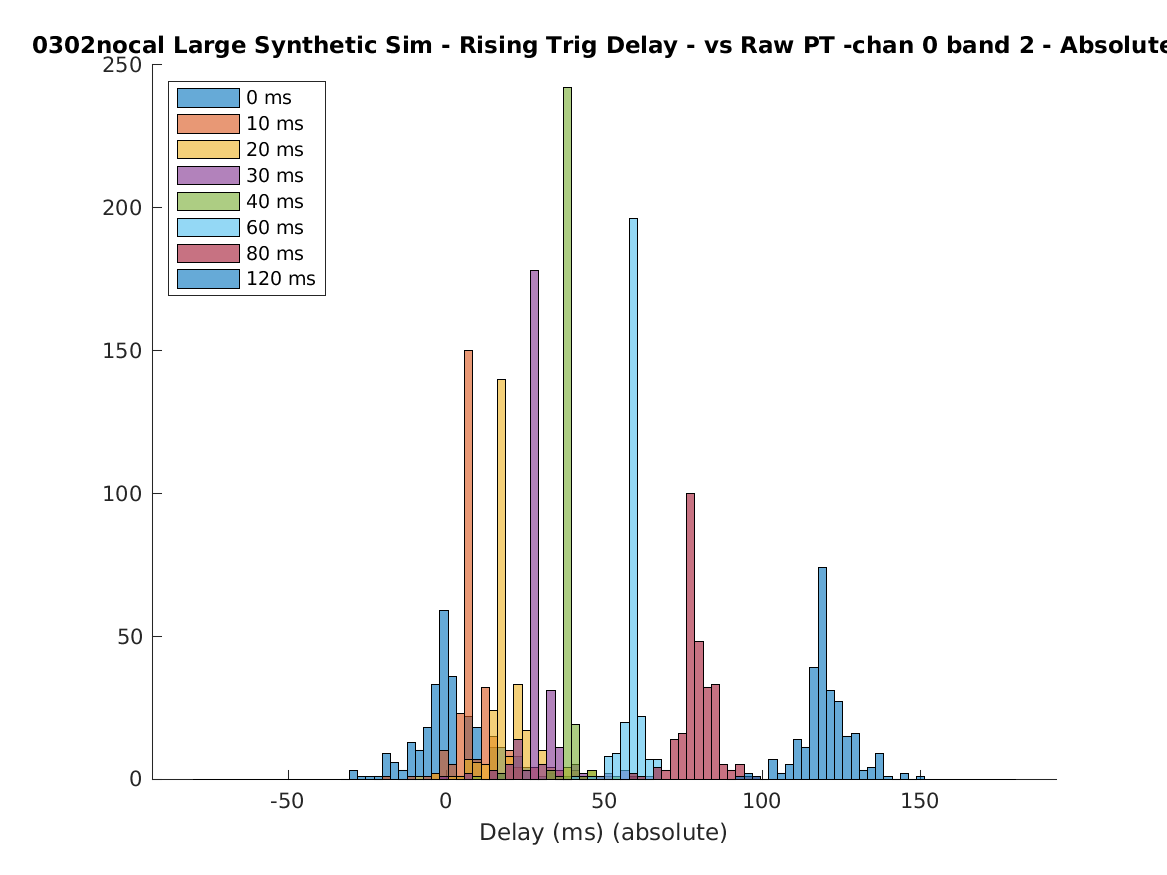}
&
\includegraphics[width=0.45\columnwidth]
{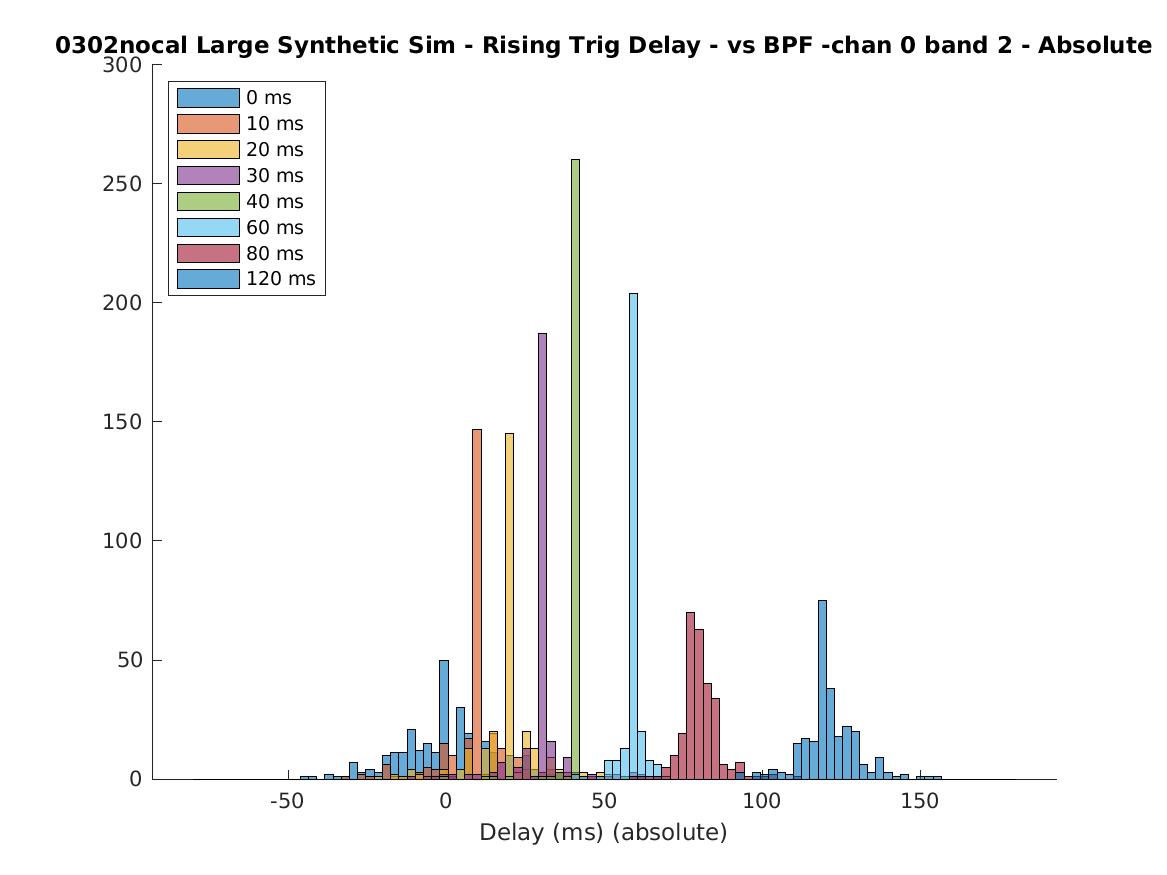}
\\
\includegraphics[width=0.45\columnwidth]
{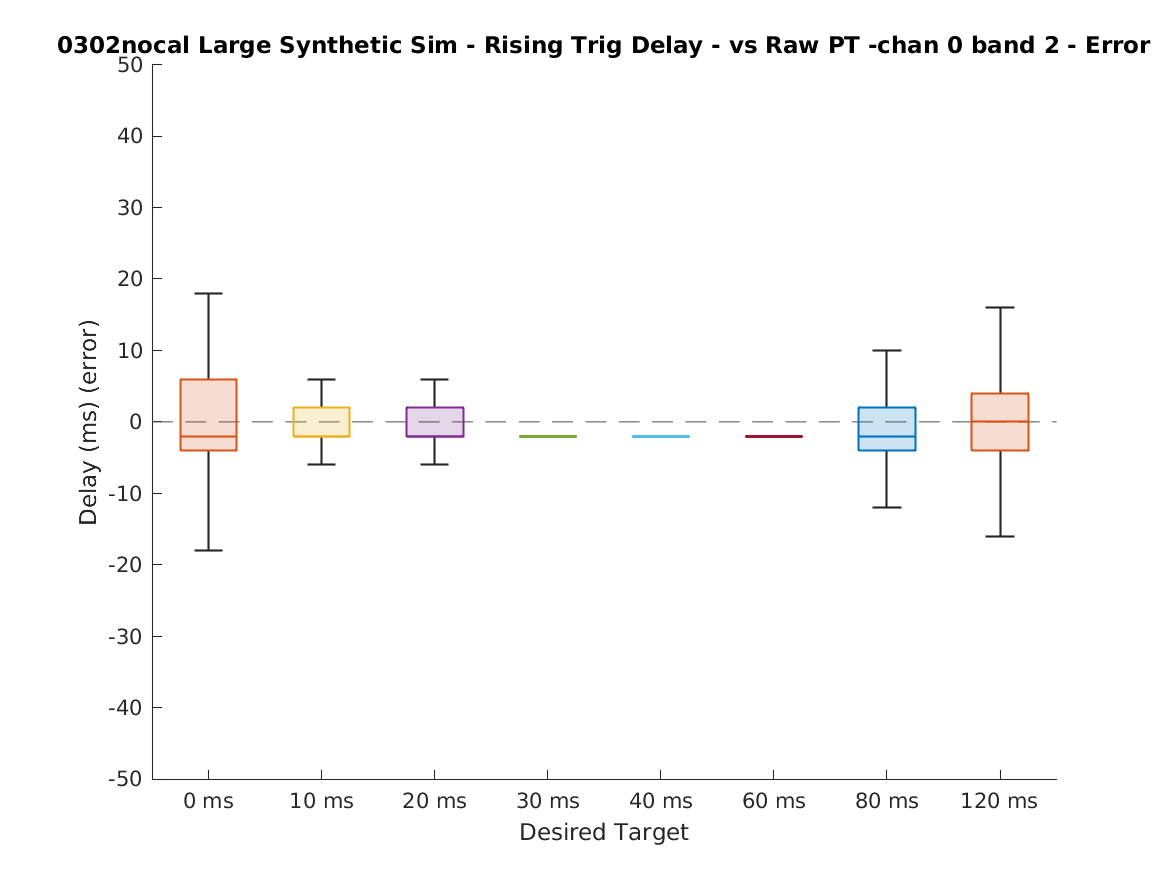}
&
\includegraphics[width=0.45\columnwidth]
{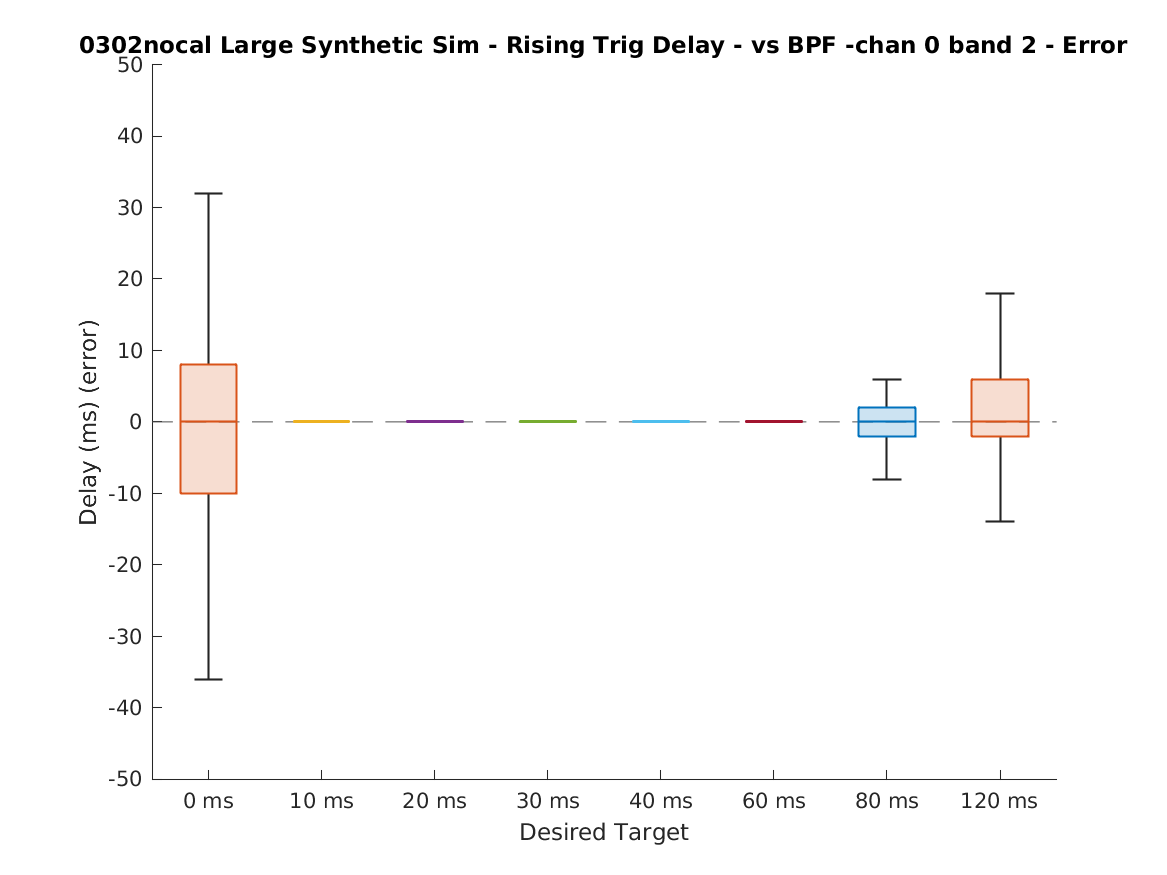}
\\
\end{tabular}
\caption{Representative plots of trigger delay (top) and delay error (bottom)
for triggers scheduled with respect to the rising zero-crossing. Left:
measurements with respect to the estimated signal. Right: measurements
with respect to the band-pass-filtered signal. Measurements are grouped
by desired delay. ``Synthetic'' dataset, beta band, IIR filters.}
\label{fig-validation-trig-risedelay}
\end{figure}

\begin{figure}[!t]
\centering
\begin{tabular}{cc}
\includegraphics[width=0.45\columnwidth]
{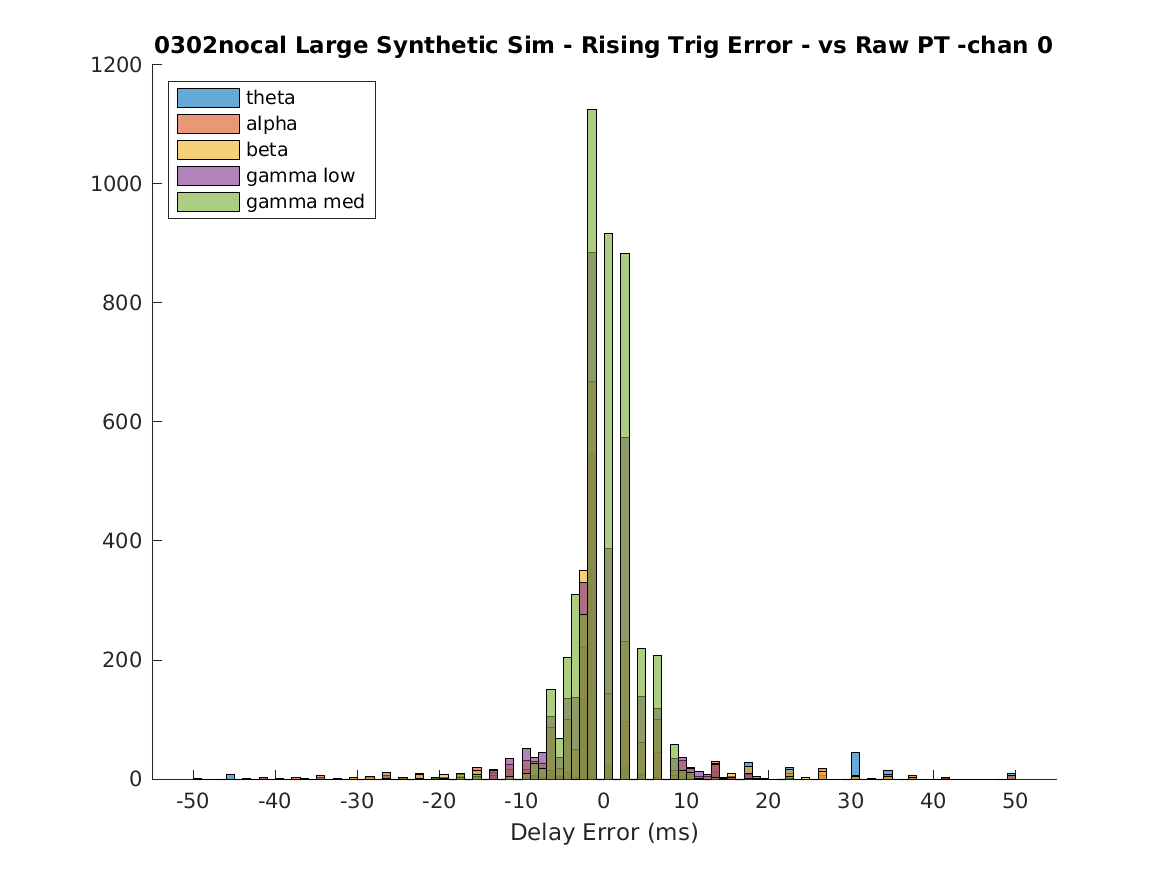}
&
\includegraphics[width=0.45\columnwidth]
{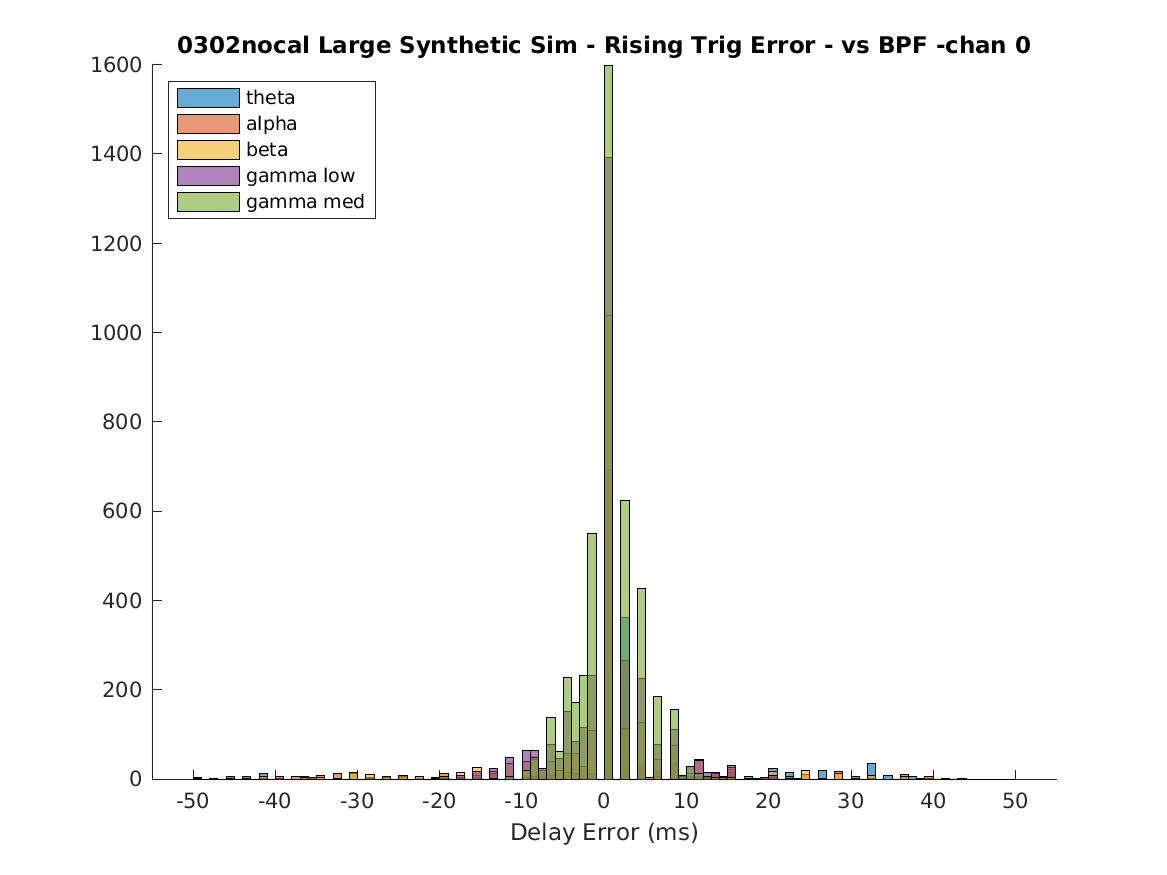}
\\
\includegraphics[width=0.45\columnwidth]
{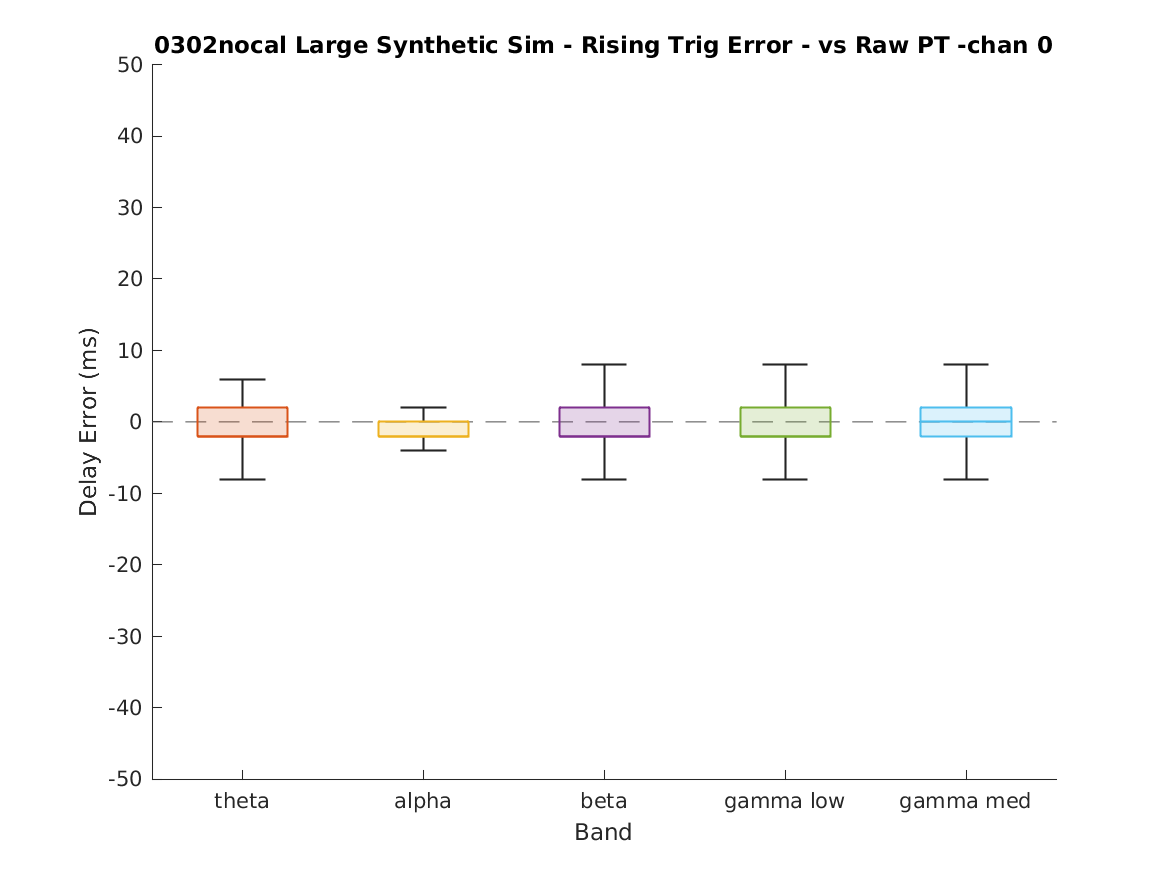}
&
\includegraphics[width=0.45\columnwidth]
{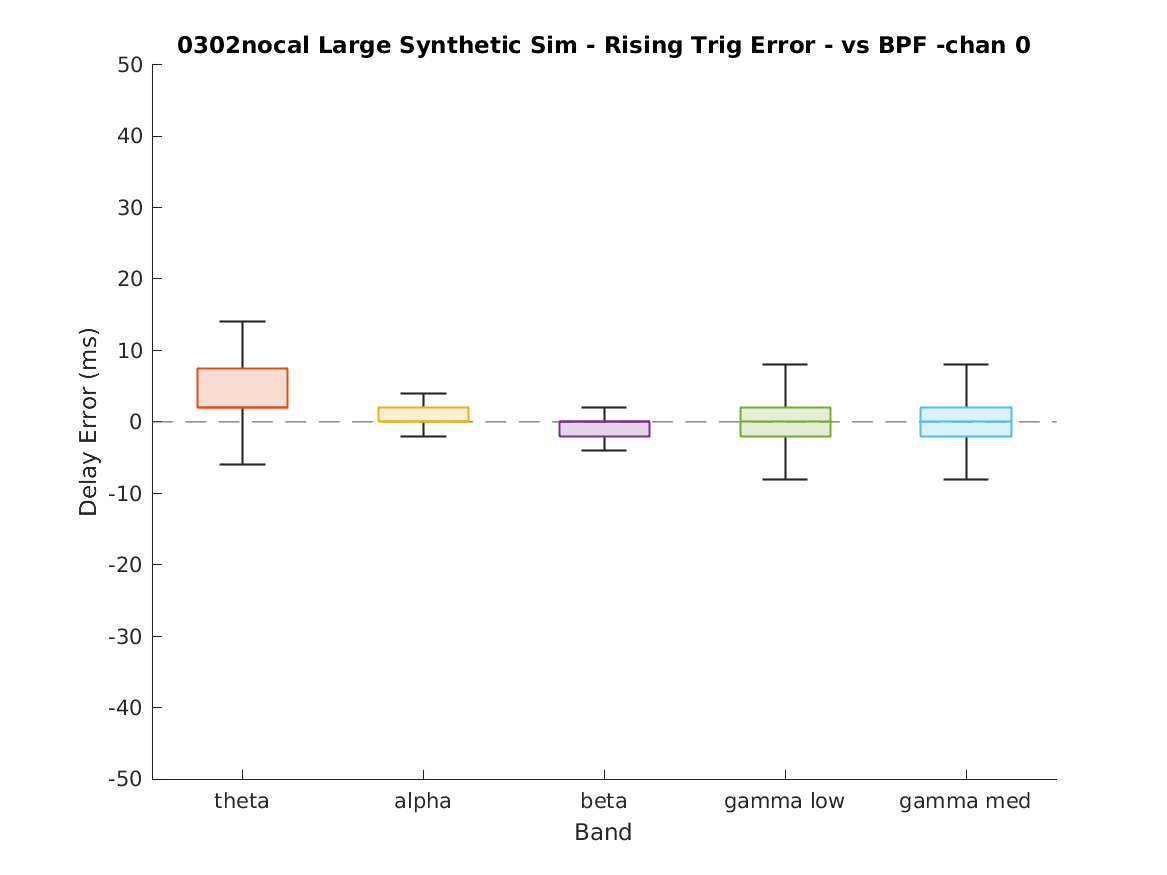}
\\
\end{tabular}
\caption{Plots of trigger delay error for triggers scheduled with respect
to the rising zero-crossing, aggregated across all target delays. Left:
measurements with respect to the estimated signal. Right: measurements
with respect to the band-pass-filtered signal. Measurements are grouped
by band. ``Synthetic'' dataset, IIR filters.}
\label{fig-validation-trig-riseband}
\end{figure}

Figure \ref{fig-validation-trig-risedelay} shows representative plots
of trigger delay (top) and of delay error (bottom) for triggers
scheduled with respect to the rising zero-crossing. The left set of plots
shows trigger delay with respect to the estimated zero-crossings, and the
right set of plots shows trigger delay with respect to the band-pass-filtered
signal's zero crossings. These measurements were taken using the beta band
IIR filter and the ``synthetic'' dataset, and are plotted for multiple
target delays. Figure \ref{fig-validation-trig-riseband} shows plots of
delay error for the same tests case aggregated by band.

\fixme{Need monkey nocal trigger plots.}{Rising trigger plots.}

From Figure \ref{fig-validation-trig-riseband}, delay-scheduled triggers
are sent with minimal delay ($\pm 1$~sample at 500~sps) with respect to
their intended trigger times. For many cases, the delay is negligible; as
illustrated in Figure \ref{fig-validation-trig-risedelay}, error is dominated
by either requested delays that are one or more periods long (beta band and
higher-frequency) or by a requested delay of zero (beta band and
lower-frequency). In the case of a delay of one period or longer,
significant frequency drift is expected to occur in the input signal. In
the case of a delay of zero, the trigger logic reschedules it for the
following period, resulting in a delay of one period.
\fixme{Frequency drift shouldn't corrupt ZC-based scheduling, though?}

\begin{figure}[!t]
\centering
\begin{tabular}{cc}
\includegraphics[width=0.45\columnwidth]
{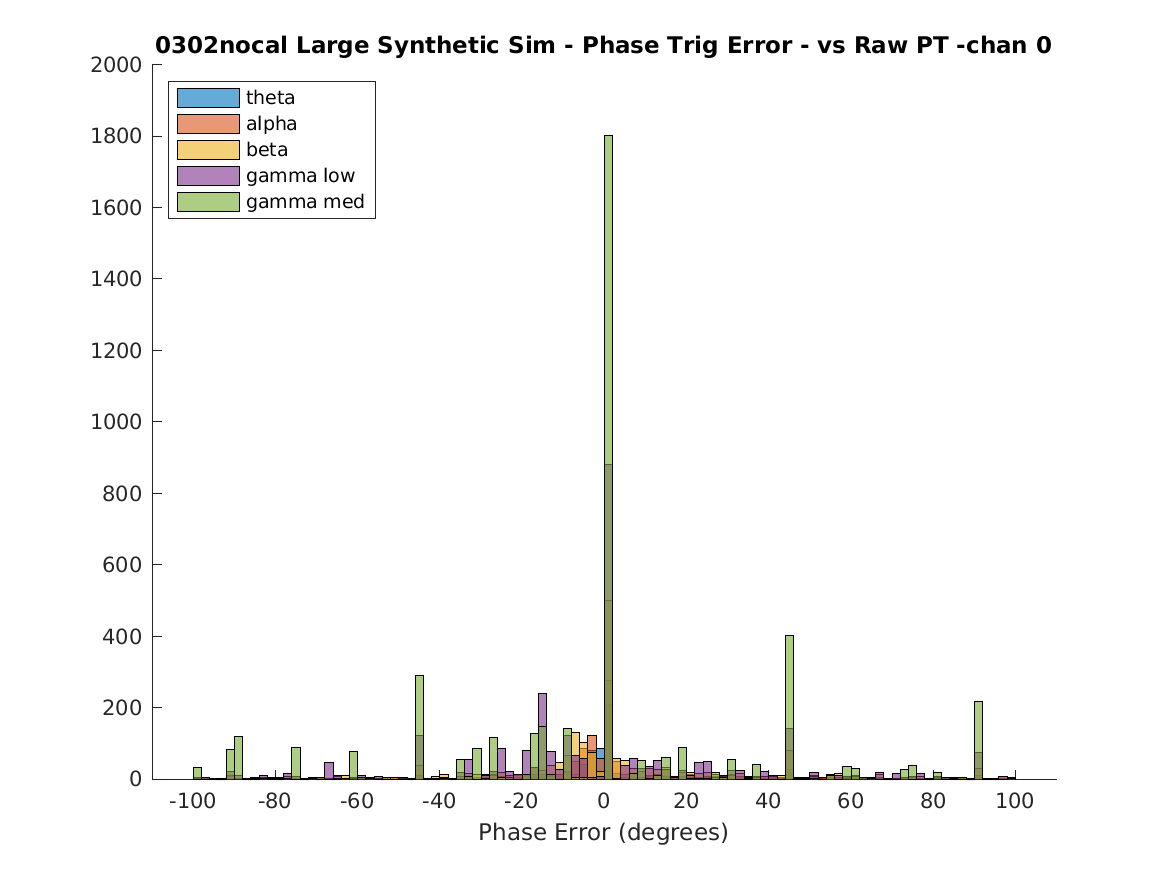}
&
\includegraphics[width=0.45\columnwidth]
{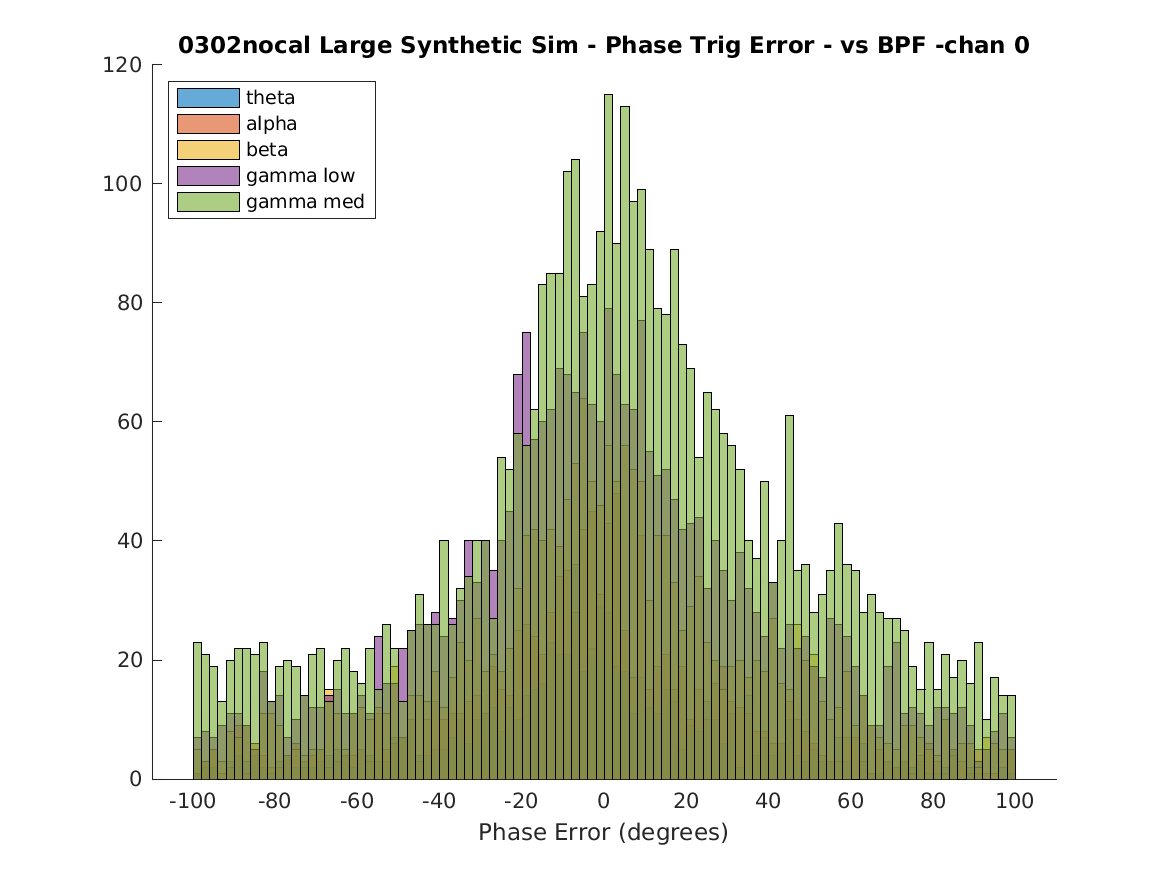}
\\
\includegraphics[width=0.45\columnwidth]
{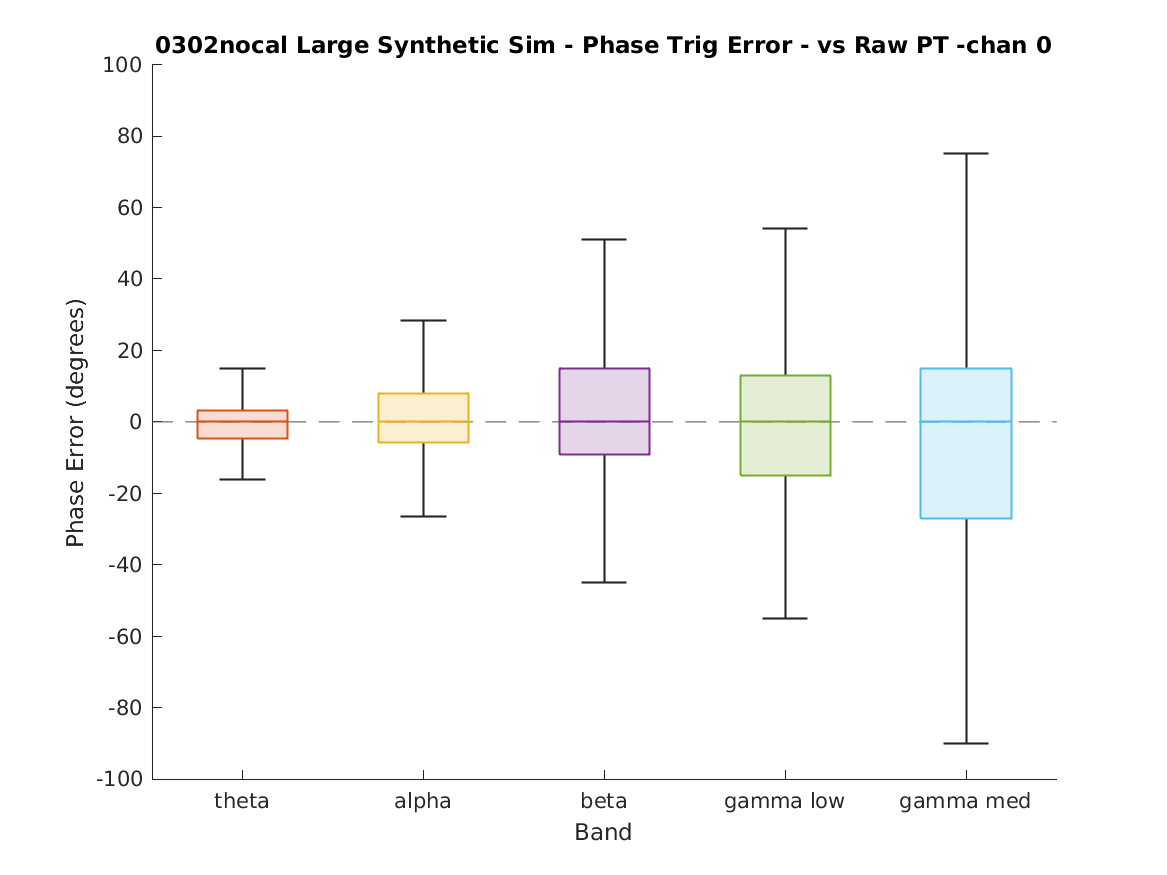}
&
\includegraphics[width=0.45\columnwidth]
{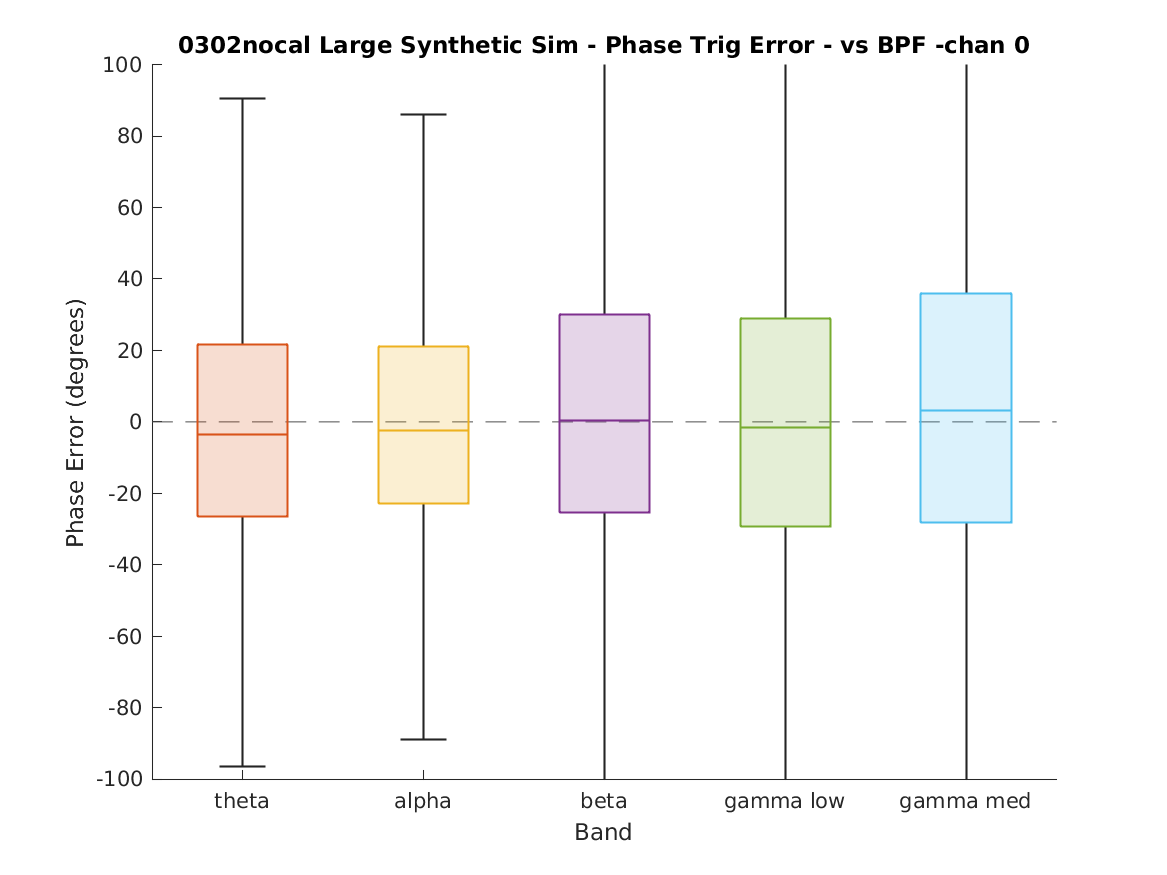}
\\
\includegraphics[width=0.45\columnwidth]
{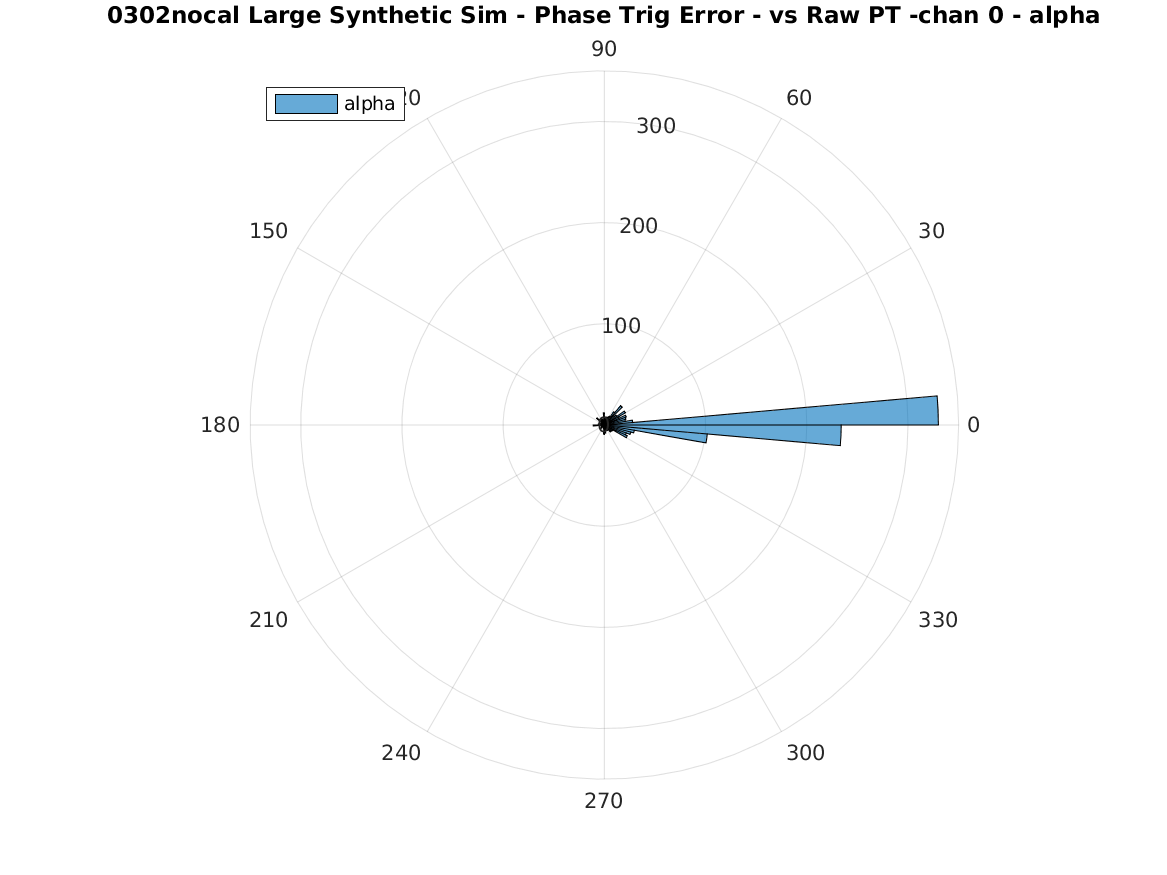}
&
\includegraphics[width=0.45\columnwidth]
{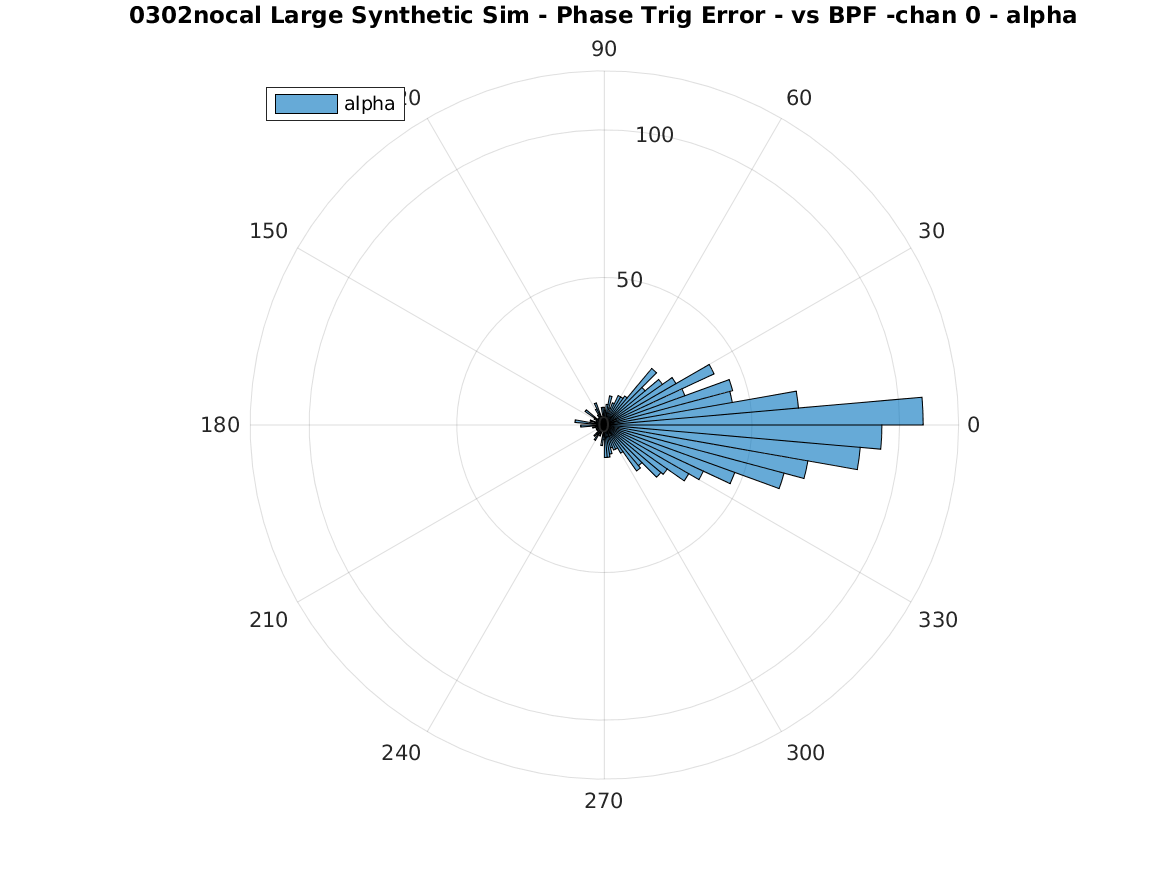}
\\
\end{tabular}
\caption{Plots of trigger phase error for triggers scheduled by phase,
aggregated across all target phases. Left: measurements with respect to the
estimated signal. Right: measurements with respect to the band-pass-filtered
signal.
Measurements are grouped by band.
Bottom row: representative rose plots of phase error (alpha band).
``Synthetic'' dataset, IIR filters.}
\label{fig-validation-trig-phaseband}
\end{figure}

Figure \ref{fig-validation-trig-phaseband} shows plots of trigger phase
error for each band (top and middle rows) and for the alpha band (bottom
row) for triggers scheduled for specific phases.
The left set of plots shows trigger phase with respect to the estimated
signal phase, and the right set of plots shows trigger delay with respect
to the phase of the analytic signal computed from the band-pass-filtered
signal.
As each oscillatory burst only resulted in one triggering event per target
phase, the number of phase error samples was not sufficient to generate
statistically meaningful plots of phase error for specific target phases.
\fixme{Maybe try a synthetic dataset 10 times larger.}{Per-phase phase error}

\fixme{Need monkey nocal trigger plots.}{Phase trigger plots.}

From Figure \ref{fig-validation-trig-phaseband}, the distribution of
phase-scheduled trigger error with respect to the estimated signal has a
narrow peak over a broad ``noise floor''. This produces unusual statistics:
the full-width half-maximum is one phase bin ($2^\circ$), but the
inter-quartile range varies from $10^\circ$ to $30^\circ$. A larger sample
size would be needed to identify the distribution (and hence source) of the
noise floor in this plot.
The distribution of phase-scheduled trigger error with respect
to the band-pass-filtered signal shows a broader peak which again has a
substantial ``noise floor'': the full-width half-maximum varies from
$20^\circ$ to $40^\circ$, with an inter-quartile range of approximately
$50^\circ$. These values are broadly consistent with the expectation that the
dominant source of trigger phase error is the phase estimation error
described in Section \ref{sect-validation-feat} (approx. $20^\circ$ FWHM),
and continue to meet the design requirement of $\le 60^\circ$~FWHM from
Section \ref{sect-background-ephys}.

\subsection{Delay Compensation for Infinite Impulse Response Filters}
\label{sect-validation-cal}

\begin{figure}[!t]
\centering
\includegraphics[width=0.95\columnwidth]
{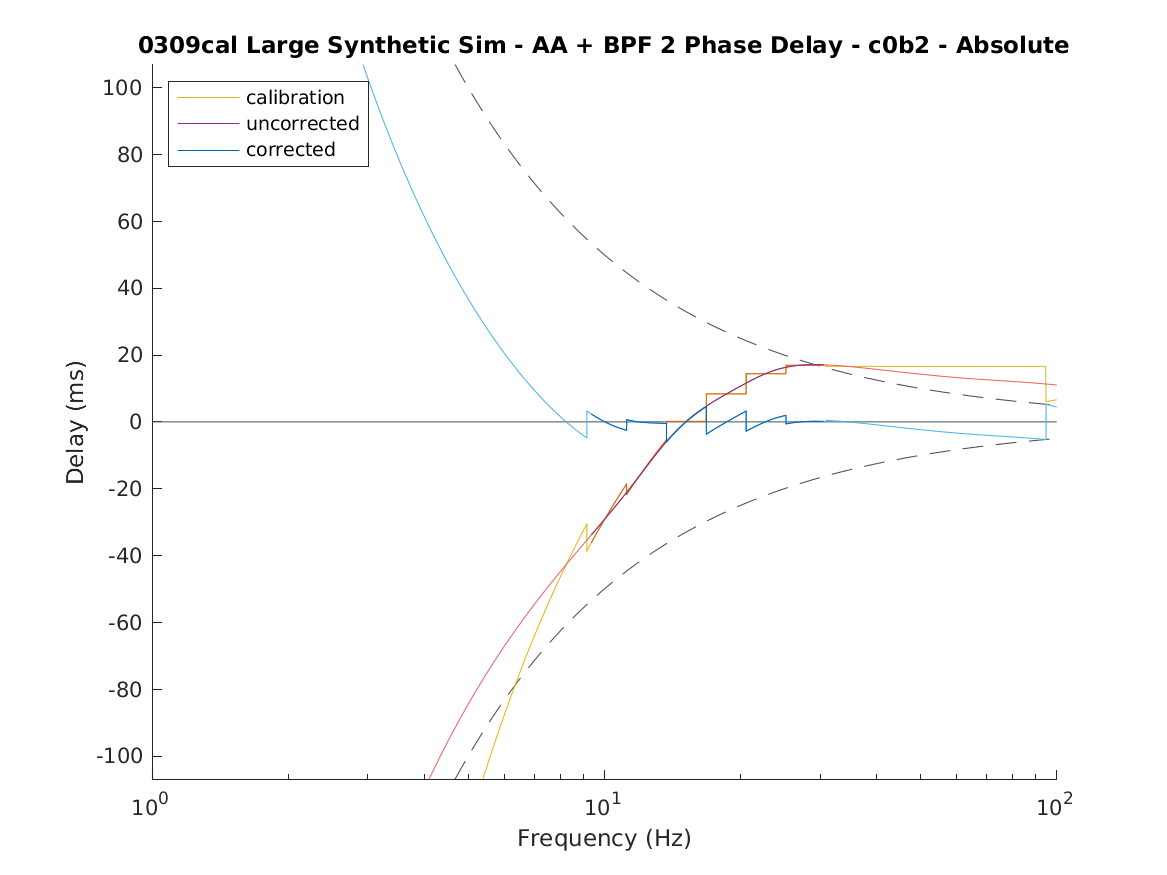}
\caption{Phase delay calibration for the anti-aliasing IIR filter followed
by the beta band IIR filter. The real phase delay (red curve) is
approximated by a lookup table of calibration delays (step-wise curve). The
resulting expected delay error after calibration is shown by the blue curve.}
\label{fig-validation-cal-iircal}
\end{figure}

All causal filters introduce delay into the filtered signal. For FIR filters,
this delay is constant, and for IIR filters, different frequency components
are delayed by different amounts. To allow later processing stages to
compensate for this, a calibration table of phase delay vs period was built.
Figure \ref{fig-validation-cal-iircal} shows an example of the lookup table
delay (step-wise curve), actual delay (red curve), and delay error after
calibration (blue curve) for beta-band output after processing by the
anti-aliasing IIR filter followed by the beta-band band-pass IIR filter
(from Figure \ref{fig-validation-filt-iirexample}). A calibration table for
a FIR filter would have only a single entry.

An ideal ``zero-shift'' version of the band-pass-filtered signal was created
for each band. This signal was computed by using the gain component
$G(\omega)$ of the filter's transfer function (from Equation
\ref{eq-validation-filt-xfer}) as a non-causal filter to transform the
wideband signal into a band-pass signal with zero time-shift (Equation
\ref{eq-validation-cal-zerofilt}). As the phase
shift of this filter is zero at all frequencies, the phase delay and group
delay of the filter are also zero. Time shift from the hardware anti-aliasing
filter, software anti-aliasing filter, and software band-pass filters can be
compensated in this manner.

\begin{equation}
\label{eq-validation-cal-zerofilt}
\begin{split}
Y_0(\omega) &= X(\omega) \cdot G(\omega) \\
y_0(t) &= \mathcal{F}^{-1}\left\{G(\omega) \cdot \mathcal{F}\{x(t)\}\right\}
\end{split}
\end{equation}

\begin{figure}[!t]
\centering
\begin{tabular}{cc}
\includegraphics[width=0.45\columnwidth]
{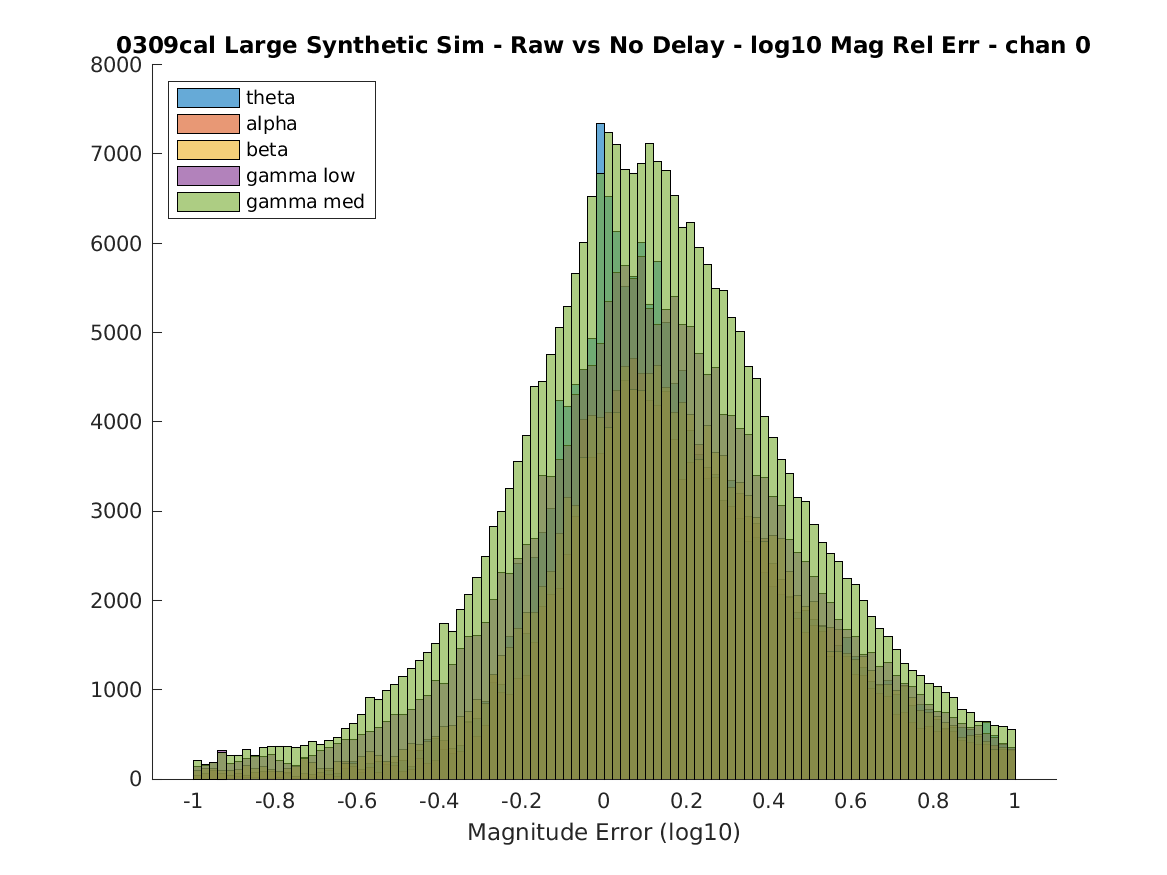}
&
\includegraphics[width=0.45\columnwidth]
{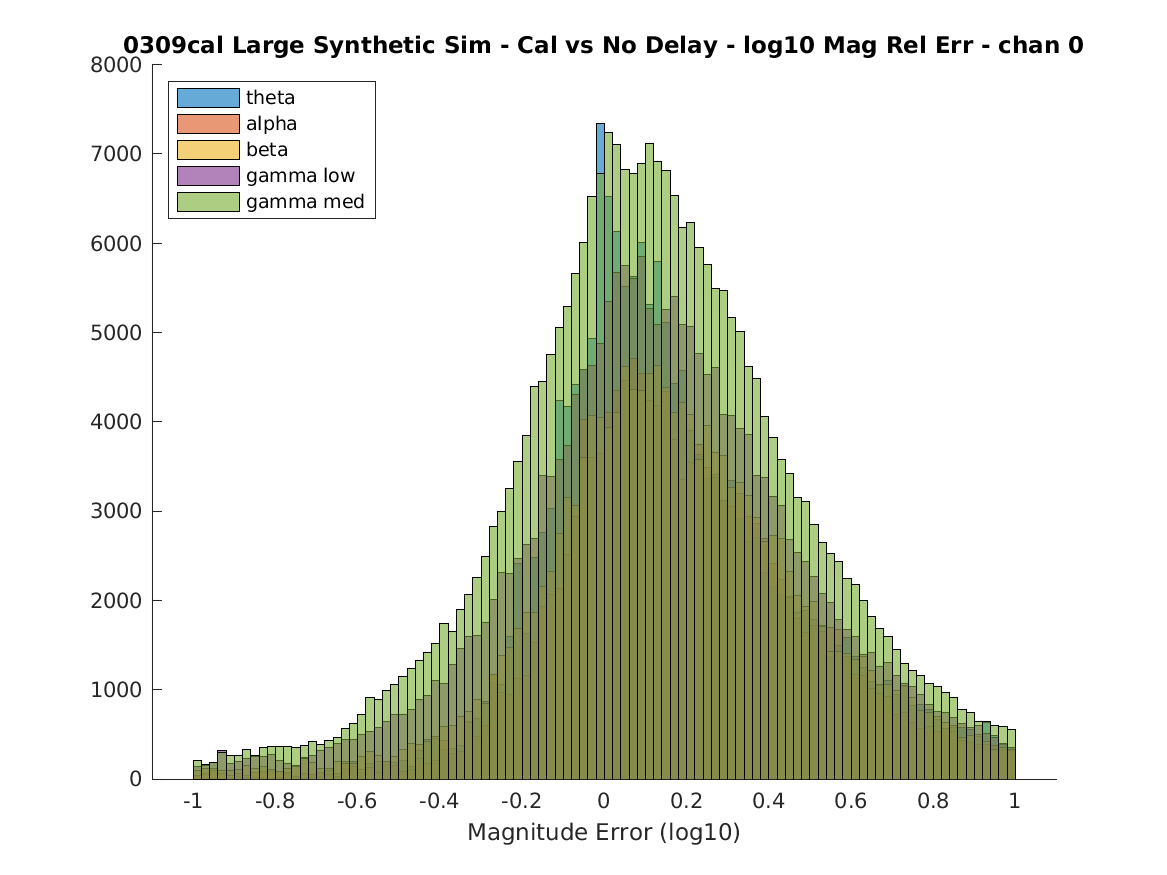}
\\
\includegraphics[width=0.45\columnwidth]
{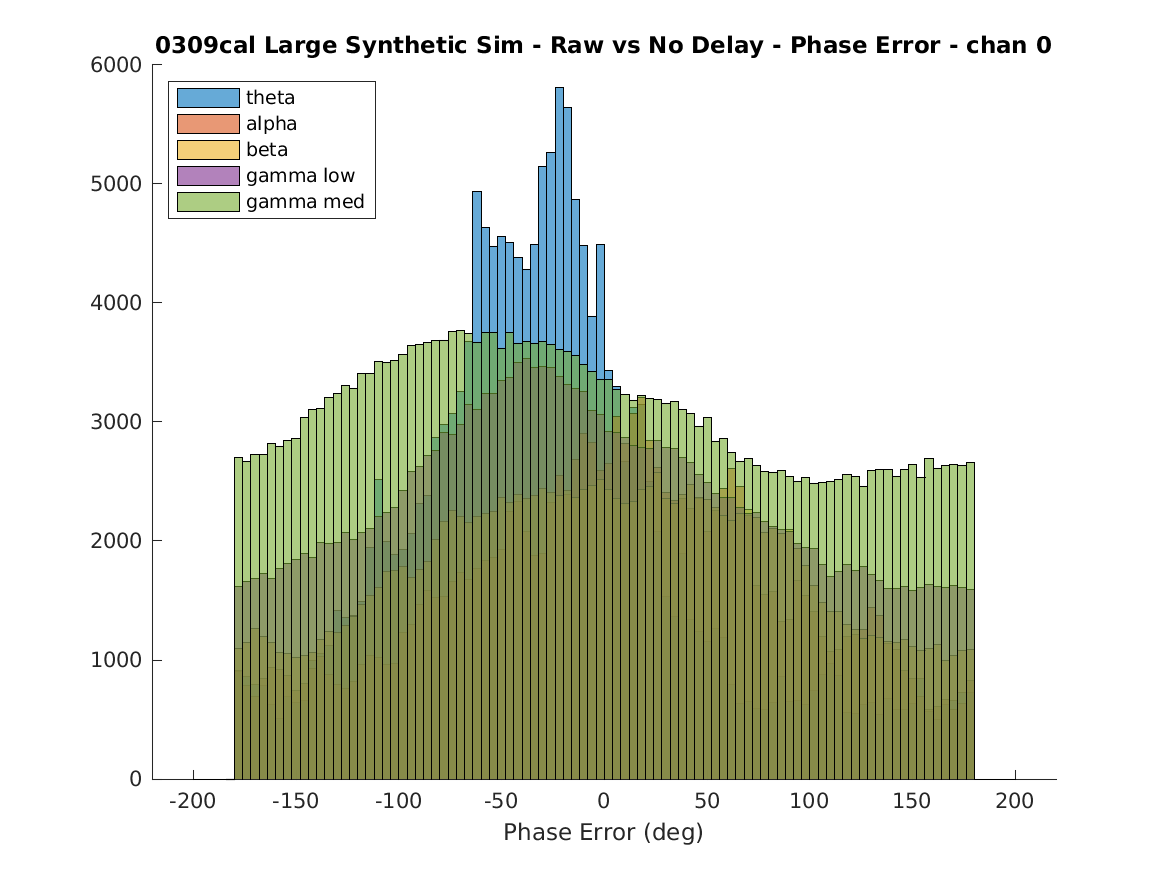}
&
\includegraphics[width=0.45\columnwidth]
{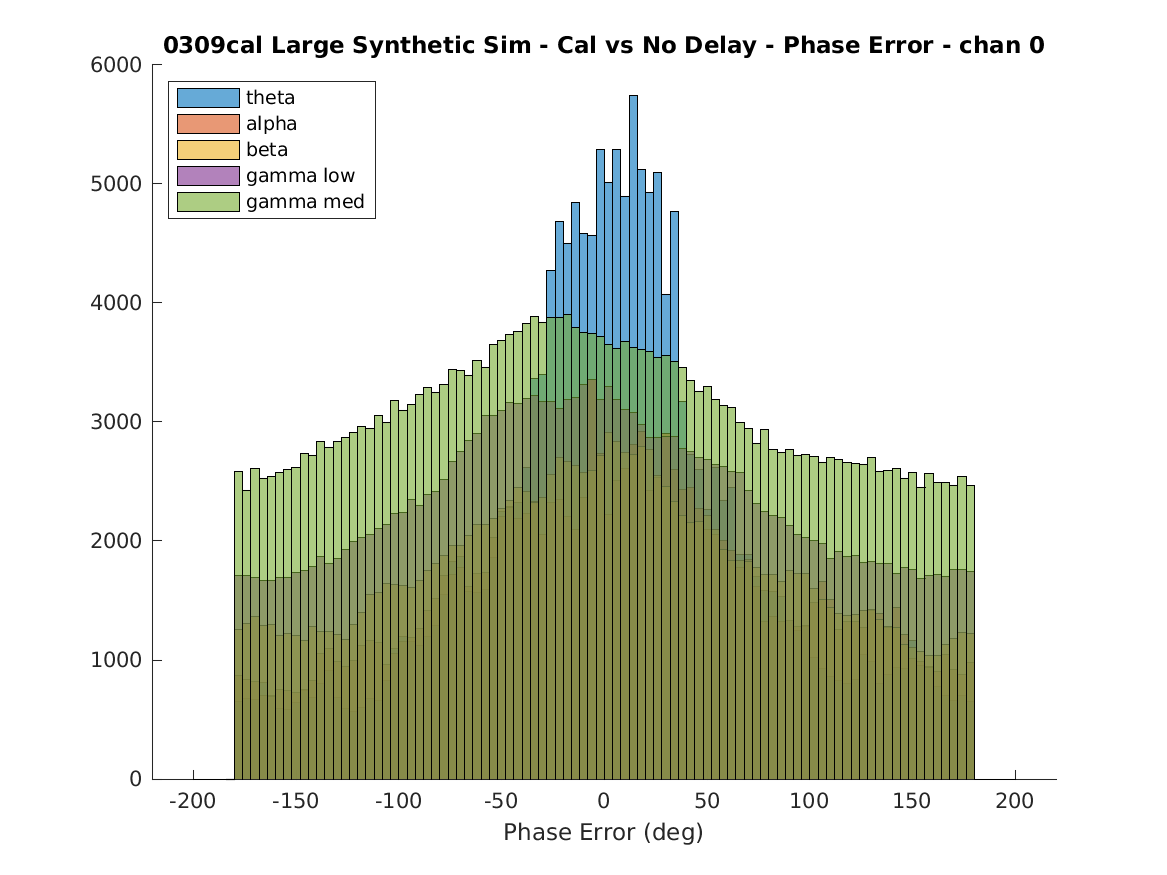}
\\
\includegraphics[width=0.45\columnwidth]
{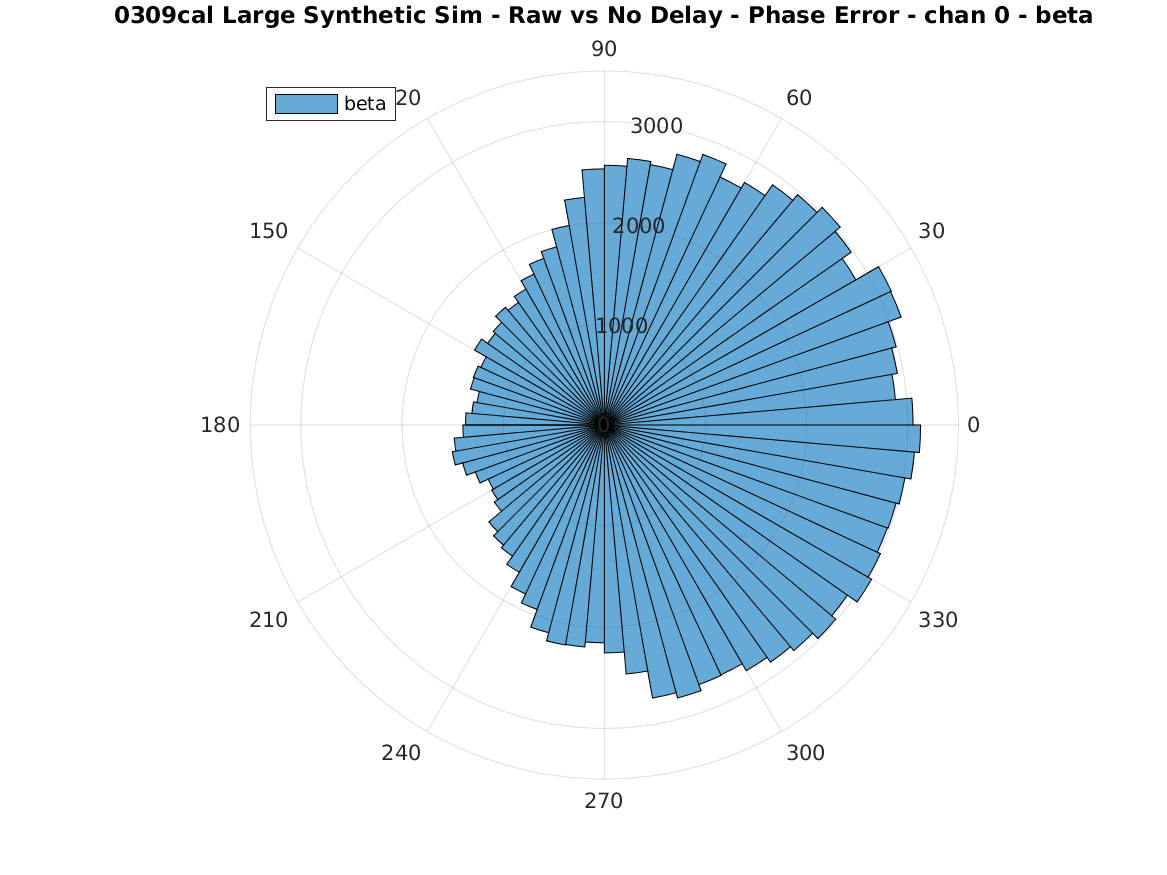}
&
\includegraphics[width=0.45\columnwidth]
{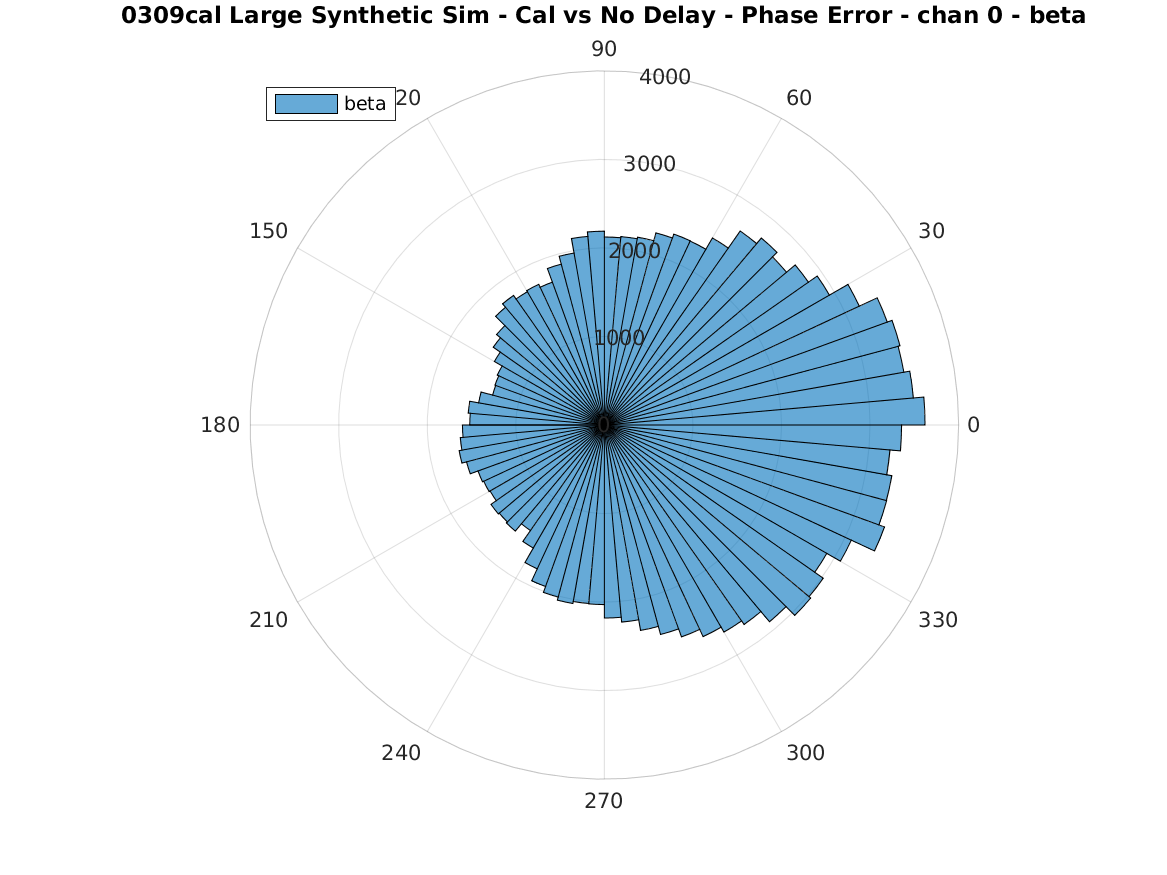}
\\
\end{tabular}
\caption{Normalized estimated magnitude error (top), absolute estimated
phase error (middle), and beta band estimated phase error (bottom) for raw
estimated magnitude and phase (left) and calibrated estimated magnitude and
phase (right). Error was measured with respect to the ``zero time shift''
band-pass-filtered signal. IIR filters, ``synthetic'' dataset.}
\label{fig-validation-cal-magphase}
\end{figure}

%
Estimated magnitude and phase after delay calibration were compared to
those of the ``zero-shift'' band-pass-filtered signal using the same
approach as in Section \ref{sect-validation-feat}.
Figure \ref{fig-validation-cal-magphase} shows histograms and box plots of
raw estimates (left) and calibrated estimates (right) of magnitude and
phase with respect to the ``zero-shift'' signal. Magnitude error (top row)
was normalized with respect to analytic signal magnitude (relative error),
and phase error (middle row) was taken as the difference between estimated
phase and the analytic signal phase. The bottom row shows a representative
rose plot of phase estimates (beta band). This analysis was performed using
the ``synthetic'' dataset and IIR filters.

The magnitude estimate is not affected by calibration, so the magnitude
estimation error is unchanged between the ``raw'' and ``calibrated''
columns in Figure \ref{fig-validation-cal-magphase}. The magnitude error
distribution is much broader when measured with respect to the ``zero-shift''
signal that when measured with respect to the delayed band-pass-filtered
signal. This can be understood as a consequence of the observation that
the signal envelope varies on a timescale comparable to the oscillation
period: with the filter's time delay exposed, magnitude has more time to
change. This results in somewhat larger variation (approx. $\pm 60\%$).

The phase error distribution measured with respect to the ``zero-shift''
signal shows three significant properties. First, aside from the distribution
in the theta band, none of the error distributions shows a sharp peak;
instead the distribution has a broad hill over a high uniform background.
This makes it difficult to meaningfully extract peak width statistics.
Second, the location of the broad peak does shift with calibration, having
significant bias before calibration and being zero-averaged afterwards.
Third, the shape of the rose plot changes substantially during calibration,
narrowing for the beta band and lower frequencies.
The primary conclusion to be drawn is that there is a large additional
source of phase estimation noise introduced by the IIR band-pass filters
that is not adequately compensated by delay calibration. This problem is
worse in the ``biological'' dataset.

%
There are several potential sources for phase estimation error. These can
broadly be classified into sources that involve rapid perturbation of
phase or frequency (due to non-stationary frequency or due to noise),
sources that cause the oscillating signal to cease to be a pure tone (such
as harmonic overtones or other strong out-of-band tones), and sources that
involve neither of these but that cause the frequency of the oscillation to
be mis-estimated, resulting in an incorrect delay calibration lookup value.
Sources of this last type can readily be investigated and might be corrected.

Mis-detection of oscillation frequency can be evaluated by plotting a
histogram of estimated vs. actual instantaneous frequency. Representative
plots showing heat-maps of estimated frequency vs analytic signal
frequency in the beta band are shown in Figure
\ref{fig-validation-cal-iirfreqfreq}. The frequency band is indicated by the
white lines. One histogram bin width represents an approximately 5\% change
in frequency. These plots confirm that for both ``biological'' and
``synthetic'' datasets a large number of samples have mis-estimated
frequency, which will lead to incorrect delay compensation.

\begin{figure}[!t]
\centering
\begin{tabular}{cc}
\includegraphics[width=0.45\columnwidth]
{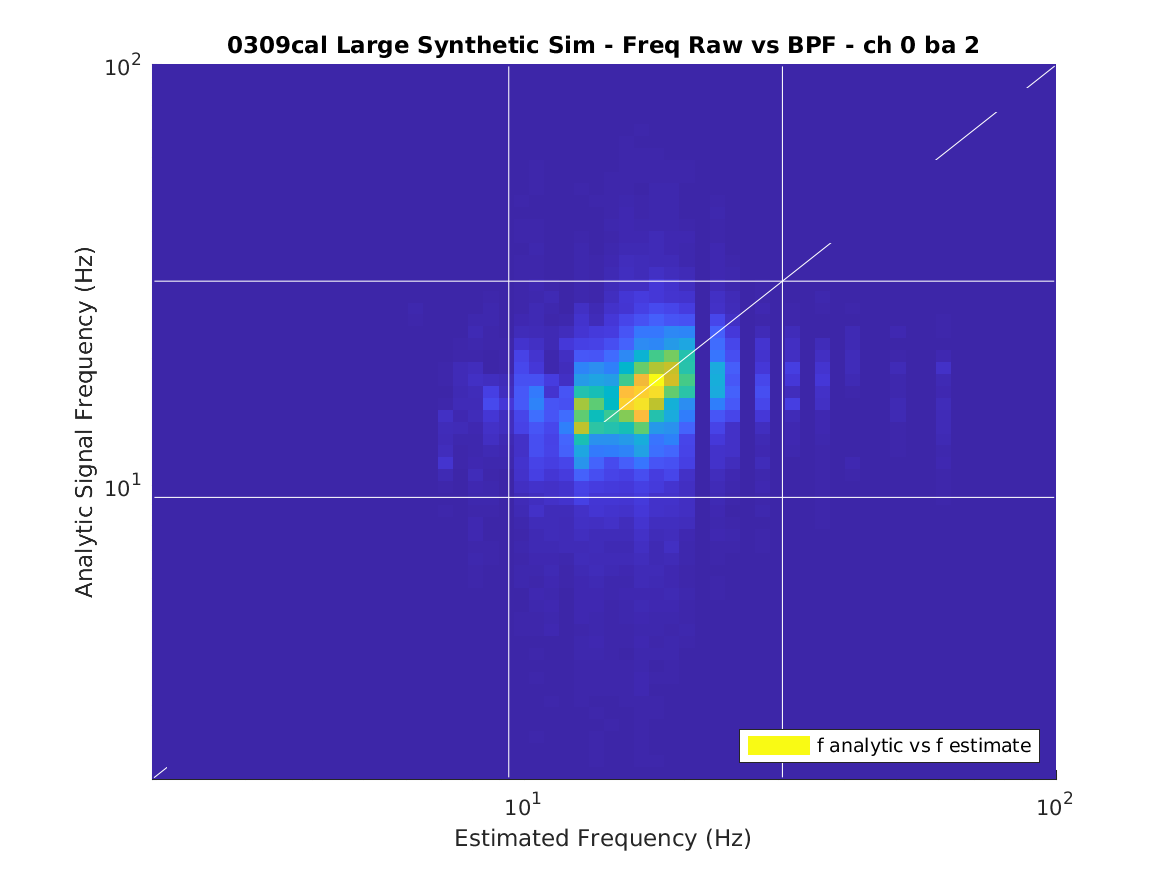}
&
\includegraphics[width=0.45\columnwidth]
{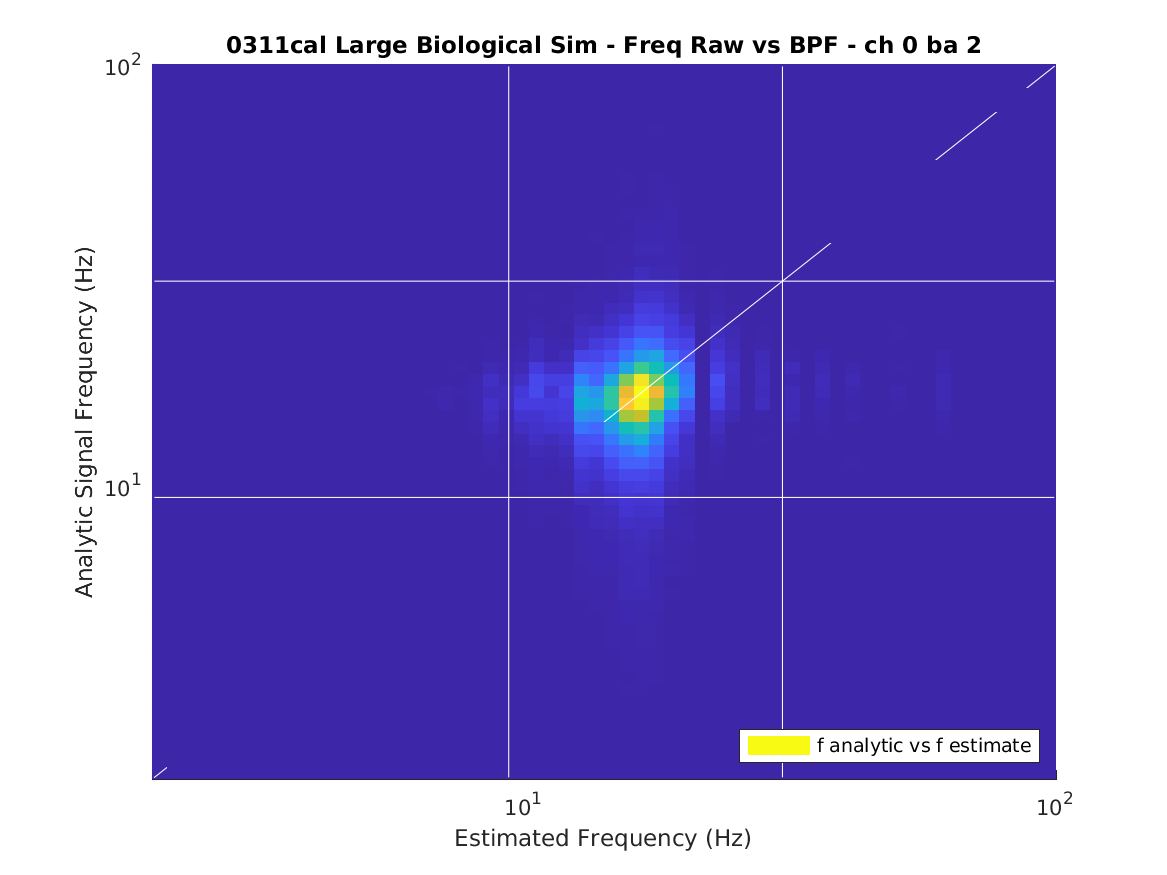}
\\
\includegraphics[width=0.45\columnwidth]
{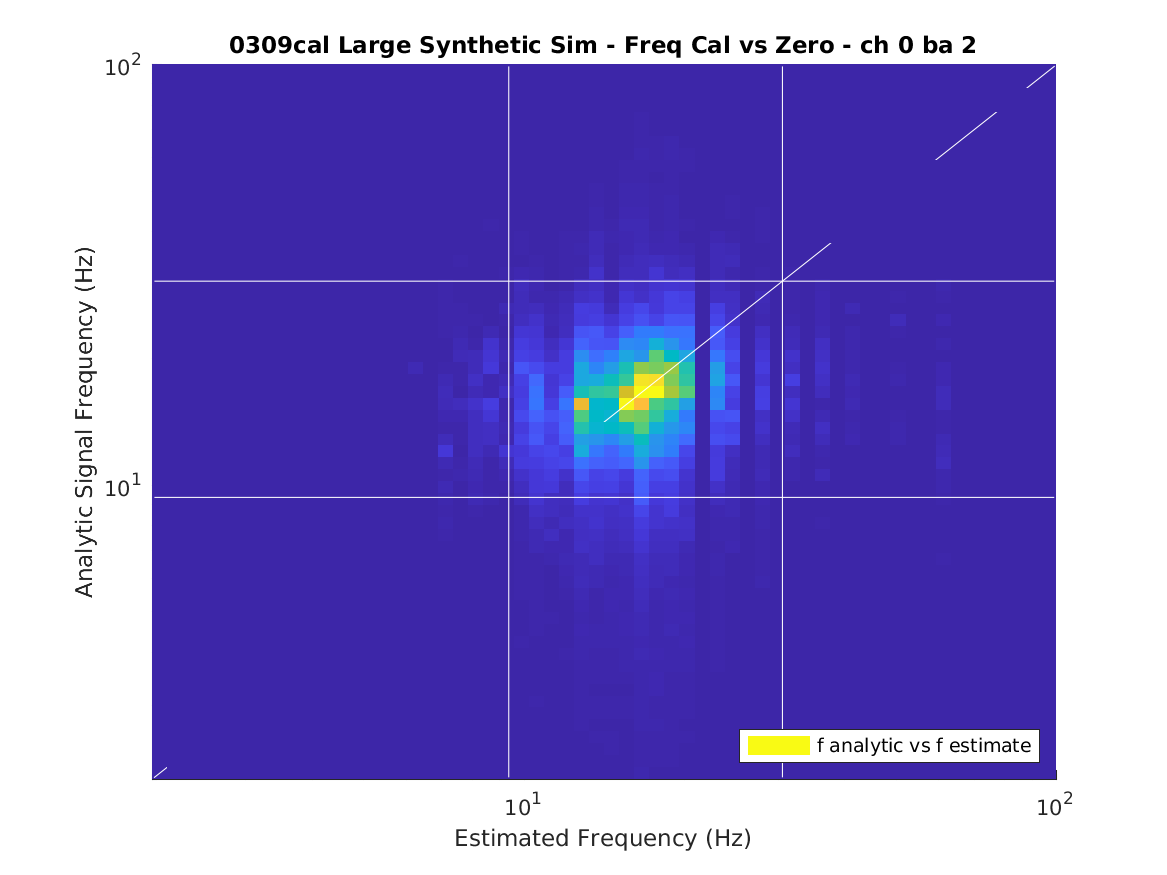}
&
\includegraphics[width=0.45\columnwidth]
{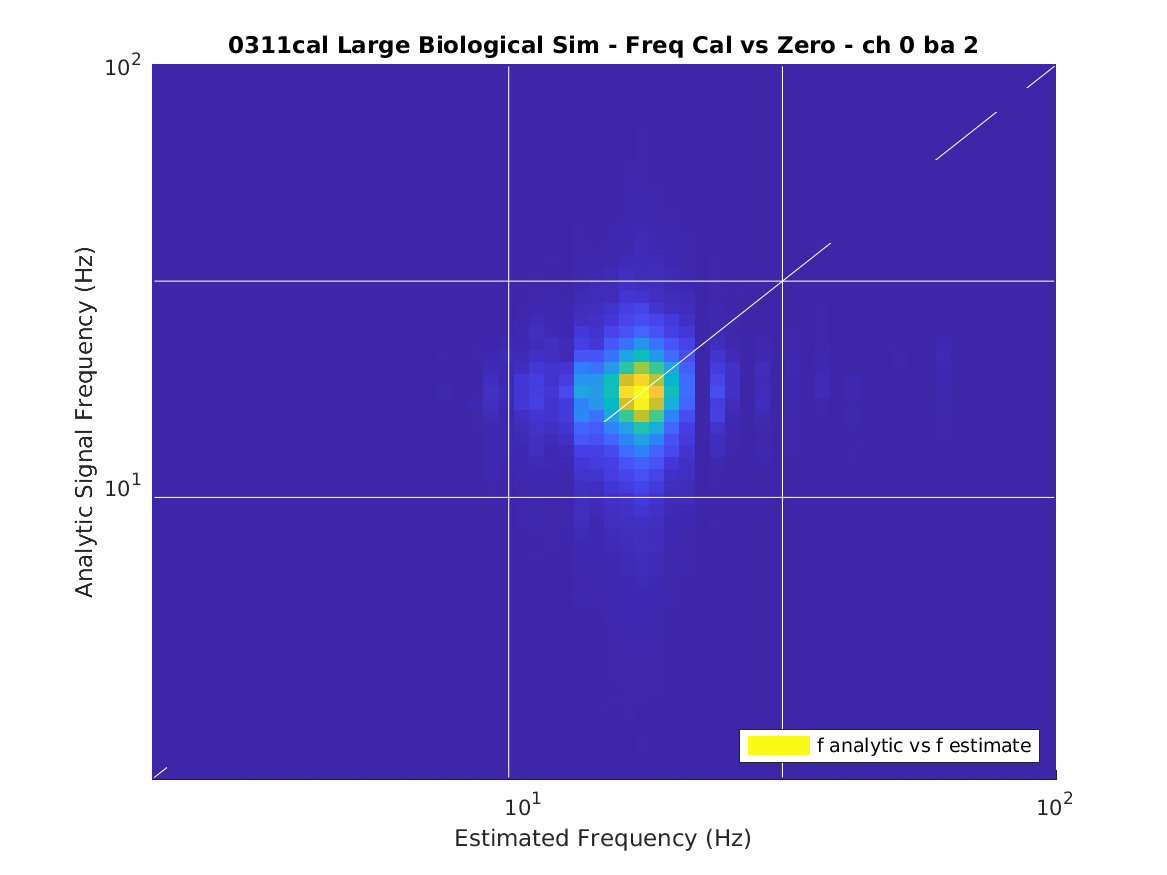}
\\
\end{tabular}
\caption{Heat map of estimated frequency vs analytic signal instantaneous
frequency, using the ``synthetic'' dataset (left) and the ``biological''
dataset (right). The top row shows the raw frequency estimate and the
delayed band-pass-filtered signal's instantaneous frequency, and the bottom
row shows the calibrated frequency estimate and the ``zero time shift''
band-pass-filtered signal's instantaneous frequency.
One bin width represents an approximately 5\% change in frequency.
IIR filters, beta band.}
\label{fig-validation-cal-iirfreqxy}
\end{figure}

\begin{figure}[!t]
\centering
\includegraphics[width=0.95\columnwidth]
{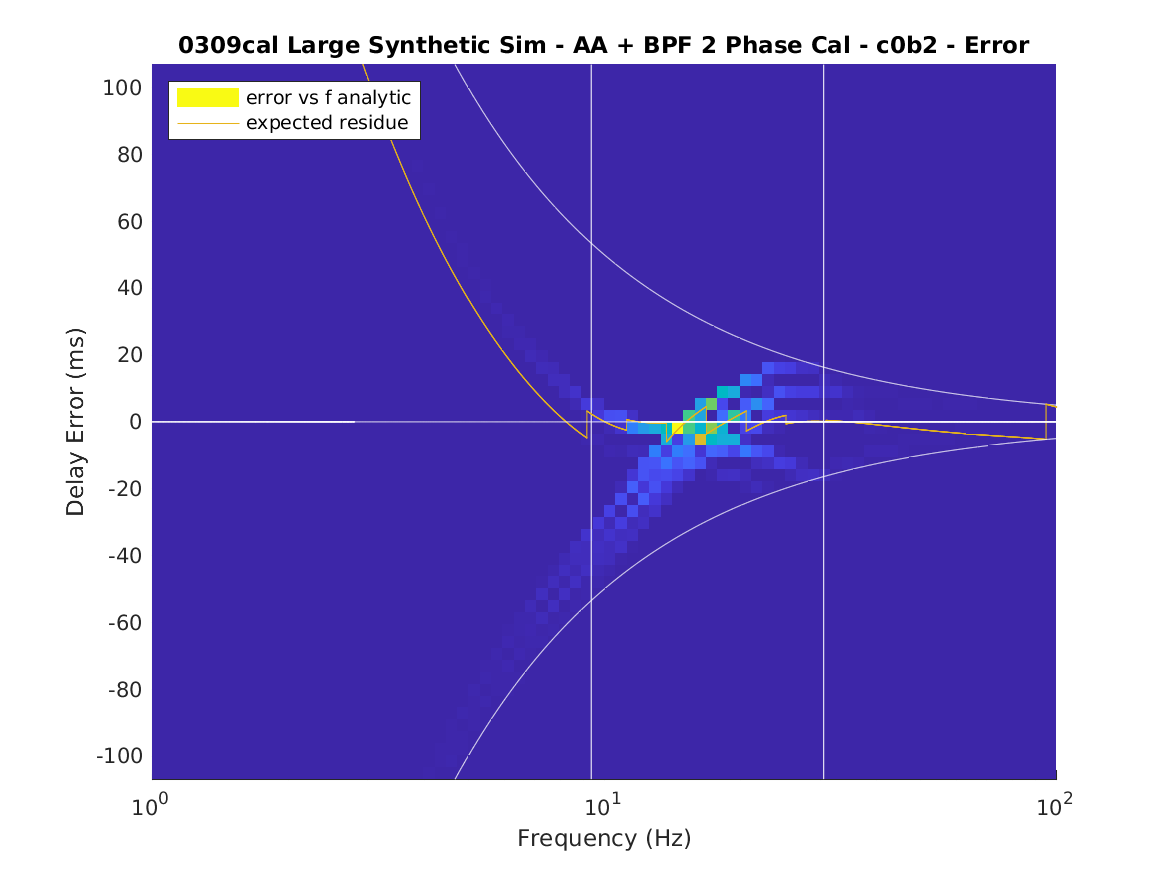}
\caption{Phase delay residue after compensating for the anti-aliasing filter
and the beta band IIR filter. The yellow curve is the
expected estimation error, the white vertical bars indicate the pass-band
edges, the white curves show the delay corresponding to $\pm$ half a period,
and the heat map shows the distribution of phase delay residue versus the
analytic signal's instantaneous frequency.
Beta band IIR filter, ``synthetic'' dataset.}
\label{fig-validation-cal-iircalerror}
\end{figure}

The effects of poorly-estimated frequency can be evaluated by plotting a
two-dimensional histogram of calibration delay applied vs the analytic signal
frequency for each sample during oscillation events. For clarity, this may
instead be plotted as the residue after the calibration delay is subtracted
from the known phase delay at that instantaneous frequency; the resulting
plot will be a phase delay error plot, with a desired value of zero. A
representative plot showing a heat-map of phase delay compensation error
is shown in Figure \ref{fig-validation-cal-iircalerror} (beta band IIR
filter, ``synthetic'' dataset).
The saw-tooth curve
in the plot represents the calibration lookup table bin size. This is
narrower than the frequency smearing distance in the plot, indicating that
bin quantization is not a significant source of error. Frequency smearing
covers a substantial fraction of the frequency band, causing phase delay
compensation errors that are large compared to the scale of the residue
expected after phase delay compensation.

\begin{figure}[!t]
\centering
\begin{tabular}{cc}
\includegraphics[width=0.45\columnwidth]
{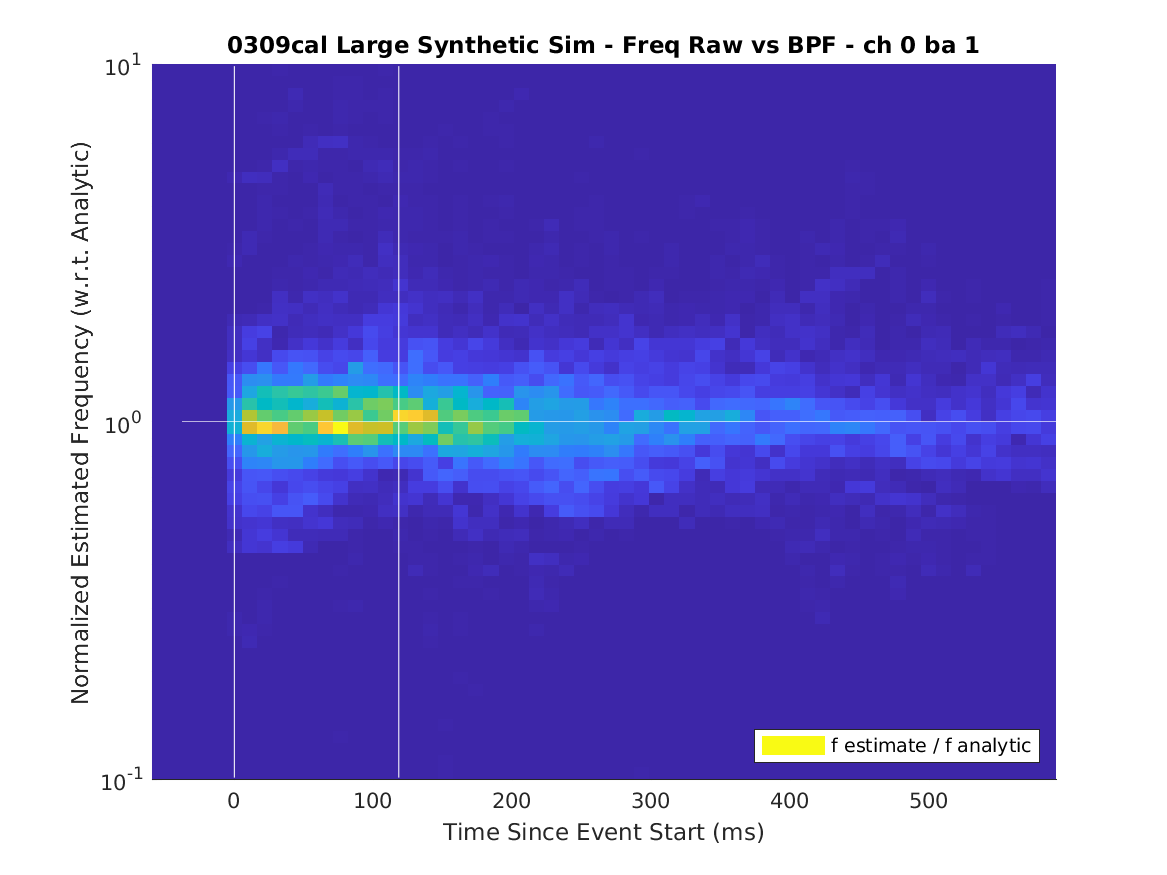}
&
\includegraphics[width=0.45\columnwidth]
{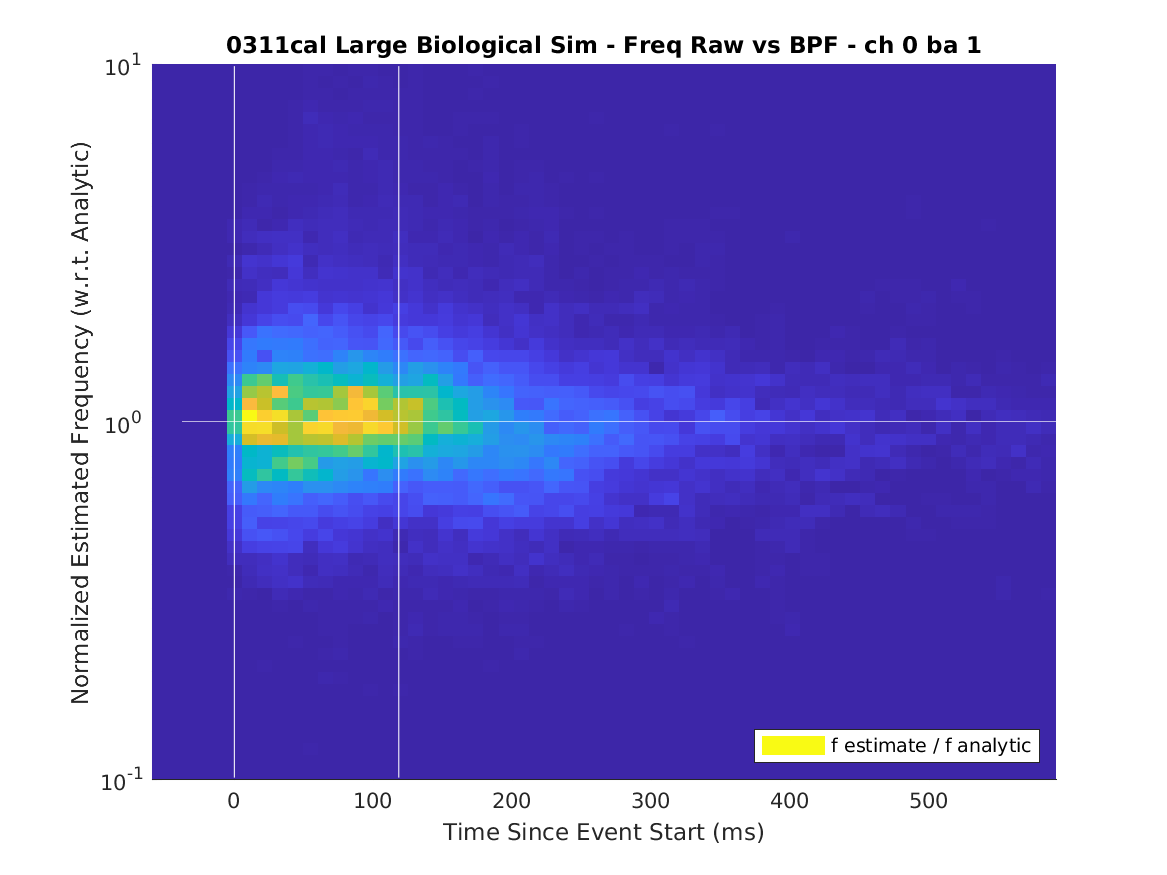}
\\
\includegraphics[width=0.45\columnwidth]
{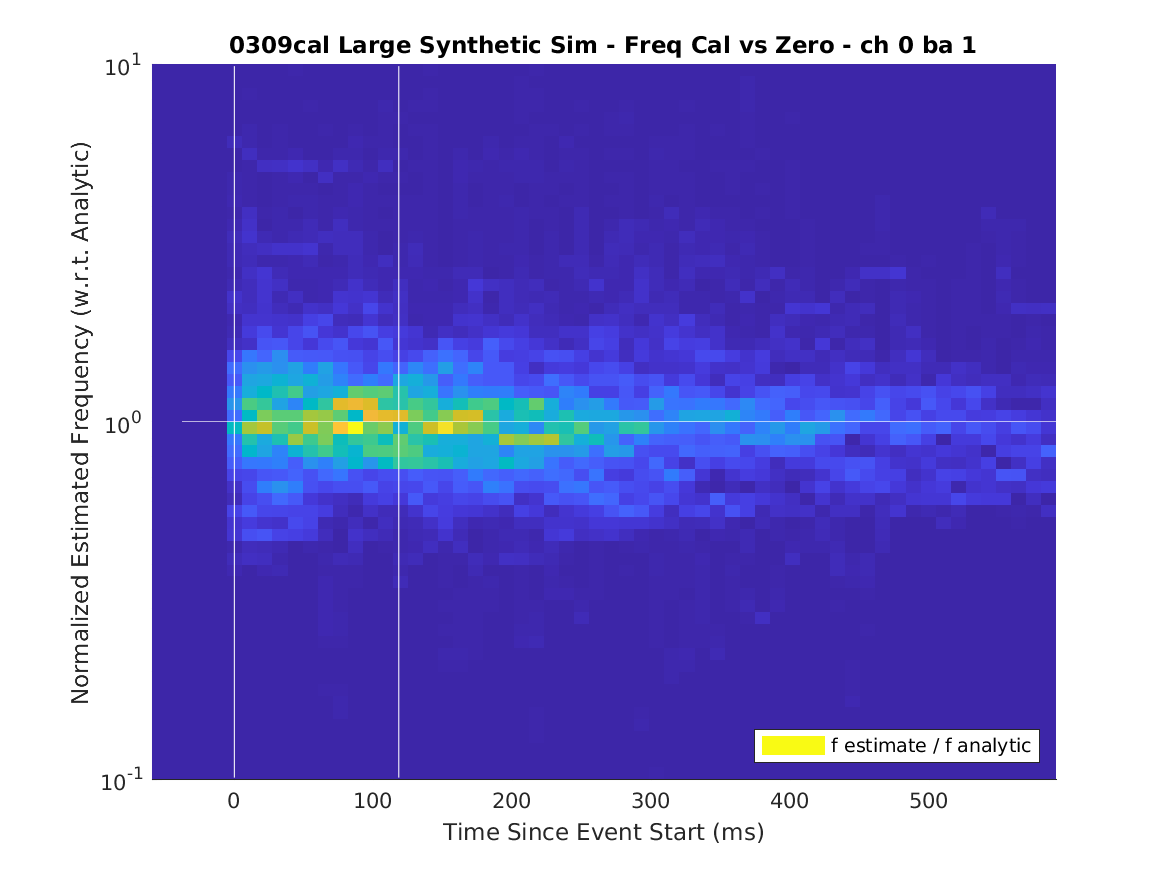}
&
\includegraphics[width=0.45\columnwidth]
{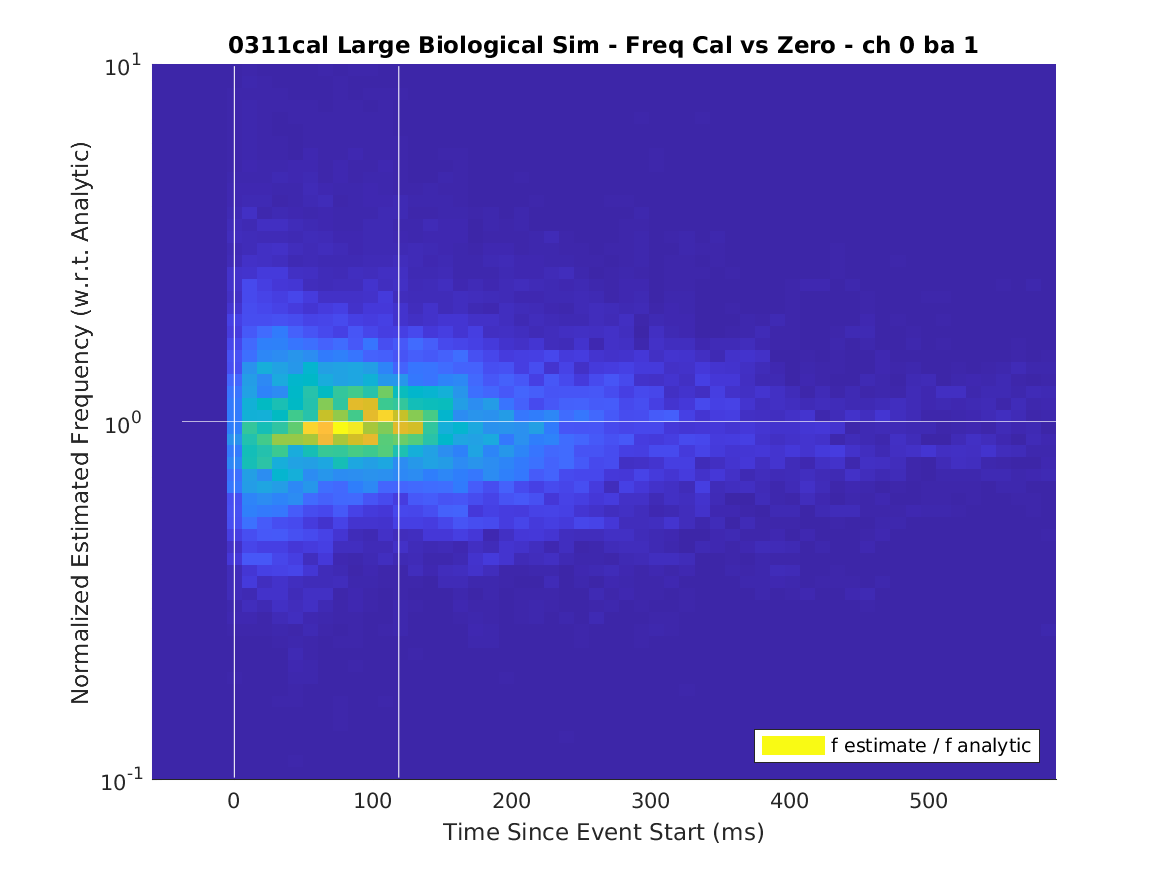}
\\
\end{tabular}
\caption{Heat map of estimated frequency error over time, using the
``synthetic'' dataset (left) and the ``biological'' dataset (right). The
top row shows the raw frequency estimate normalized to the delayed
band-pass-filtered signal's instantaneous frequency, and the bottom row
shows the calibrated frequency estimate normalized to the ``zero time shift''
band-pass-filtered signal's instantaneous frequency.
IIR filters, alpha band.}
\label{fig-validation-cal-iirfreqtime}
\end{figure}

One source of potentially incorrect frequency estimates is a bad estimate
of oscillation frequency during the first half-period of the oscillation. As
described in Section \ref{sect-design}, the frequency estimate is delayed
by up to half an oscillation period. When an oscillation is first detected,
the frequency estimate is from a time interval prior to the oscillation's
magnitude excursion, when the signal-to-noise ratio may have been poor.
This can be evaluated by plotting a two-dimensional histogram of estimated
frequency error over time. Representative plots showing heat-maps of
normalized estimated frequency (relative error) for the alpha band are shown
in Figure \ref{fig-validation-cal-iirfreqtime}. While there is a substantial
amount of mis-detection at all times, the alpha band and other low-frequency
bands do have more error during the first half-period of the oscillation in
both the ``synthetic'' and ``biological'' traces.
\fixme{Evaluate this more rigorously.}{Frequency detection error vs time.}
\fixme{Shuffle this again after implementing a fix.}{Waiting a half-period.}

%
%

%% file: burst-box-sigproc-conc.tex
%
\section{Conclusion}
\label{sect-conc}

\fixme{Distinguish monkey vs synthetic results.}{Conclusion data.}
\fixme{Describe per-band results.}{Conclusion data.}
\fixme{Need updated burst box results.}{Conclusion data.}

A modular, scalable signal processing framework has been presented that is
capable of detecting and characterizing oscillations on the local field
potential of neural signals, and of generating trigger signals to allow
phase-aligned and delay-aligned stimulation to be performed.
As a case study, this framework was used to prototype an ``off-line''
workstation-based stimulation controller and an ``on-line''
microcontroller-based stimulation controller.

The workstation-based ``off-line'' implementation
was used to validate the performance of the oscillation detection and
stimulation control architecture. Phase could be estimated within
$\frac{3}{4}$~period of oscillation onset, with an error distribution FWHM
of $20^\circ$ for synthetic data and $30^\circ$ for biological data, with
respect to the instantaneous frequency of the band-pass-filtered signal.

Stimulation pulses with timing specified relative
to zero-crossings in the signal waveform could be generated with
an error distribution FWHM of 2~ms with respect to the band-pass-filtered
signal.
Stimulation pulses timed to specific phases in the
signal waveform could be generated with an error distribution FWHM of
$40^\circ$ with respect to the band-pass-filtered signal. This meets the
minimum requirements for closed-loop experiments studying phase-aligned
stimulation.

The microcontroller-based ``on-line'' implementation was used to demonstrate
that a usable oscillation detector and stimulation controller could be built
with severely constrained hardware resources, making the case for
application of this framework to systems with thousands of channels and/or
hundreds of bands on readily-available hardware. A microcontroller-based
implementation was presented that duplicated the functionality of the 
`off-line'' implementation in real-time with a single channel, a single
band, and using infinite impulse response filters. This was done with an
estimated processing budget of 100~kMAC/sec, compared to an estimated
processing budget of at least 3~MMAC/sec$\cdot$channel provided by
readily available user-configurable electrophysiology equipment.

In conclusion, the signal processing framework presented was shown to be
sufficient for rapid prototyping of controllers for phase-aligned neural
stimulation and is readily adapted to large-scale implementations using
FPGA-based or DSP-based electrophysiology controllers.

%
%